\documentclass{aa}  

\usepackage{graphicx}
\usepackage{txfonts}
\usepackage{subcaption}
\usepackage{longtable}
\usepackage{lscape}
\usepackage{placeins}

\usepackage{xcolor}
\definecolor{blue} {rgb} {0.1,0.1,0.8}
\definecolor{green} {rgb} {0.0,0.6,0.0}
\definecolor{magenta} {rgb} {0.6,0.1,0.6}
 \renewcommand{\arraystretch}{1.15}
\begin{document}

   \title{Spatial distribution and clustering properties of the young stellar populations in the Carina Nebula complex and Car~OB1}

   \author{C. G{\"o}ppl
          \inst{1,2}
          \and
          T. Preibisch\inst{1}
          }

   \institute{Universit\"ats-Sternwarte M\"unchen, 
              Ludwig-Maximilians-Universit\"at,
              Scheinerstr.~1, 81679 M\"unchen, Germany
              \and
              Excellence Cluster ORIGINS, Boltzmannstr.~2, 85748 Garching, Germany\\
              \email{cgoeppl@usm.uni-muenchen.de}
             }

\titlerunning{Young stellar clusters in the Carina Nebula complex and Car~OB1}
\authorrunning{G{\"o}ppl \& Preibisch}

   \date{Received 4 November 2024; accepted 16 January 2025}

  \abstract
   {}
   {We use \textit{Gaia}~DR3 astrometry and photometry to analyze the spatial distribution of the young stellar populations and stellar clusters and to search for new OB star candidates in the Carina Nebula complex and the full extent ($\sim 5\degr$, corresponding to $\;\sim 200$\;pc) of the Car~OB1 association.
   }
   {We first performed a new census of high-mass stars in Car~OB1 and compiled a comprehensive 
   catalog of 517 stars with known spectral types (128 O-type, WR, and supergiant stars, and 389 B-type stars) that have \textit{Gaia}~DR3 parallaxes consistent with membership in the association. We applied the clustering algorithm DBSCAN on the \textit{Gaia}~DR3 data of the region to find stellar clusters, determine their distances and kinematics, and estimate ages. 
   We also used \textit{Gaia} astrometry and the additional \texttt{astrophysical\_parameters} table to perform a spatially unbiased search for further high-mass members of Car~OB1 over the full area of the association. 
   }
   {Our DBSCAN analysis finds 15 stellar clusters and groups in Car~OB1, four of which were not known before. Most clusters (80\%) show signs of expansion or contraction, four of them with a $\ge 2\sigma$ significance. We find a global expansion of the Car~OB1 association with a velocity of $v_{\rm out} = 5.25\pm0.02~\rm km\,s^{-1}$. A kinematic traceback of the high-mass stars shows that the spatial extent of the association was at a minimum 3--4~Myr ago. Using astrophysical parameters by \textit{Gaia}~DR3, we identified 15 new O-type and 589 new B-type star candidates in Car~OB1. The majority ($\gtrsim 54\%$) of the high-mass stars constitute a non-clustered distributed stellar population. Based on our sample of high-mass stars, we estimate a total stellar population of at least $\sim 8\times10^4$ stars in Car~OB1. 
   }
   {Our study is the first systematic astrometric analysis that covers the full spatial extent of the Car~OB1 association, and it therefore substantially increases the knowledge of the distributed stellar population and spatial evolution of the entire association. Our results suggest suggests Car~OB1 to be the most massive known star-forming complex in our Galaxy.} 
   
\maketitle

\section{Introduction}

Most stars form in large complexes consisting of giant molecular clouds, open clusters,
and OB associations \citep[see, e.g.,][for a recent review]{2023ASPC..534..129W}.
The spatial distribution of the young stellar populations in these complexes
is often highly substructured. The young stars concentrated in 
open clusters, which are often prominently featured in optical or infrared images, are rather 
easy to identify,
but the population of young stars in a
``widely distributed'' configuration with a low average stellar space density
is much more difficult to identify and study.

The relation between the clusters and the distributed population
 bears important
implications about the formation and evolution of OB associations
\citep[see, e.g.,][]{2020MNRAS.495..663W}.
The spatial configuration of the stellar populations also sets the environment in which
stars form and their protoplanetary disks evolve.
The disk evolution and planet formation processes can be strongly affected by
environmental influences such as photoevaporation driven by
external UV irradiation from nearby O-type stars \citep[e.g.,][]{2022EPJP..137.1132W,2023ApJ...958L..30R}.
Even for models of the origin of the Solar System, the
spatial distribution of stellar populations in the birth environment of the Sun
plays an important role
\citep[e.g.,][]{2010ARA&A..48...47A,2023A&A...670A.105A}.

\begin{figure*} 
\includegraphics[width=18.0cm,trim=4.2cm 9.05cm 4.8cm 8.2cm, clip]{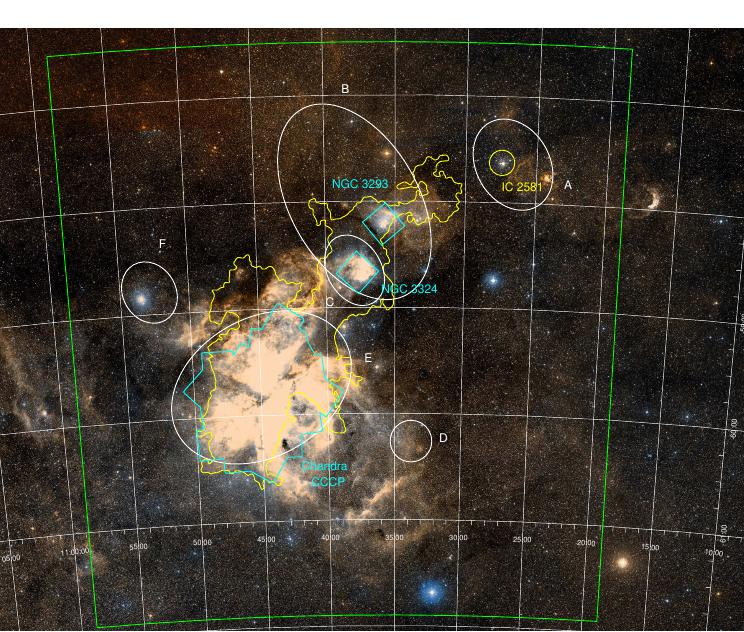}
\caption{DSS2 optical image of the area around the Carina Nebula, with outlines illustrating the spatial extent of the cloud complex (CNC, yellow) and the
subgroups of the Car~OB1 association.
North is up and east to the left. The white grid shows the J2000
celestial coordinates. 
The yellow line shows the boundary of the
cloud emission related to the CNC as traced by the contour
for a  surface brightness of  $\ge 7.4\;\rm mJy/(arcsec)^2$ in the \textit{Herschel} $160\;\mu \rm m$ map.
The X-ray survey region of the \textit{Chandra} Carina Complex
Project (CCCP) as well as the areas of the additional \textit{Chandra} observations of NGC~3324
and NGC~3293 are marked with cyan outlines and labeled, the cluster IC~2581 is highlighted in yellow and the white ellipses show the subgroups in Car~OB1 as defined in \citet{1995AstL...21...10M}. 
The area for our cluster search with DBSCAN is marked in green.}
\label{fig:Grid}
\end{figure*}

The Car~OB1 association \citep[see][]{1995AstL...21...10M} 
contains several hundred OB stars and
extends over $\approx 5$\;degrees on the sky, corresponding to a physical
diameter of  $\approx 200$\;pc (see Fig.~\ref{fig:Grid}).
The optically most prominent part of Car~OB1 is the  famous Carina Nebula \citep[NGC~3372, see, e.g.,][]{SB08},
a large \ion{H}{II} region that is located in a
giant molecular cloud complex \citep{2000ApJ...532L.145S,2016MNRAS.456.2406R}, which we refer to as 
the Carina Nebula complex (CNC, hereafter).
The CNC has a total cloud mass of $\approx 10^6\;M_\odot$ \citep{SB07,Preibisch12},
extends over $\approx 3$\;square-degrees on the sky,
and harbors more than 100 high-mass (i.e., O-type, WR, and supergiant) stars
\citep[e.g.,][]{Smith06,2023A&A...671A..20B}.

Our knowledge of the stellar populations in the CNC and 
Car~OB1 is still very
inhomogeneous. For the central area of the CNC, the young stellar
populations down to $\approx 1\;M_\odot$ were uncovered by
the X-ray imaging survey of the
{\it Chandra} Carina Complex Project \citep[CCCP; see][for an overview]{CCCP-intro}.
Outside the CCCP field, studies of the stellar populations are largely restricted
to patchy regions around a few prominent clusters.
The clusters 
Trumpler 14, 15, 16 in the central Carina Nebula,
NGC~3324 to the north, and NGC~3293 near the northwestern edge of the cloud complex,
have been well studied at optical, infrared 
\citep[e.g.,][]{Smith10b,CCCP-HAWKI,2017A&A...603A..81D,Povich19,2023JKAS...56...97H, 2024A&A...685A.100I},
and X-ray wavelengths \citep{CCCP-intro,2011ApJS..194...12W,2011ApJS..194...11W,Preibisch14,2017A&A...605A..85P}.
Analysis of the \textit{Gaia}~DR3 data for the known OB stars and the X-ray selected stars in the CNC
has yielded a common and well-defined distance of $(2.35\pm 0.05)$\;kpc 
 \citep{2022A&A...660A..11G}.
From the above mentioned studies, ages ranging from 
$\approx\,$1--2\;Myr  (Tr~14 and NGC~3324), over $\approx\,$3--4\;Myr (Tr~16),
and up to $\approx\,$8--10\;Myr  (Tr~15 and NGC~3293) 
have been estimated.
In some parts of the CNC, such as the ``Southern Pillars'' region south of Tr~16
\citep[e.g.,][]{Smith10b} or the \ion{H}{II} bubble rim\footnote{including the 
``Cosmic Cliffs'' region imaged as part of the Webb Early Release Observations
\citep[see][]{2024AJ....168...63C}.}  
around NGC~3324 \citep{Ohlendorf13},
star formation is still ongoing.

The cluster IC~2581 (subgroup A of Car~OB1, at the northeastern edge of the association) is located outside the CNC and has
received much less attention so far. Its distance of $\simeq 2.4$~kpc and age of $\simeq 12$\;Myr derived from 
\textit{Gaia} data by \citet{2021A&A...647A..19T}, and confirmed in this study,
fits well within the range of values for the
large clusters in the CNC.

In addition to these prominent and rather well-studied clusters,
a variety of further smaller clusters and stellar groups has been found
in searches of selected regions in the CNC
\citep[e.g.,][]{2005AJ....129..888S,Smith10b,CCCP-Clusters,VISTA1,Tr16-SE-KMOS}.
However, most of these clusters and groups have been identified by visual inspection 
of various image data, and thus the resulting
cluster sample is very inhomogeneous.
No systematic and homogeneous search for clusters over the full spatial extent
of the CNC or Car~OB1 has been performed so far.

A further complication arises from the fact that the CNC is located directly on the galactic plane,
and near the tangent point of the Carina-Sagittarius spiral arm. The sky field around Car~OB1
therefore shows galactic clusters over a wide range of different distances. To mention just two examples,
the famous star cluster Westerlund~2 in the galactic background \citep[$D \approx 4.2 - 6$\;kpc, see][]{2015MNRAS.446.3797H,2018AJ....156..211Z} is seen just  next to Car~OB1 subgroup A, and
the foreground cluster vdBH~99
($D = 440$\;pc) is seen in front of the western part of the CNC \citep[see][]{2022A&A...660A..11G}.

The astrometric data from \textit{Gaia}, for the first time, now allow
for a systematic and unbiased search for clusters to be performed
over the full area of Car~OB1 ($\approx 5 \times 5$~square-degrees).
The approach of identifying clusters in the five-dimensional
[ position -- proper motion -- distance ] space is completely independent of the previous
cluster searches by visual image inspection.
This search also provides an objective basis for the distinction between the stellar populations 
in clusters and the distributed stellar population.
Furthermore, the spectral type information available for a subset of the stars in the \textit{Gaia} 
catalog can be used to search for further and yet unidentified high-mass stars over the 
full spatial extent of Car~OB1.

\bigskip

In this paper, we want to address the following questions:

\begin{enumerate}
\item How many known clusters and stellar groups (identified by visual inspection) can be confirmed as coherent stellar groupings, and how many of them are only chance-alignments of actually unrelated stars. \medskip

\item Which of the clusters seen in the Car~OB1 area are actually in Car~OB1 (i.e.,~at a distance of 2.35\;kpc), and which are in the galactic background. \medskip 

\item If we can find further clusters in Car~OB1. \medskip 

\item If we can find evidence for expansion of the individual clusters. \medskip 

\item If we can confirm and quantify the expansion of the whole OB association. \medskip 

\item If we can get more quantitative information about the distributed population of young stars, in particular how large this population (in comparison to the population of clustered stars) is, and how far it extends in comparison to the extent of the cloud complex and the currently known association subgroups. \medskip 

\item And if we can improve the census of high-mass stars in the CNC and Car~OB1 by using \textit{Gaia} data. 
\end{enumerate}

To clarify the designations used in this paper, we denote as ``Carina Nebula'' the central visually bright  nebulosity in Fig.~1 (diameter $\approx 2^\circ$),
we identify the ``CNC'' as the extended (diameter $\approx 3.7^\circ$) cloud complex as approximately traced by the far-infrared emission contour shown in Fig.~1, and we designate as ``Car OB1'' the area comprising the five OB subgroups marked in Fig.~1 (diameter $\approx 5^\circ$).

\section{A catalog of spectroscopically identified high-mass stars in Car OB1}
\label{sec:CarOB1Sample}

As an initial step of our analysis, we collected literature data in order to compile a comprehensive catalog of stars with 
spectroscopically determined spectral types in the area of Car~OB1, and checked their \textit{Gaia}~DR3 parallaxes for consistency of being members of Car~OB1. Our aim was to create a catalog
of high-mass stars that can serve as an update and an extension
(in search area) of the stellar census of high-mass stars in the 
Carina Nebula by \citet{Smith06}.
We therefore considered stars with 
spectral types O and B, Wolf-Rayet (WR) stars,  as well as supergiants.

For this, we combined the list of 241 previously known OB stars in the area of the CNC collected from the
literature in \citet{2022A&A...660A..11G} (comprising 80 O-type, 2 WR, 154 B-type stars, and 5 supergiants)
with 203 OB stars from \citet{2023A&A...671A..20B} (14 O-type, 1 LBV, 181 B-type stars, and 7 supergiants), 35 OB stars (3 O-type, 31 B-type stars, and 1 supergiant) from \citet{2016AJ....152..190A}, 82 stars (15 O-type, 41 B-type stars and 26 supergiants) from \citet{2020MNRAS.493.2339M}, 40 stars (38 B-type stars and 2 supergiants) from \citet{2017A&A...603A..81D}, 10 stars (9 B-type stars and 1 supergiant) from \citet{2019MNRAS.490..440L}, 2 M-type supergiants from \citet{1978ApJS...38..309H}, and 5 stars (3 B-type stars and 2 supergiants) from \citet{1995AstL...21...10M}. This resulted in a sample of  618 OB stars. The use of the word "star" here refers to resolved objects and does not differentiate between single stars and spectroscopic binaries. If both parts in a spectroscopic binary are taken into account (as in Sect.~\ref{sec:IMF}), it is explicitly mentioned.
For 617 of these stars we could identify a reliable \textit{Gaia} DR3 match (in almost all cases at a matching radius of $\leq 1\arcsec$), and 613 of these have a full astrometric solution in \textit{Gaia} DR3. 

As found in \citet{2022A&A...660A..11G}, the stellar population in the
CNC has a well-defined and common distance of 2.35\;kpc.
At this distance, the $\approx 5\degr$ angular diameter of Car~OB1
corresponds to a physical diameter of $\approx 200$\;pc.
We therefore define the interval of $\pm 100$\;pc around 2.35\;kpc
(i.e.,~from 2.25\;kpc to 2.45\;kpc) as the likely distance interval for
Car~OB1.
We exclude stars from the Car~OB1 sample if their $\pm 2\sigma$ distance uncertainty interval (calculated from their inverted parallax) is incompatible with the $[2.25\,, 2.45]$\;kpc distance interval for Car~OB1.
This selection leads to a final sample of 517 stars (including 22 spectroscopic binaries), which we denote as Car~OB1 high-mass star sample in the following. 
This sample includes 88 O-type, 3 WR, 36 supergiant stars, and 1 luminous blue variable ($\eta$ Car).
The complete list of these stars is provided in Table~\ref{tab:CarOB1HighMass} labeled with 'L' in the column 'Selection' (also available at the CDS).

\section{A systematic search for clusters over the full extent of Car OB1}

\subsection{Cluster search with DBSCAN}
\label{sec:FoundClusters}
\textit{Gaia} astrometry can be utilized for identification of star clusters, distance estimations, and kinematic and structural analysis of the found clusters. 
We used the algorithm DBSCAN \citep[Density Based Spatial Clustering of Applications with Noise,][]{10.5555/3001460.3001507} to search for clusters
in the five-dimensional space defined by position, 
proper motion, and parallax. The algorithm identifies star clusters based on their overdensity in the 5D-space compared to the field star density. It has two input parameters: $minPts$ and the density parameter $\epsilon$. DBSCAN assigns each source one of three designations: core point, non-core point, or noise point. Sources are classified as a core point if there are at least $minPts-1$ other points inside a hypersphere with radius $\epsilon$ around that source. Non-core points can reach at least one core point in a hypersphere with radius $\epsilon$, but not enough other sources to be considered a core point themselves. Noise points have no core points lying inside a hypersphere with radius $\epsilon$ centered around them.
Advantages of DBSCAN are that it can identify clusters with arbitrary shape and that the number of clusters does not need to be specified beforehand, but the results depend heavily on its input parameters, especially $\epsilon$.

We applied the method devised by \citet{2018A&A...618A..59C} to determine $\epsilon$.  
This method exploits the fact that cluster members have shorter nearest neighbor distances compared to field stars, which is then used to determine the density parameter $\epsilon$ that separates the cluster and field populations.

First the $k$th nearest neighbor distance (with $k = minPts -1$) for all stars in the sample is calculated and the minimum saved as $\epsilon_{kNN}$. Then the sources are resampled using a Gaussian kernel density estimator that creates a new sample with the same amount of stars that mimics 
a field without any clusters and only field stars. Then the $k$th nearest neighbor distance is again computed and the minimum saved. Since each resampling leads to a different realization, this step is carried out thirty times in order to minimize the impact of outliers. The mean of the resampled minimum $k$th nearest neighbor distance is then taken as $\epsilon_{GKDE}$. The final $\epsilon$ value is calculated as $\epsilon = (\epsilon_{kNN}+\epsilon_{GKDE})/2$.

For our analysis, we used a five-dimensional approach based on position ($l,b$),
parallax ($\varpi$), and proper motion ($\mu^*_\alpha, \mu_\delta$) of the stars. We rescaled our data to have a mean of zero and standard deviation of one. The input parameter $minPts$ was chosen as $minPts = 10$, which is twice the number of dimensions as recommended in \citet{1998DMKD....2..169S}. 

For our cluster search in Car~OB1, we utilized the \textit{Gaia} DR3 catalog \citep{Gaia, Gaia-DR3}.
We chose a rectangular field centered around $(\alpha, \delta) = (10^\mathrm{h}38^\mathrm{m}48^\mathrm{s}, -59\degr12\arcmin)$
with dimensions of $\approx 5 \degr \times 5.4 \degr$ to analyze the whole Car~OB1 region. The chosen search field can be seen as a green box in Fig.~\ref{fig:Grid}.
This area contains 9\,295\,328 objects in the \textit{Gaia}~DR3 database, out of which 8\,236\,565 (88.6\%) have full astrometry (position, proper motions and parallax) available. 
Following the results by \citet{Lindegren.2018}, we excluded all sources with $\rm RUWE>1.4$, which removes sources with very uncertain astrometry and minimizes the impact of bad astrometric fits.
This left us with 7\,807\,017 (84\%) sources to which we applied the parallax offset correction provided by \citet{Lindegren.2021}.
Since only 1.6\% of the objects in our sample have a radial velocity measurement in the \textit{Gaia} DR3 catalog, a cluster search in full position-velocity space (6D) was not feasible. 

We divided the selected field into 90 individual sections with $L \times H \approx 0.5\degr \times 0.6\degr$. In order to not miss any clusters at the edges of a section or overlapping between several sections, we vary the sections by \textit{L}/2 and/or \textit{H}/2, and run DBSCAN on the additional fields. 
Finally, we compiled a merged sample of 462 unique 
clusterings, identified by DBSCAN as overdensities,
in the full area.

\begin{figure}
 \resizebox{\hsize}{!}{\includegraphics{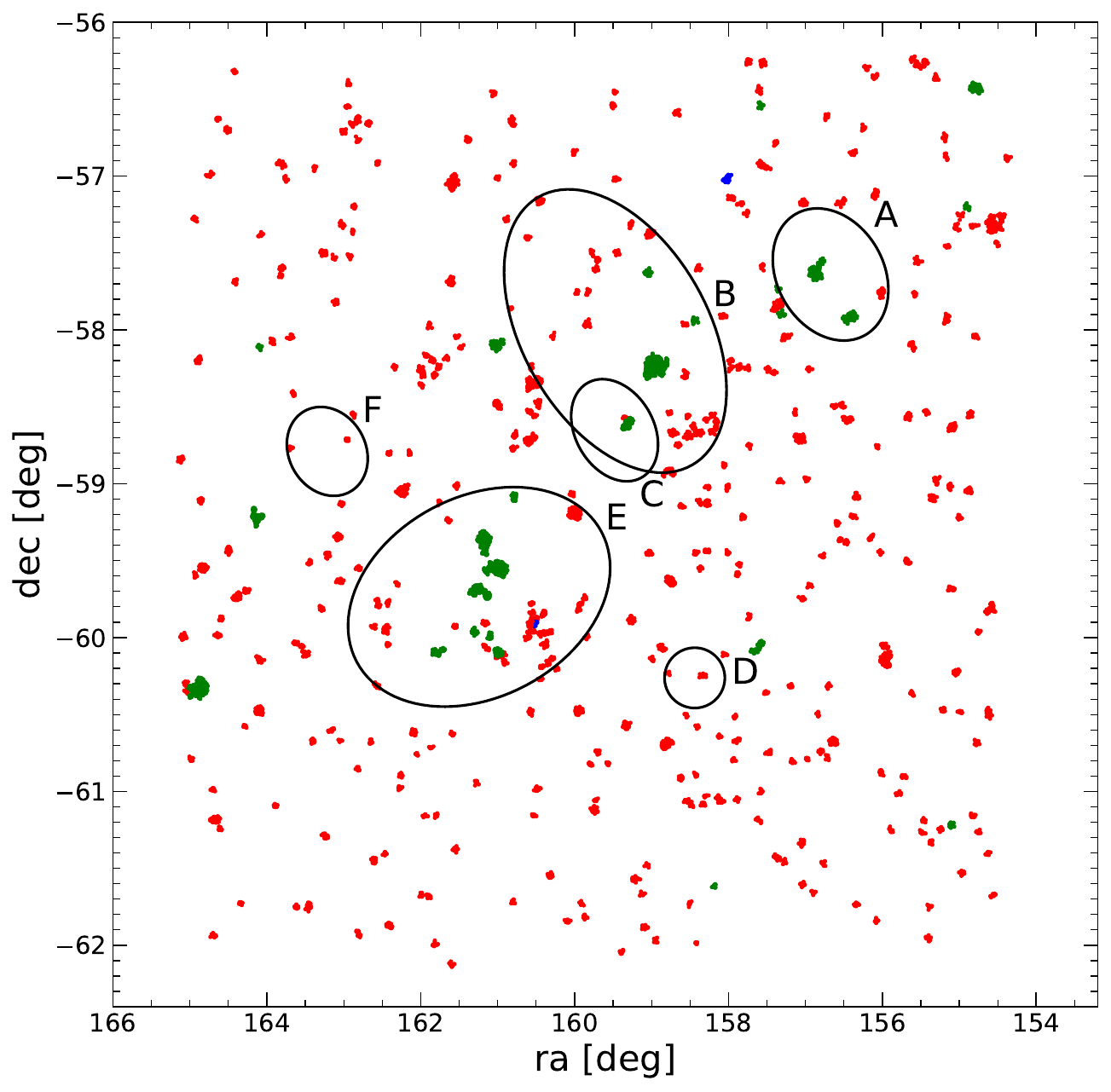}}
\caption{Map of all clusters and groups identified with DBSCAN in the investigated region. Clusters with a mean parallax that is consistent with the distance of the CNC (2.35~kpc) are shown in green, clusters in the background in red, and clusters in the foreground in blue. The ellipses show the Car~OB1 subgroups as defined by \citet{1995AstL...21...10M}.}
\label{fig:AllClusters}
\end{figure}

We classify the 365 overdensities with fewer than 20 members as stellar groups, and the 97 overdensities with 20 or more members as stellar clusters. 
The locations of all identified groups and clusters are visualized in Fig.~\ref{fig:AllClusters}. 

Distances of the groups and clusters were determined in a two-step procedure. In the first step, we computed a maximum likelihood estimate of the mean distance, that is, the inverted error-weighted mean value of the parallaxes of the member stars.
In a second step, we then applied the Bayesian inference code \textit{Kalkayotl} \citep{2020A&A...644A...7O}, which determines the mean distance values and associated confidence intervals for the clusters, using the Maximum Likelihood estimate as a prior for the distance.
In the subsequent analysis, we used the distance estimates based on \textit{Kalkayotl} for all cluster with distances less than
5~kpc; for clusters with larger distances, we used the Maximum Likelihood inferred mean distance since \textit{Kalkayotl} is less reliable for cluster distances greater than 5~kpc \citep[see discussion in ][]{2020A&A...644A...7O}.
 
A substantial number of groups or clusters are located very close to another group or cluster, and are therefore probably subparts of a larger group/cluster.
We assume that a group or cluster is a substructure of another cluster if it lies inside a cluster's $2\times r_{90}$ radius (with $r_{90}$ being the radius in which 90\% of the DBSCAN members of the cluster lie) and if their $\pm 1\sigma$ distance intervals (derived from inverting the mean parallax) overlap.

This leads to 38 clusters, including Trumpler~14, 15, 16, Bochum~11, NGC~3324, and IC~2581,  containing in total 93 subclusters.
Cluster Trumpler~16 was found to be divided into six individual subclusters by DBSCAN, similar to the result of the cluster analysis in \citet{CCCP-Clusters}.
The full list of found stellar clusters and groups can be found in the Appendix in Table~\ref{tab:AllClusters}.

\subsection{Relation to previously known clusters} 

We cross-checked our list of DBSCAN clusters and groups with a list of previously known cluster in the area compiled from the literature \citep{Cantat-Gaudin.2018,2020A&A...640A...1C,2018A&A...618A..59C, 2020A&A...635A..45C, 2022A&A...661A.118C, 2022A&A...660A...4H, CCCP-Clusters, 2014ApJ...787..107K, Smith10b, 2023A&A...673A.114H, 2019AJ....157...12B}. We found literature counterparts for 61 (63\%) of our DBSCAN clusters and for 31 (8\%) of our DBSCAN groups. 
Almost all of the well-known CNC star clusters (in particular, Trumpler~14--16, Collinder~228 and 232, Bochum~11, NGC~3324, and NGC~3293) were also identified as clusters in our DBSCAN search.

Four previously known clusters were not recovered in our DBSCAN cluster search for various reasons: 

(1) The Treasure Chest cluster, a very young ($\lesssim 0.1$\;Myr \citep{2005AJ....129..888S}) cluster still embedded in its natal cloud. With typical visual extinctions ranging from
$A_V \ga 5$~mag up to $A_V \approx 50$~mag for individual members \citep{2005AJ....129..888S, HAWKI-survey}, most cluster members are too faint at optical wavelengths to be detected by \textit{Gaia}. 

(2) Tr16-SE, a strongly obscured cluster of stars
south-east of Tr~16, which was  discovered in infrared and X-ray observations \citep{2007ApJ...656..462S, Tr16-SE-KMOS}. Due to the high extinction, only very few stars in this cluster were detected by \textit{Gaia}.

(3) Bochum~10, which constitutes a quite sparse stellar group with a low star density, was not identified with DBSCAN. \citet{2001MNRAS.325.1591P} and \citet{2004A&A...418..525C} 
had already suggested that Bochum~10 may not be a real physical cluster. 

(4) Collinder~234, a group of stars located just $\approx 1.5\arcmin$ south-east from the edge of Trumpler~16;
we classify it here as one of the subclusters of Trumpler~16, as already suggested by \citet{2004A&A...418..525C} and \citet{CCCP-Clusters}.

\subsection{New clusters in Car OB1}
Our first criterion for assuming a cluster or a group to be part of Car~OB1 is that the 90\% confidence interval for its distance (determined with \textit{Kalkayotl}) is compatible with the distance interval of [2.25, 2.45]\;kpc for Car~OB1 as mentioned in Sect.~\ref{sec:CarOB1Sample}. 
Out of the 462 clusters and groups, 47 fulfill this criterion. Group 339, with $D_{\rm ML} =  15.9^{+61.9}_{-7.1}$~kpc, satisfies this requirement only due to its extremely large distance uncertainty; we therefore exclude this cluster from our list of Car~OB1 clusters. 
The positions of all clusters and groups in the investigated area are shown in Fig.~\ref{fig:AllClusters}, where also the subgroups of the Car~OB1 association \citep[as defined by][]{1995AstL...21...10M} are shown as black ellipses. We exclude groups 23, 32, 130, 132, and 424 from being a part of Car~OB1 due to their large spatial separation from any of the  Car~OB1 subgroups and the cloud complex.

\subsection{Cluster age estimates} 

\begin{figure*} 
\centering
\begin{subfigure}[t]{5.8cm}
   \resizebox{5.8cm}{!}{\includegraphics{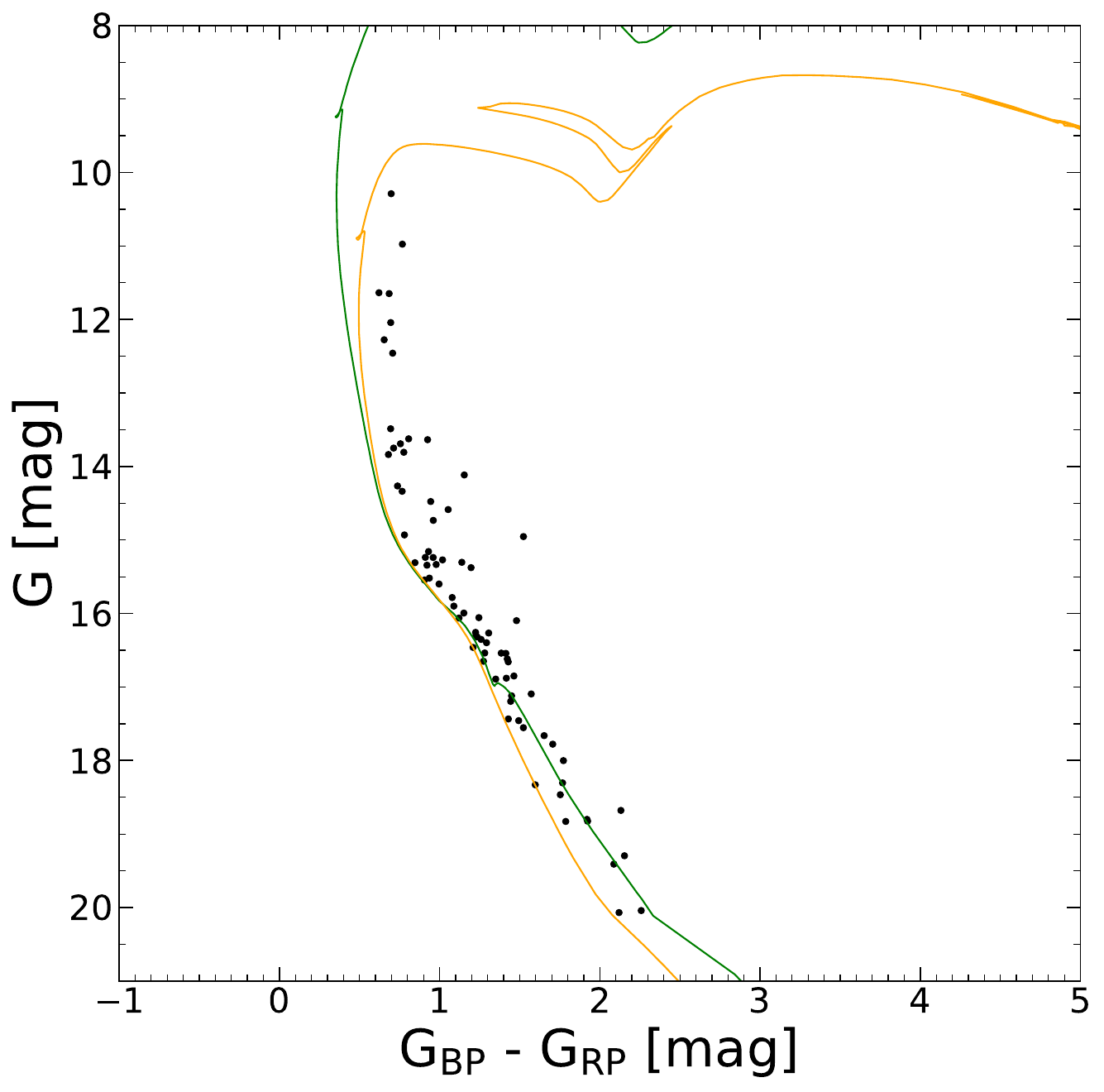}}
\end{subfigure}
\hspace{0.2cm}
\begin{subfigure}[t]{5.8cm}
   \resizebox{5.8cm}{!}{\includegraphics{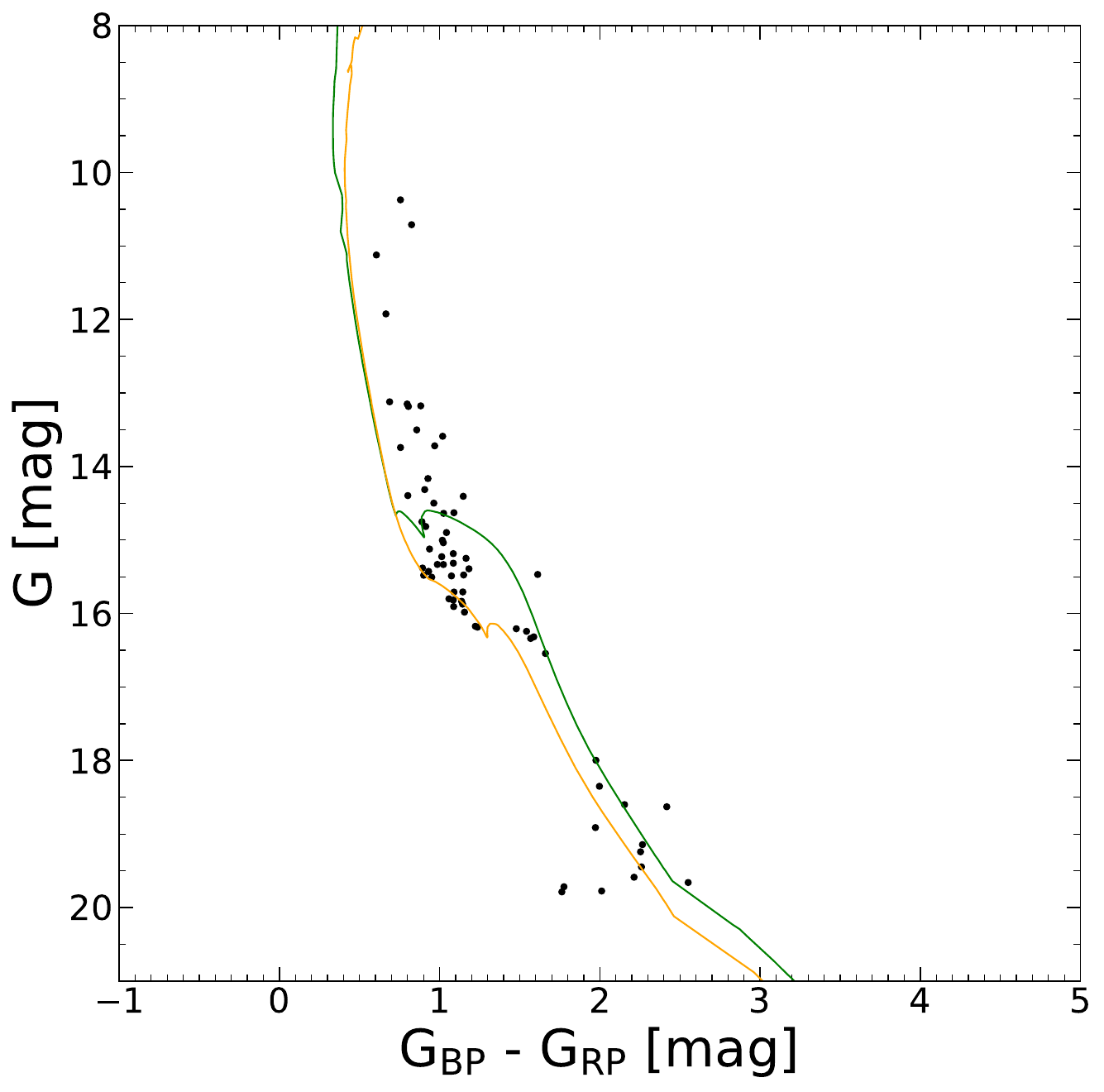}}
\end{subfigure}
\hspace{0.2cm}
\begin{subfigure}[t]{5.8cm}
   \resizebox{5.8cm}{!}{\includegraphics{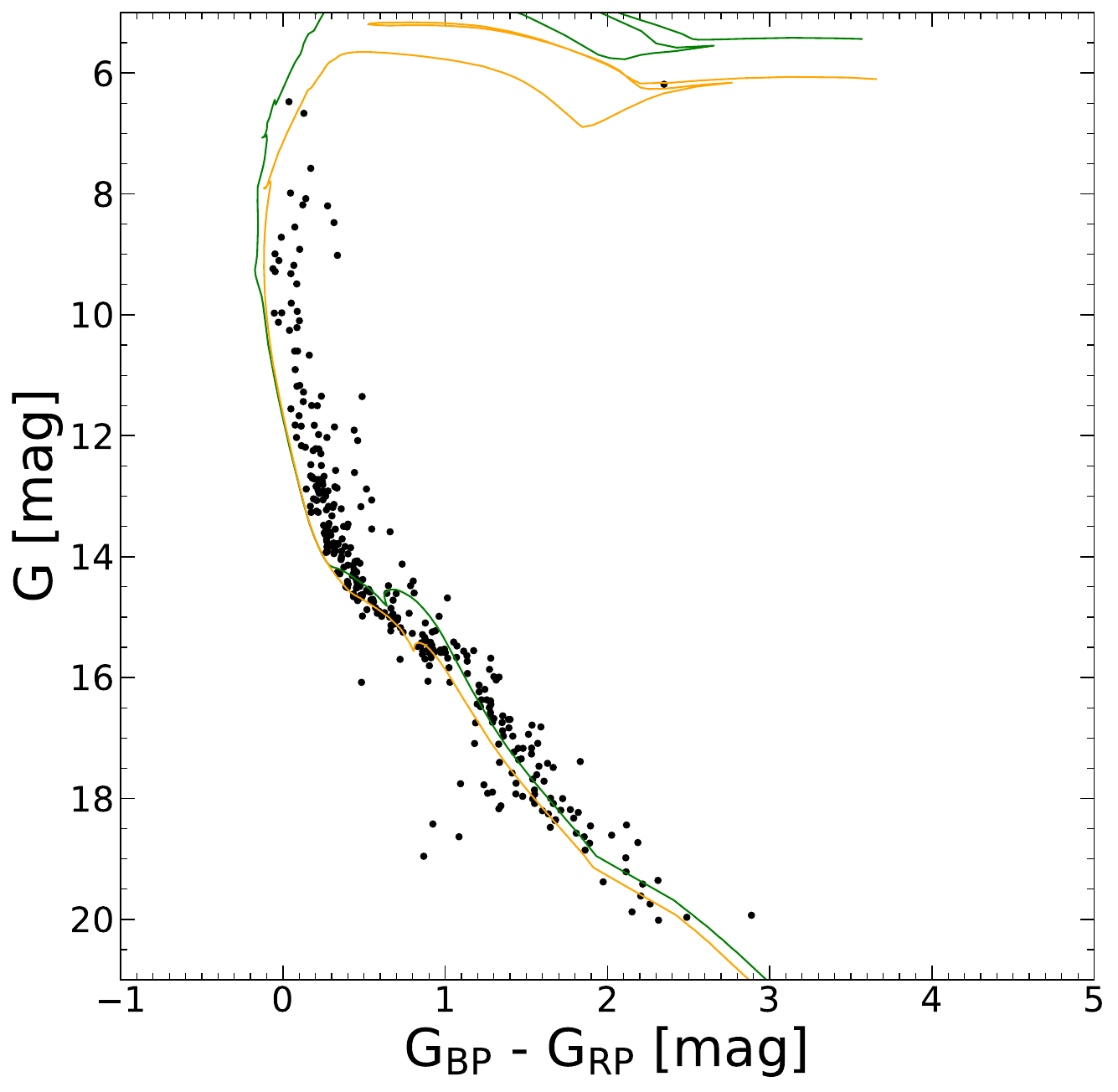}}
\end{subfigure}
\caption{Color-magnitude diagrams of the DBSCAN-selected members of clusters with their literature age and their best-fitting PARSEC v3.7 isochrone \citep{2012MNRAS.427..127B}. Left panel: Trumpler~17 with 20~Myr \citep{2020A&A...640A...1C} isochrone in green and 80~Myr isochrone in orange, assuming an extinction of $A_V = 1.6$~mag for both isochrones. Middle panel: UBC~501 with 6~Myr \citep{2019ApJS..245...32L} isochrone in green and 13~Myr isochrone in orange, assuming an extinction of $A_V = 1.8$~mag for both isochrones. Right panel: NGC~3293 with 10~Myr \citep{2020A&A...640A...1C} isochrone in green and 15~Myr isochrone in orange, assuming an extinction of $A_V = 0.6$~mag for both isochrones.}
\label{fig:ClustersCMD}
\end{figure*}

\begin{figure*} 
\centering
\begin{subfigure}[b]{8.5cm}
  \includegraphics[width=8.5cm,trim=6.1cm 10.3cm 6.65cm 11.7cm, clip]{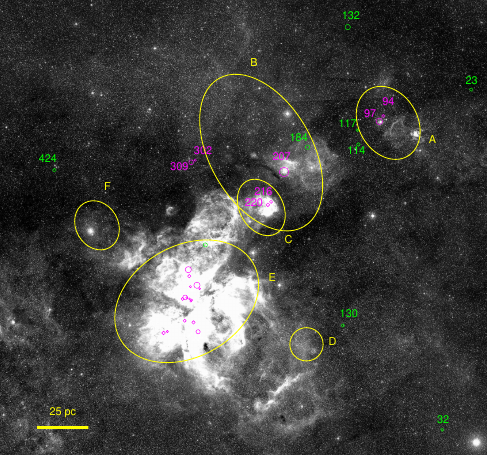}
\end{subfigure}
\begin{subfigure}[b]{8.5cm}
\includegraphics[width=8.5cm,trim=5.18cm 10.3cm 7.68cm 11.8cm, clip]{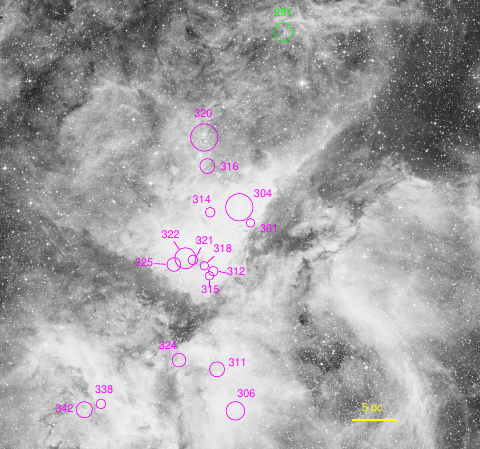}
\end{subfigure}
\caption{Optical image of the CNC.
(Left panel: DSS2 red in grayscale; right panel: DSS2 red in grayscale {\tt www.eso.org/public/images/eso0905b/}; image credit:
ESO/Digitized Sky Survey 2, Davide De Martin). North is up and east to the left. The circles show the clusters and groups whose distance and age is compatible with being a part of Car~OB1. Circles in magenta represent clusters with counterparts in the literature, circles in green clusters without counterparts. The clusters' radii are chosen as the radii in which 75\% of their cluster members reside.}
\label{fig:ClustersAge}
\end{figure*}

\begin{figure*} 
\centering
\includegraphics[width=17cm,trim=3.3cm 9.3cm 5.3cm 9.1cm, clip]{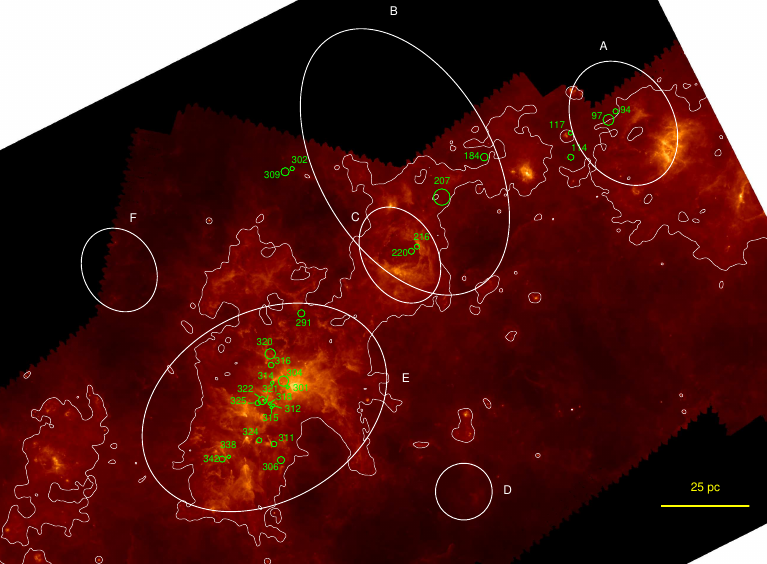}
\caption{\textit{Herschel} $160~\mu \rm m$ image \citep[from \protect\url{https://www.cosmos.esa.int/web/herschel/pacs-jscanam-mosaics-hpdp}; see][]{2017ASPC..512..379G} of the Car~OB1 region.
A white contour line is drawn at an intensity of 
$7.4~\rm mJy/(arcsec)^2$. The locations of clusters with distances and ages $\leq 20$~Myr are shown in green. North is up and east to the left.}
\label{fig:ClustersHerschel}
\end{figure*}

As a second criterion for assuming a cluster to be related to Car~OB1, we considered the clusters' ages.
Table~\ref{tab:ClustersCarOB1Ages} lists the age estimates for previously known clusters in Car~OB1 collected from the literature. 
Since we are interested in the young ($\la 30$~Myr old) populations of the Car~OB1 association, we
excluded the six previously known clusters/groups with literature ages of $\ga 40$\;Myr.

To determine cluster ages, we used the PARSEC v3.7 stellar evolution models \citep[see][]{2012MNRAS.427..127B} and fit them to a color-magnitude diagram of the DBSCAN selected members of all potential Car~OB1 clusters. For most clusters, we find values that agree well with the literature based age estimations. We find deviating ages for UBC~501 and Trumpler~17, which can be seen in Fig.~\ref{fig:ClustersCMD}. No age confirmation was possible for [DBS2003]~53 due to low number of members. 
For UBC~501 we find a slightly older age than listed in the literature, but still under the maximum age requirement for Car~OB1.
For Trumpler~17, our age estimate of $\approx 80$~Myr, together with the cluster's location well outside the cloud complex and the known association subgroups,
excludes this cluster from the association. With fewer than 20 members, it was not possible to confirm ages for stellar groups using CMD fitting. We therefore included groups 114, 117, 184, and 291 in our Car~OB1 stellar cluster and group sample.

\begin{table}
    \centering
     \resizebox{\hsize}{!}{\begin{tabular}{cccc}
    \hline\hline
       Cluster & Age in Literature& Reference & Age estimate\\
        & [Myr]& &this work [Myr]\\
        \hline
         Trumpler 14 & $\approx 1$ & (1)\\
        Trumpler 16 & $\approx$ 3--4 & (2)\\
        Bochum 11 & $\approx$ 3--10 & (2)\\
        Trumpler 15 & $\approx$ 5--8 & (2)\\
        UBC 501 & $\approx 6$ &(5) & $\approx 13$ \\
        NGC 3293 & $\approx 10$, 20 & (2,3), (8) & $\approx 15$\\
        NGC 3324 & $\approx 11$& (4)\\
        IC 2581 & $\approx 12$ &(4) &\\
        \hline
        Trumpler 17 & $\approx 20$ &(4) & $\approx 80$\\        
        UBC 258 & $\approx 40$& (4) \\
        {[}DBS2003] 53  & $\approx 126$ &(7) \\
        UBC 259 & $\approx 126$& (4) \\
        Cl vdBH 92 & $\approx 275$ &(4) \\
        Collinder 220 & $\approx 282$ &(4) \\
        NGC 3496 & $\approx 537$& (4)\\        
    \end{tabular}}
    \caption{Age estimates of the clusters in and around Car~OB1.\\
    (1): \citet{2024A&A...685A.100I}, (2): \citet{HAWKI-survey}, (3): \citet{2017A&A...605A..85P}, (4): \citet{2020A&A...640A...1C}, (5): \citet{2019ApJS..245...32L}, (6): \citet{2022A&A...660A...4H}, (7): \citet{2016A&A...585A.101K}, (8): \citet{2022A&A...665A.108M}}
    \label{tab:ClustersCarOB1Ages}
\end{table}

The locations of the clusters/groups we consider being parts of the
Car~OB1 association are marked in Fig.~\ref{fig:ClustersAge};
their basic properties are listed in Table~\ref{tab:CarOB1clusters}. In total, we find 15 groups and clusters (split into their 27 (sub)clusters and groups in Table~\ref{tab:CarOB1clusters}) to be part of Car~OB1.

In order to find out which of these clusters and groups are related to the clouds of the CNC, we used a mosaic of \textit{Herschel} $160~\mu \rm m$ maps that was produced as 
part of the Herschel High Level Images (HHLI) by \citet{2017ASPC..512..379G}; this map is shown in Fig.~\ref{fig:ClustersHerschel}. 
We used the contour corresponding to a surface brightness of $7.4\;\rm mJy/(arcsec)^2$
to define the boundaries of the cloud emission related to the CNC. 
19 of the 27 groups and clusters  in Car~OB1 are located inside the boundaries of the cloud complex.

\begin{table*}
    \centering
        \caption{Clusters and groups with a mean distance compatible with Car~OB1 and an age $\leq 30$~Myr. (1): \citet{2018MNRAS.476..842O}, (2): \citet{2014ApJ...787..107K}, (3): \citet{Smith10b}, (4): \citet{CCCP-Clusters}.}
    \label{tab:CarOB1clusters}
\resizebox{\hsize}{!}{ \begin{tabular}{llccccll}
    \hline \hline 
DBSCAN & N & R.A. & Dec.  &  $D_{\rm Kalkayotl}$ & Central 68.3\% quant. & literature name & age\\
Cluster/Group &  & [J2000] & [J2000]  & [kpc] & [kpc] & & Myr\\
       \hline
94& 10 &10:27:08.99& $-57$:33:41.5 &2.432 & [2.352, 2.513] & part of IC 2581 & $\approx 12$\\
97& 109 &10:27:28.73& $-57$:37:31.3 &2.438 &[2.369, 2.507] & IC 2581& $\approx 12$\\
114& 17 &10:29:16.01& $-57$:54:11.4 &2.483 &[2.381, 2.585] & \\
117& 10 &10:29:22.74& $-57$:44:13.8 &2.537 &[2.446, 2.629] & \\
184& 11 &10:33:46.12& $-57$:56:40.1 &2.343 & [2.262, 2.425] & \\
207& 358 &10:35:51.67& $-58$:14:13.4 &2.330 &[2.268, 2.392] & NGC 3293& $\approx 15$\\
216& 24 &10:37:02.12& $-58$:35:16.3 &2.505 &[2.414, 2.596] & part of NGC 3324& $\approx 11$\\
220& 59 &10:37:19.36& $-58$:37:17.6 &2.379 &[2.308, 2.450] & NGC 3324& $\approx 11$\\
291& 19 &10:43:04.92& $-59$:04:46.6 &2.551 &[2.454, 2.651] & \\
301& 14 &10:43:43.11& $-59$:35:33.9 &2.333 &[2.256, 2.410] & [OBB2018] 1 (1), Carina A (2), part of Trumpler 14& $\approx 1$\\
302& 11 &10:43:45.59& $-58$:04:48.6 &2.413 &[2.345, 2.479] & UBC 501& $\approx 13$\\
304& 416 &10:43:57.52& $-59$:33:03.9 &2.368& [2.302 ,2.434] & Trumpler 14& $\approx 1$\\
306& 40 &10:43:59.66& $-60$:05:52.6 &2.353 &[2.280, 2.427] & Collinder 228\\
309& 56 &10:44:07.56& $-58$:06:12.7 &2.410 &[2.341, 2.478] & UBC 501& $\approx 13$\\
311& 19 &10:44:23.89& $-59$:59:12.9 &2.383 &[2.305, 2.461] & Spitzer B (3), CCCP-Gp 13, 14, 16 (4)\\
312& 11 &10:44:29.97& $-59$:43:24.1 &2.255 &[2.194, 2.316] & part of Trumpler 16& $\approx$ 3--4\\
314& 18 &10:44:34.47& $-59$:33:55.2 &2.325 &[2.263, 2.387] & Collinder 232, part of Trumpler 14& $\approx 1$\\
315& 15 &10:44:34.70& $-59$:44:09.3 &2.192 &[2.133, 2.251] & part of Trumpler 16& $\approx$ 3--4\\
316& 16 &10:44:38.58& $-59$:26:28.7 &2.463 &[2.379, 2.546] & Carina F (2), CCCP-Cl 7 (4), part of Trumpler 15&$\approx$ 5--8\\
318& 10 &10:44:41.35& $-59$:42:32.6 &2.311 &[2.240, 2.382] & part of Trumpler 16&$\approx$ 3--4\\
320& 238 &10:44:42.72& $-59$:21:52.8 &2.366& [2.300, 2.433] & Trumpler 15& $\approx$ 5--8\\
321& 9 &10:44:56.05& $-59$:41:36.9 &2.295 &[2.220, 2.368] & part of Trumpler 16& $\approx$ 3--4\\
322& 70 &10:45:05.64& $-59$:41:21.5 &2.359& [2.293, 2.424] & part of Trumpler 16& $\approx$ 3--4\\
324& 24 &10:45:13.04& $-59$:57:46.7 &2.373& [2.292, 2.453] & Carina M (2), Spitzer F (3), CCCP-Cl 13 (4)\\
325& 16 &10:45:20.24& $-59$:42:22.5 &2.266& [2.209, 2.323] & part of Trumpler 16& $\approx$ 3--4\\
338& 15 &10:46:53.48& $-60$:04:51.7 &2.417& [2.337, 2.495] & Carina S (2), CCCP-Gp 28, 30 (4), part of Bochum 11& $\approx$ 3--10\\
342& 31 &10:47:15.05& $-60$:05:48.6 &2.325& [2.257, 2.392] & Bochum 11& $\approx$ 3--10\\
    \end{tabular}}
\end{table*}

\subsection{Quantification of the clustered versus distributed populations of high-mass stars in Car~OB1}
\label{sec:clusterprop}

\begin{figure}
    \centering
       \resizebox{\hsize}{!}{\includegraphics[trim=7.cm 9.4cm 5.cm 8.7cm, clip]{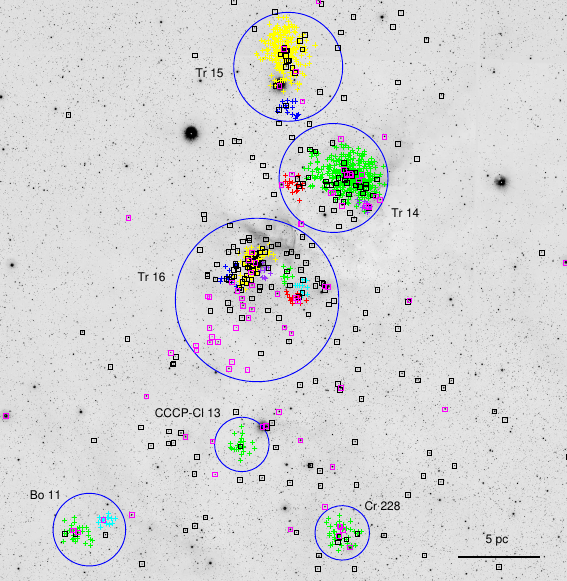}}
    \caption{VISTA $J$-band image of the central region of the Carina Nebula.
    The DBSCAN members of the clusters Tr14, 15, and 16 (split into subclusters) are marked by crosses. O- and WR stars and supergiants from the Car~OB1 high-mass star sample are marked by magenta, B-type stars by black open boxes. Stars in clusters have different colors in order to differentiate the subclusters.
    }
    \label{fig:VISTAClusterCenter}
\end{figure}

\begin{figure}
    \centering
       \resizebox{\hsize}{!}{\includegraphics[trim=8.cm 9.7cm 4.cm 11.1cm, clip]{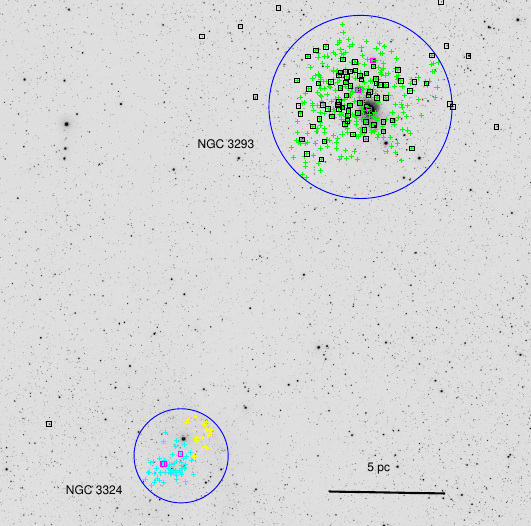}}
    \caption{VISTA J-band image of the northwestern part of the CNC.
    The DBSCAN members of the clusters NGC~3293 and NGC~3324 are marked by crosses. O-type stars and supergiants from the Car~OB1 high-mass star sample are marked by magenta, B-type stars by black open boxes. Stars in clusters have different colors in order to differentiate the subclusters.
    }
    \label{fig:VISTAClusterNGC3293}
\end{figure}

\begin{figure}
    \centering
       \resizebox{\hsize}{!}{\includegraphics[trim=7.6cm 11.7cm 8.4cm 11.1cm, clip]{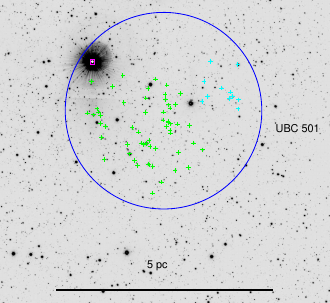}}
    \caption{VISTA J-band image of the northwestern part of the CNC.
    The DBSCAN members of the cluster UBC 501 are marked by crosses. The red supergiant HD~303250 (M3Iab) is marked by the magenta box. Stars in clusters have different colors in order to differentiate the subclusters.
    }
    \label{fig:VISTAClusterUBC501}
\end{figure}

\begin{figure}
    \centering
       \resizebox{\hsize}{!}{\includegraphics[trim=6.1cm 11.2cm 7.6cm 10.6cm, clip]{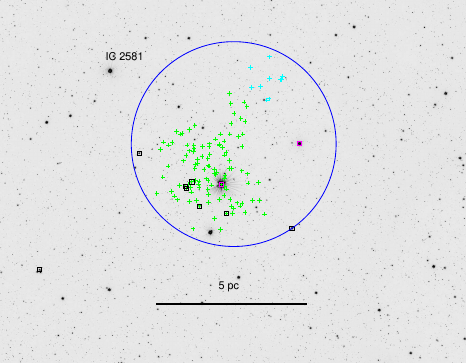}}
    \caption{VISTA $K$-band image of the northwestern part of Car~OB1.
    The DBSCAN members of the cluster IC~2581 are marked by crosses. The two supergiants V399~Car (F0Ia) and HD~90706 (B2.5Ia) are marked by magenta, B-type stars by black open boxes. Stars in clusters have different colors in order to differentiate the subclusters.
    }
    \label{fig:VISTAClusterIC2581}
\end{figure}

For a quantification of the clustered versus distributed
populations of high-mass stars in Car~OB1, we need an efficient criterion for considering a star as either being a member of one of the clusters or belonging to the distributed
(i.e.,~non-clustered) population.
Out of the 517 stars in our Car~OB1 high-mass star sample, 180 (35\%) were identified as members of one of the clusters found by DBSCAN. Furthermore, several stars from our Car~OB1 high-mass sample are located in the area of one of the found clusters, but were not classified as cluster members by DBSCAN (e.g., because they have slightly deviating parallaxes or proper motions, which precluded their classification as cluster members by DBSCAN). We therefore defined circles around the clusters that include all DBSCAN members\footnote{The circle around Trumpler~16 extends beyond the DBSCAN member selection to include the cluster Tr16-SE, which was not identified by DBSCAN.} and assumed that all Car~OB1 high-mass stars inside these circles belong to the clustered population.
Figures~\ref{fig:VISTAClusterCenter} -- \ref{fig:VISTAClusterIC2581} show the clusters and their chosen perimeter.
With this definition, we find that
301 (58\%) of the Car~OB1 high-mass  stars are in clusters,
while the remaining 216 (42\%) Car~OB1 high-mass stars are part of the non-clustered, distributed population.
Considering only O-type, WR, and supergiant stars, the
ratio of the clustered versus distributed population changes only very slightly to 55\,:\,45.
The 58\% fraction of clustered high-mass stars is slightly higher, but similar to
the $\approx 51\%$ clustered population fraction among the
X-ray selected young stars in the CNC \citep{CCCP-Clusters},
and the $\approx 47\%$ fraction of clustered young stars from the analysis of the spatial distribution of infrared-excess selected young star candidates \citep{VISTA2}.

This demonstrates that a very substantial fraction of about 40\% of the
young high-mass stars with known spectral type in the complex are in a non-clustered, distributed population.
While this should not be surprising, since the Carina complex is an OB association, it suggests that the census of OB stars may still be seriously incomplete, as spectroscopic searches for OB stars have been primarily focused on the central regions of the Carina Nebula and the well-known and prominent clusters in the CNC, whereas the more widely spread distributed population has received much less attention.
In Sect.~\ref{sec:GaiaCarOB1} we use the \textit{Gaia} data to identify candidates for further, yet unidentified OB stars in the CNC, in order to move toward a more complete census of the stellar population.

\clearpage

\section{Kinematic analysis of the individual clusters}
\label{sec:Expansion}

We analyzed the proper motions and radial velocities of the stars in the 10 Car~OB1 DBSCAN clusters (Table~\ref{tab:CarOB1clusters}) with $\geq 20$ members, to gain more information about the internal kinematics of these clusters. 
We treated subclusters separately from their cluster if their mean weighted proper motions in either right ascension or declination lies one standard deviation outside of the full cluster's proper motion distribution in right ascension or declination. This concerns Trumpler~14, 15, and 16 where Groups 301, 316, and 312 showed diverging proper motion from the rest of their cluster and were treated separately.

For our kinematic analysis we use proper motions and radial velocities from \textit{Gaia} DR3. Since only a few stars have radial velocities in \textit{Gaia} available, we complement these with radial velocity measurements from the Gaia-ESO survey (GES) \citep{2012Msngr.147...25G,2023yCat..36760129H}.
GES is a public spectroscopic survey whose observations include stars in the Carina Nebula and NGC~3293 and that provides radial velocity measurements for 342 stars in our cluster member list. 

First, we calculated correction factors for the proper motions using radial velocities in order to account for the effect of the cluster moving toward or away from us, which can mimic expansion or contraction. For this, we used equation~13 from \citet{2009A&A...497..209V}:
\begin{equation}
\label{eq:mualpha}
    \Delta \mu_{\alpha^*,\rm per} \approx \Delta \alpha_i \left(\mu_{\delta,c} \, \mathrm{sin} \, \delta_c - \frac{v_r \varpi_c}{\kappa}\, \mathrm{cos} \,\delta_c \right)
\end{equation}
and
\begin{equation}
\label{mudelta}
    \Delta \mu_{\delta, \rm per} \approx -\Delta \alpha_i \mu_{\alpha^*,c}  \, \mathrm{sin} \,  \delta_c - \Delta \delta_i \frac{v_r \varpi_c}{\kappa},
\end{equation}
with $\alpha_c, \delta_c, \mu_{\alpha^*,c}, \mu_{\delta,c}, \varpi_c$ as the weighted mean cluster properties, $\Delta \alpha_i = \alpha_i - \alpha_c$ (analogous for $\Delta \delta_i$), and $\kappa = 4.74$ as the conversion factor from $\rm milliarcsecond\, year^{-1}$ to $\rm kilometer \, second^{-1}$ at a distance of 1~kpc.
For the cluster radial velocity, we used the median value of each cluster. If there were fewer than five radial velocity measurements for a cluster available, we did not calculate the correction factor. This was the case for five of the ten clusters. The correction factors were $< 0.02~\frac{\rm mas}{\rm yr}$ for all stars in the remaining five clusters, which is sufficiently small that we can assume that the lack of correction for the other five clusters has negligible influence on their results. To compare this, we have performed the kinematic analysis with and without the correction factor if possible. The results are shown in Table~\ref{tab:compare} and show only small differences.

Next, we applied the correction factors (if available) as in equations (3) and (4) in  \citet{2019ApJ...870...32K} and calculated the velocities $v_{\alpha}$ and $v_{\delta}$, which are parallel to RA and Dec, respectively:
\begin{equation}
\label{eq:valpha}
    v_{\alpha} \approx \kappa \left(\frac{\Delta \mu_{\alpha^*, \rm obs}-\Delta \mu_{\alpha^*, \rm per}}{\varpi_c} \right) 
\end{equation}
and 
\begin{equation}
\label{eq:vdelta}
    v_{\delta} \approx \kappa \left(\frac{\Delta \mu_{\delta, \rm obs}- \Delta \mu_{\delta, \rm per}}{\varpi_c} \right).
\end{equation}

In order to characterize possible expansion motions in all directions (not only along right ascension and the declination axis), we rotated the coordinate system by a sequence of angles from $\theta = 0\degr$
to $\theta = 179\degr$ in steps of
$1\degr$, and translated the position coordinates accordingly in parsec (centered at the cluster center).

\begin{equation}
    x(\theta) = \frac{1000}{\varpi_c} (-\Delta \alpha\,\cos(\theta)+\Delta\delta\,\sin(\theta))
\end{equation}
and 
\begin{equation}
    v_x(\theta) = -v_{\alpha}\,\cos(\theta)+v_{\delta}\,\sin(\theta).
\end{equation}

For all values of the rotation angle $\theta$, we then plotted \textit{x} versus the velocity in \textit{x} ($v_x$) and checked for correlation. A positive correlation indicates expansion, while a negative correlation indicates contraction of the cluster. 
In order to quantify the significance of the expansion/contraction we first 
computed the Pearson's correlation coefficient $r_{xv_x}$, 
then determined the Student's \textit{t}-value, 
and used it to quantify the significance of the correlation.
Using bootstrap analysis, we carried out this step 20\,000 times for each cluster while adding errors with random weights based on the stars' normally distributed uncertainties in velocity and then took the \textit{t}-value of the mean correlation coefficient to determine the significance, which can be seen in Table~\ref{tab:tdyn}. The uncertainties in the velocity of each star were estimated using error propagation for equations~\ref{eq:mualpha}--\ref{eq:vdelta} and a covariance matrix to take the rotation into account.

We did this for $\theta \in [0\degr, 179\degr]$ and determined the angle for which the Student's \textit{t}-value is maximized, which is the direction of the most significant expansion or contraction.
Figure~\ref{fig:expansionplots2} shows this for all clusters that have correlation at a significance higher than $1.5\,\sigma$.

In total, we find indications of expansion/contraction with at least $1 \sigma$ significance for eight investigated clusters. Four clusters show a significance $\geq 2\,\sigma$, with Trumpler~14 yielding the highest
significance ($5.2\,\sigma$). The clusters generally show signs of expansion; the only exception is NGC~3293, which shows indications of contraction and expansion. 

Our expansion analysis results for Trumpler~14 and 16 agree with the conclusions by \citet{2019ApJ...870...32K} who found expansion in the two clusters as well based on \textit{Gaia} DR2 data. They find low-level expansion for at least $\sim 75\%$ of their analyzed clusters, which agrees with our results that 80\% of our investigated clusters show signs of expansion.

\citet{2024MNRAS.533..705W} analyzed 18 groups and stellar clusters (including Trumpler~14 and 16) using data from GES and \textit{Gaia}~EDR3. They also found indications of expansion for Trumpler~16, while they did not find significant expansion or contraction in Trumpler~14. The differences in their analysis results could be due different membership selection criteria of stars since the GES observations of Trumpler~14 go beyond our spatial definition of Trumpler~14. 

The kinematic age of clusters can be calculated by inverting the slope of the fit in the velocity-position plot (see Fig.~\ref{fig:expansionplots2}). The expansion time and its uncertainty was estimated via bootstrapping (as the significance before) and inverting the mean of the gradients with added/subtracted standard deviation of the gradients. 
The kinematic ages are generally rather well consistent with age estimates for the stellar populations of these clusters in the literature (see Table~\ref{tab:tdyn}).

\begingroup
\renewcommand{\arraystretch}{1.5}
\begin{table}
     \caption{Carina Nebula complex clusters that show indications of expansion or contraction at a least $1\,\sigma$ significance and their kinematic age $\tau_{\rm kin}$.}
    \label{tab:tdyn}
     \resizebox{\hsize}{!}{\begin{tabular}{lccccc}
    \hline \hline
    Cluster & Angle [$\degr$]& \multicolumn{2}{c}{Expansion} &  & Contraction \\
     &   & Significance & $\tau_{\rm kin}$~[Myr]&  $\tau_{\rm stellar}$ & Significance \\
        \hline
NGC~3293 &38 & & & $\approx 10$ &$ 2.3\,\sigma$  \\ 
Trumpler~14& 90 & $5.2\,\sigma$ & $1.07^{+0.22}_{-0.16}$ & $\approx 1$ & \\
Trumpler~16 & 168 & $2.8\,\sigma$ &$5.19^{+3.14}_{-1.42}$ & $\approx 3.5$ & \\
CCCP-Cl 13 & 5 &$1.5\,\sigma$ & $1.09^{+1.38}_{-0.39}$ &  & \\ 
Bochum~11 & 99 &  $1.9\,\sigma$ & $0.62^{+0.48}_{-0.19}$ & $\approx$ 3--10\\ 
IC 2581 &170 &  $2.3\,\sigma$ &$4.80^{+3.92}_{-1.49}$ & $\approx 12$\\
NGC 3324&135 & $1.2\,\sigma$ &  $1.98^{+10.30}_{-0.90}$ & $\approx 11$\\
UBC 501&48 &  $1.3\,\sigma$ &  $4.91^{+7.40}_{-1.84}$ & $\approx 13$\\
\end{tabular}}
\end{table}
\endgroup

\begin{figure}[b]
\centering
\begin{subfigure}[b]{8.5cm}
   \includegraphics[width=8.0cm]{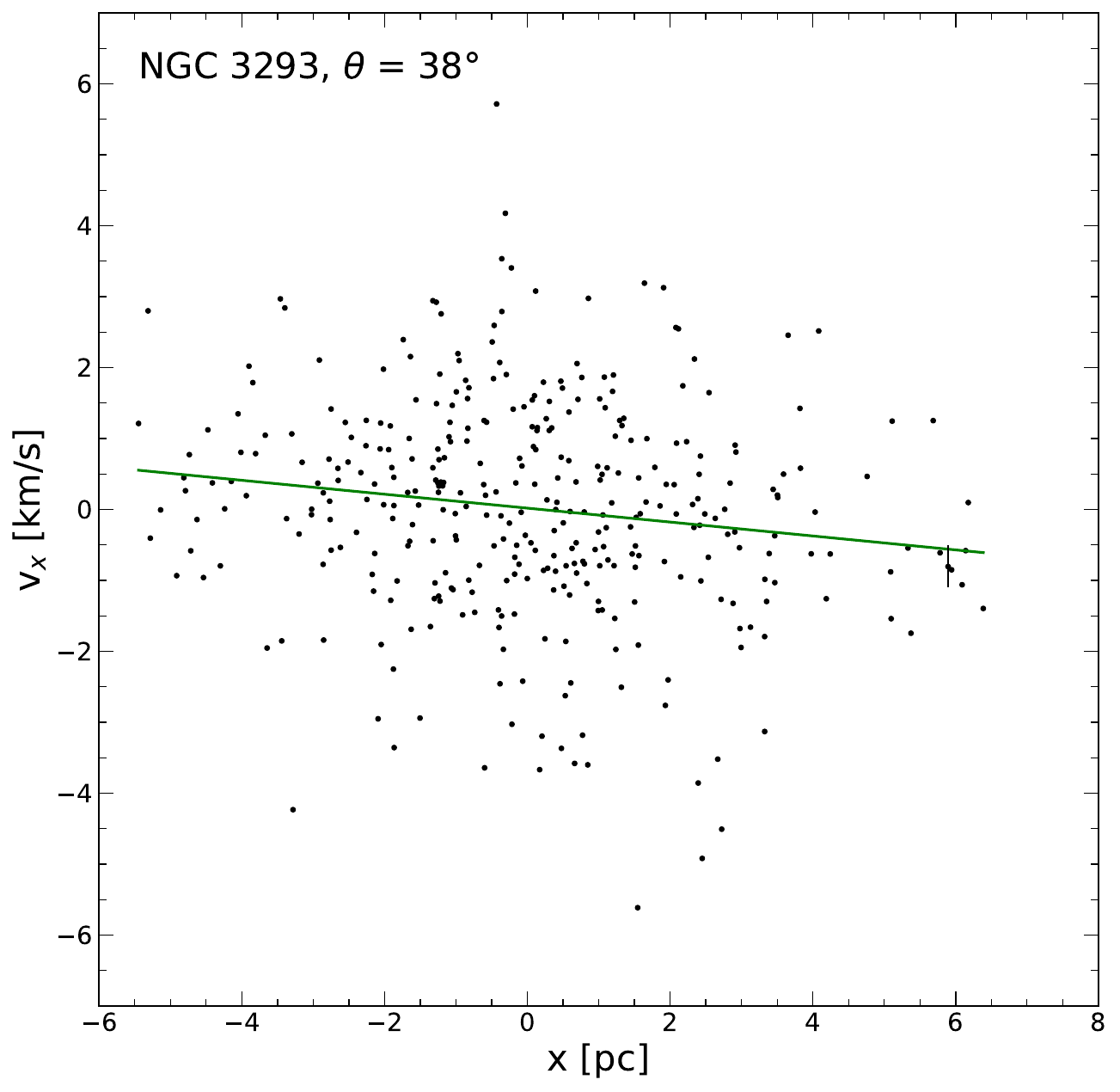}
\end{subfigure}
\begin{subfigure}[b]{8.5cm}
   \includegraphics[width=8.0cm]{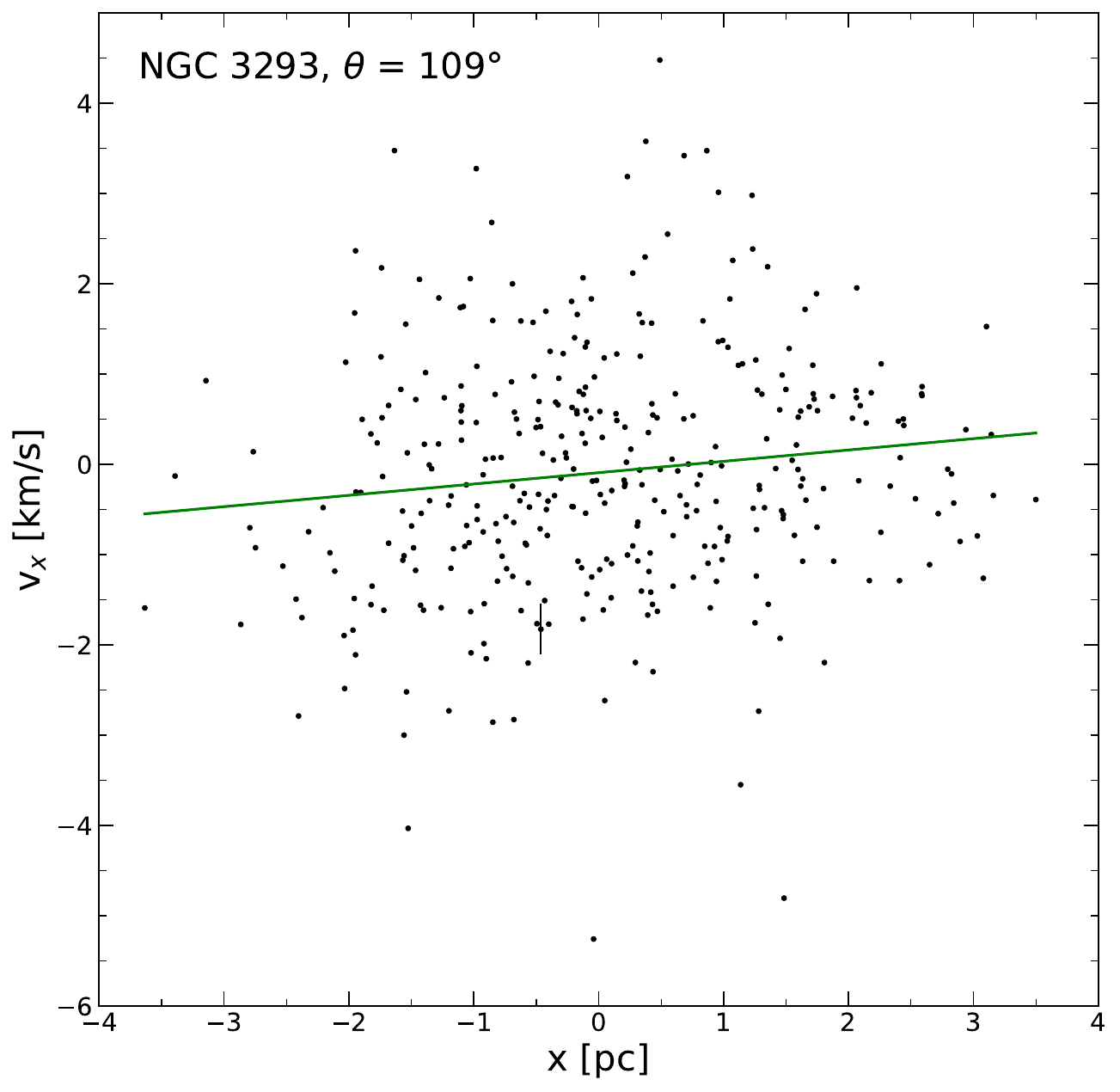}
\end{subfigure}
\caption{Relative position versus relative velocity with its slope for members of NGC 3293 at angles of the primary ($\theta=38\degr$,  left panel) and secondary significance peak ($\theta=109\degr$, right panel).}
\label{fig:NGC3293SecondaryPeak}
\end{figure}

The kinematic analysis for NGC~3293 yields interesting results. Figure~\ref{fig:polarplots2}~(f) shows the significance of expansion/contraction according to the angle for NGC~3292. While most clusters only show one peak with a significance higher than $1.5\sigma$, only NGC 3293 shows a secondary peak above this threshold at $\theta=111\degr$ with a significance of $1.9\sigma$. But while the proper motions show contraction at the primary peak with a significance of $2.3\sigma$ (as seen on the top in Fig.~\ref{fig:NGC3293SecondaryPeak}), the proper motions indicate expansion at an angle of $\theta=109\degr$ as shown on the bottom in Fig.~\ref{fig:NGC3293SecondaryPeak}. The directions of contraction and expansion are almost perpendicular to each other and show that NGC~3293 is contracting and expanding at the same time along different axes.

Simulations predict that clusters contract at the beginning of their lifetime and are in their most compact form in their first few Megayear after which cluster expansion follows \citep{2024MNRAS.527.6732F}. 
This is also seen in observational kinematic studies as \citet{2019ApJ...870...32K} conclude that 75\% of their investigated clusters show signs of expansion with only the cluster M17 contracting. As M17 is very young ($\sim1$~Myr) they conclude that it is still in the first evolutionary phase of contraction. 
\citet{2024A&A...683A..10D} kinematically analyzed young clusters and found expansion for 80\% of their young ($t < 30$~Myr) clusters. This makes the kinematic result for NGC~3293 very interesting as its at an age of $\sim10 - 15$~Myr and is therefore expected only to expand.

\citet{2020MNRAS.495..663W} analyzed the kinematics of stars in a region of $\sim 0.6\degr \times \sim 0.45\degr$ around the stellar cluster NGC~3923 with \textit{Gaia}~DR2 and determined a contraction as well at a contraction velocity of $2.67\pm6.63~\rm km\,s^{-1}$, which is slightly higher than our contraction velocity of $1.32\pm0.07~\rm km\,s^{-1}$ for NGC~3293 at an angle of $38\degr$ that was calculated using only stars with proper motions showing contraction. 

\clearpage

\section{New OB star candidates in Car OB1 from \textit{Gaia}}
\label{sec:GaiaCarOB1}

\begin{figure*}
    \centering
    \begin{subfigure}[t]{8.5cm}
        \includegraphics[width=8.0cm]{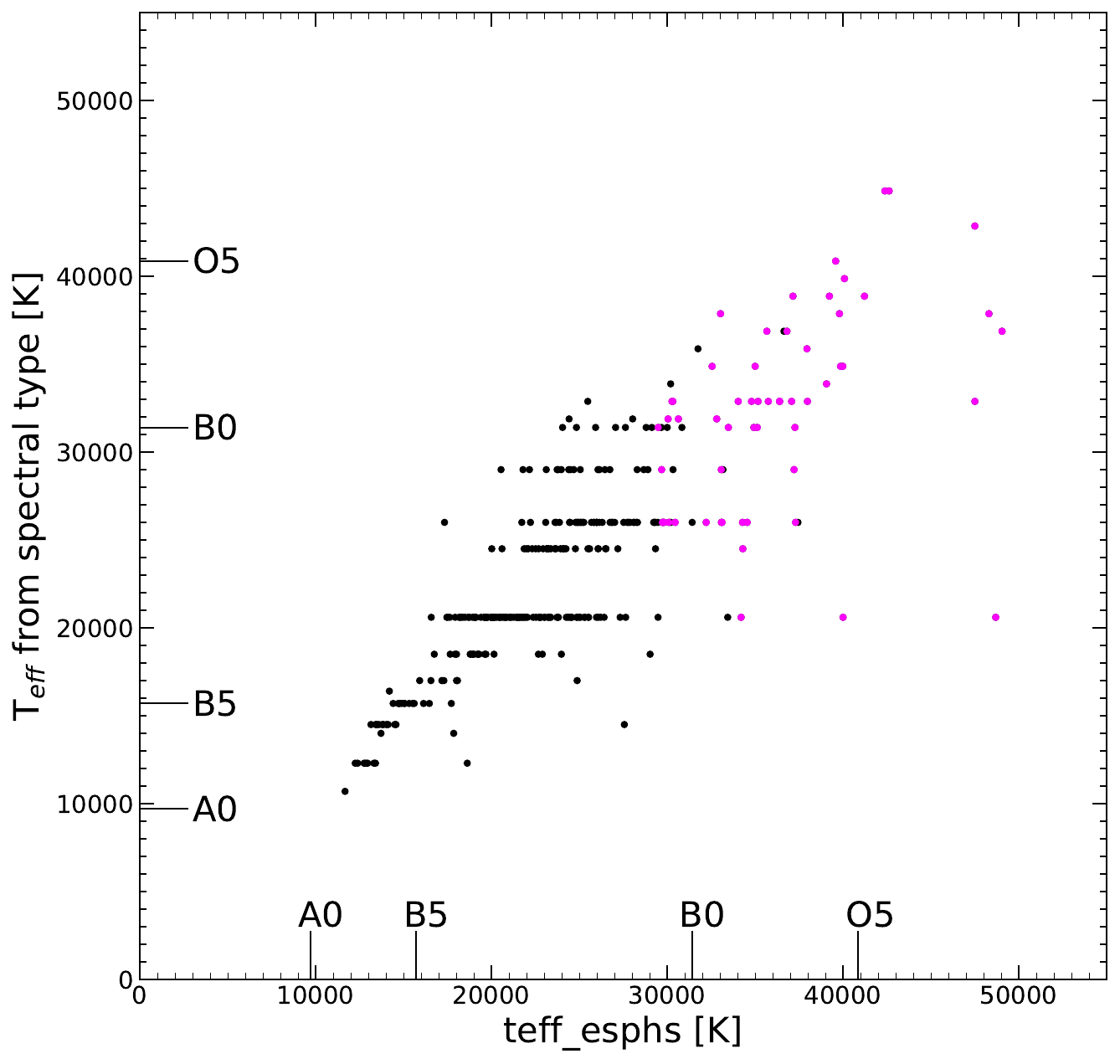}
    \end{subfigure}
    \hspace{0.25cm}
    \begin{subfigure}[t]{8.5cm}
        \includegraphics[width=8.0cm]{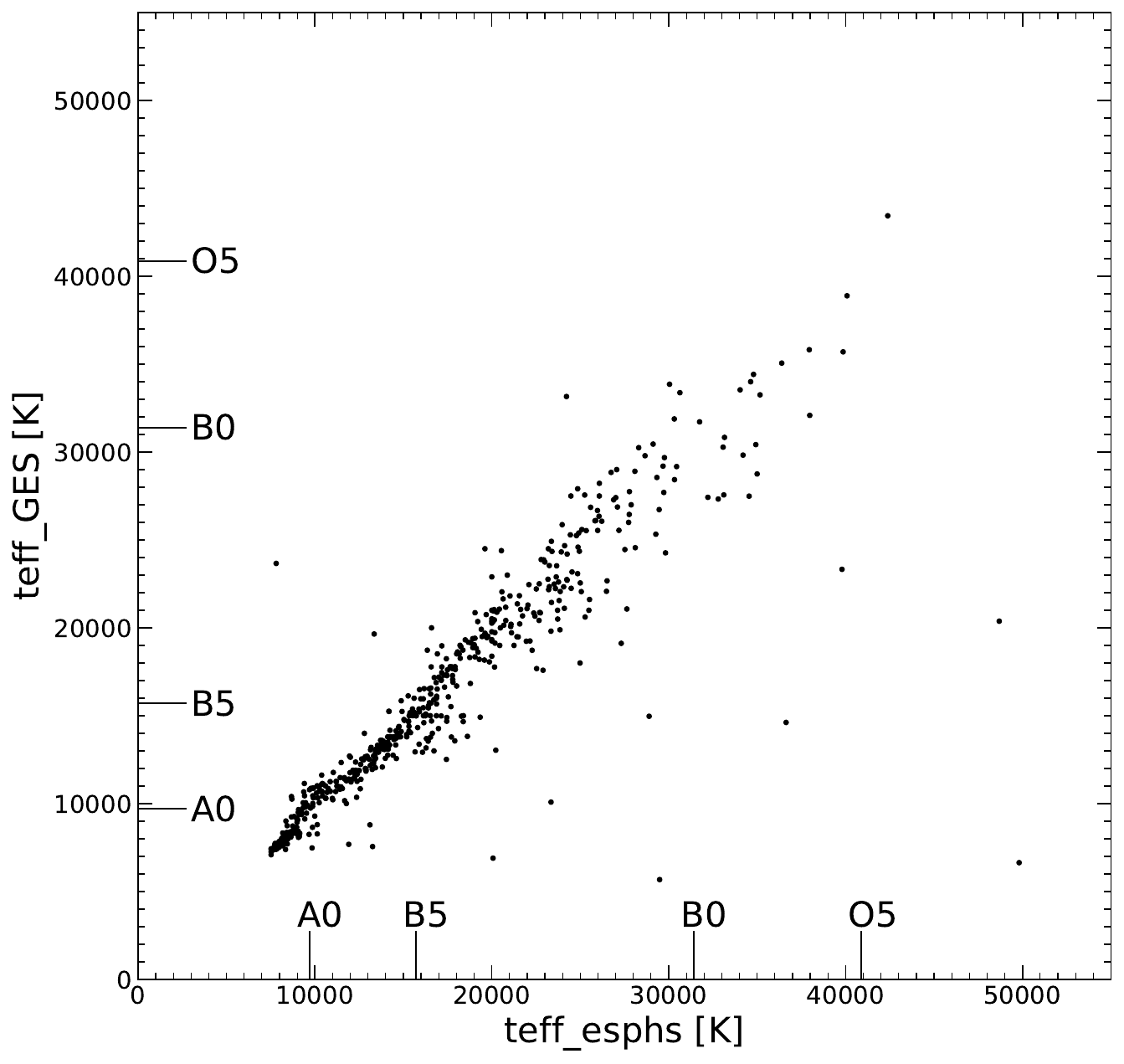}
    \end{subfigure}
    \caption{Comparison of \texttt{teff\_esphs} with the effective temperature expected from the spectral type and the effective temperature given in GES. Left panel: Effective temperature from spectral types versus the \textit{Gaia} DR3 effective temperature \texttt{teff\_esphs} for the Car~OB1 high-mass star sample. The stars highlighted in magenta have \texttt{spectraltype\_esphs} O. Right panel: \textit{Gaia} DR3 effective temperature \texttt{teff\_esphs} versus effective temperature from GES for stars in the region of Car~OB1.}
    \label{fig:TeffVergleich}
\end{figure*}

As mentioned in the introduction, the sample of known high-mass members of Car~OB1 is not yet complete, especially in the regions between and around the well-studied known clusters and the peripheral regions of the association.
We therefore used the \textit{Gaia} main catalog and the additional \texttt{astrophysical\_parameters} catalog to perform a spatially unbiased search for further high-mass members of Car~OB1 over the full area of the association. Our aim was to identify still unidentified OB stars in Car~OB1, to find out how large the total population of high-mass stars is, to ascertain how far the OB association extends, and to analyze the distributed OB star population. 

The \textit{Gaia} database provides estimates of spectral types and stellar parameters for some stars  based on \textit{Gaia} BP/RP spectra with five effective temperature estimates based on different models and data. Here we only use \texttt{teff\_gspphot} and \texttt{teff\_esphs} as \texttt{teff\_gspphot} is the temperature estimation that is available for most stars in Car~OB1 while \texttt{teff\_esphs} is focused on stars with spectral types O, B, and A.

To assess how useful and reliable the \textit{Gaia} temperature estimates are, we compare them to the effective temperatures according to their spectral type for our sample of spectroscopically identified
OB stars in Car~OB1  (see Sect.~\ref{sec:clusterprop}).
For the effective temperatures from the spectral type, we used the observational scales provided by \citet{2005A&A...436.1049M} for O-type stars and the scale from \citet{2013ApJS..208....9P} for B-type stars.

Figure~\ref{fig:TeffVergleich} shows the comparison for \texttt{teff\_esphs}. We find a best agreement between the spectral type and derived effective temperatures for the parameter \texttt{teff\_esphs} (compared to the other \textit{Gaia} derived effective temperatures), which was to be expected because only stars that were classified as O, B, or A stars by \textit{Gaia} were processed. \citet{2022gdr3.reptE..11U} also report that \texttt{teff\_esphs} performs better than \texttt{teff\_gspphot} for hot stars since it takes corrections for the $T_{\rm eff}$-extinction degeneracy into account. 

We also compare \texttt{teff\_esphs} with effective temperatures from the \textit{Gaia}-ESO Survey (GES), which is shown in Fig.~\ref{fig:TeffVergleich} on the right. In the context of the GES, optical high-resolution spectra were obtained for 2875 stars in the region of Car~OB1, and effective temperatures for 1817 of these stars were derived. 572 of these stars also have a \texttt{teff\_esphs} value in the \textit{Gaia} database available. We can see that there is generally good agreement (with a median absolute difference of 668~K) between the effective temperatures derived by \textit{Gaia} and GES, but there are some outliers. In general, we consider effective temperatures derived in the GES to be more reliable than \textit{Gaia} \texttt{teff\_esphs} values, since their GES parameters are based on spectra with higher resolution. For stars that have both temperature estimations, we therefore favor the effective temperature in GES. 
Four stars with \mbox{\texttt{teff\_esphs} $> 35\,000$\;K} (i.e.,~suggesting an O spectral type) have $T_{\rm eff, GES} < 25\,000$\;K (suggesting a mid-late B spectral type).

\subsection{Identifying new OB star candidates in the Car OB1 association}
\label{sec:NewOBDBSCAN}

We used two approaches to find new candidate OB stars: In the first approach, we applied DBSCAN on the sample of stars with \texttt{spectraltype\_esphs} O or B, and \texttt{teff\_esphs} $ \ge 17\,000$~K, in order to see whether we can find further association members by the coherent parallaxes and proper motions as expected for members of an association.

\begin{figure} 
 \resizebox{\hsize}{!}{\includegraphics{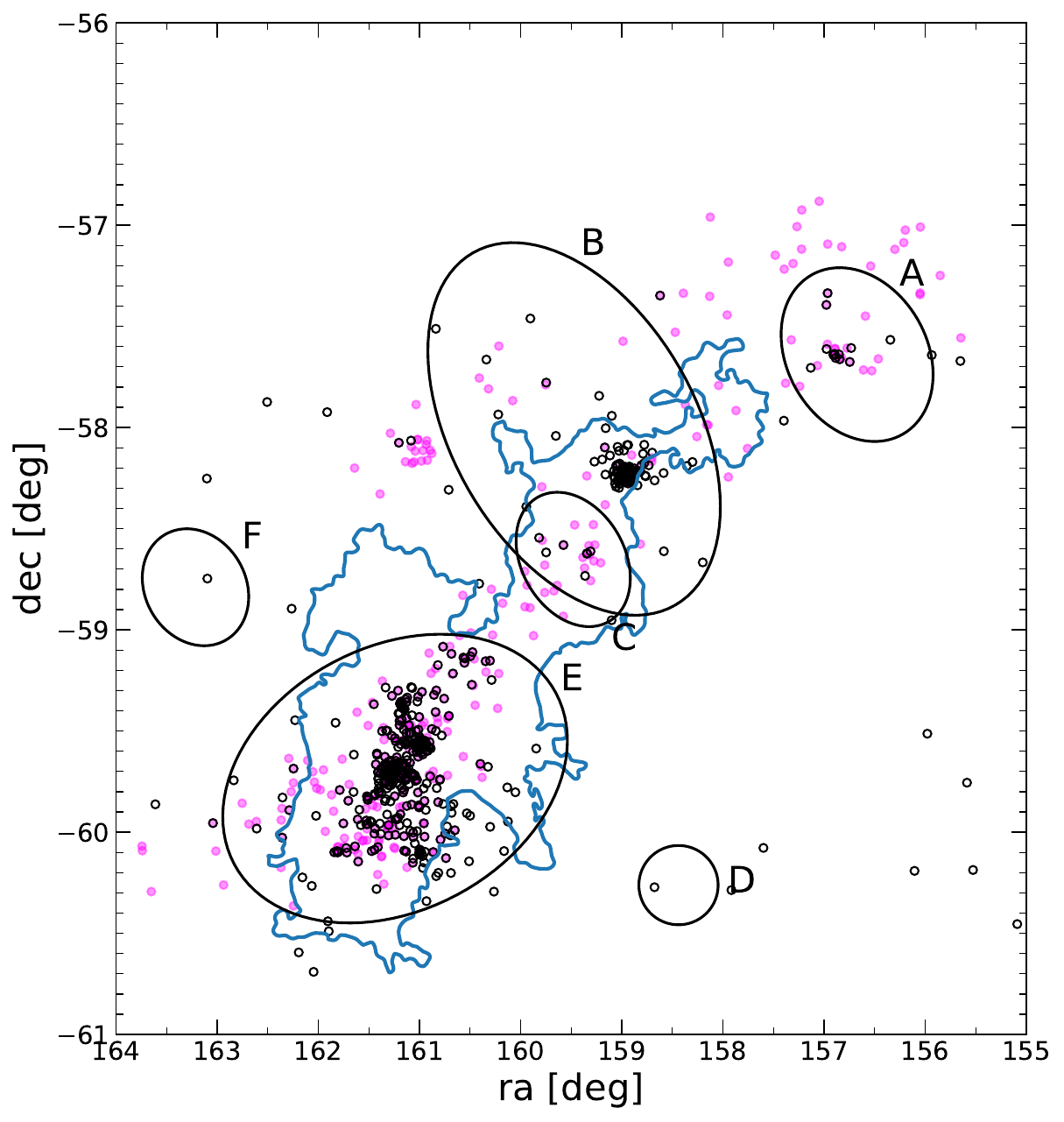}}
\caption{Map of the Car~OB1 high-mass stars and OB candidates. The Car~OB1 high-mass star sample is in black and \textit{Gaia} stars with $\texttt{teff\_esphs} \ge 17\,000$~K, which were identified by DBSCAN in the region of Car~OB1, are in magenta. The Car~OB1 subgroups from \citet{1995AstL...21...10M} are shown as ellipses.}
\label{fig:GaiaSpTDBSCAN}
\end{figure}

\begin{figure} 
 \resizebox{\hsize}{!}{\includegraphics{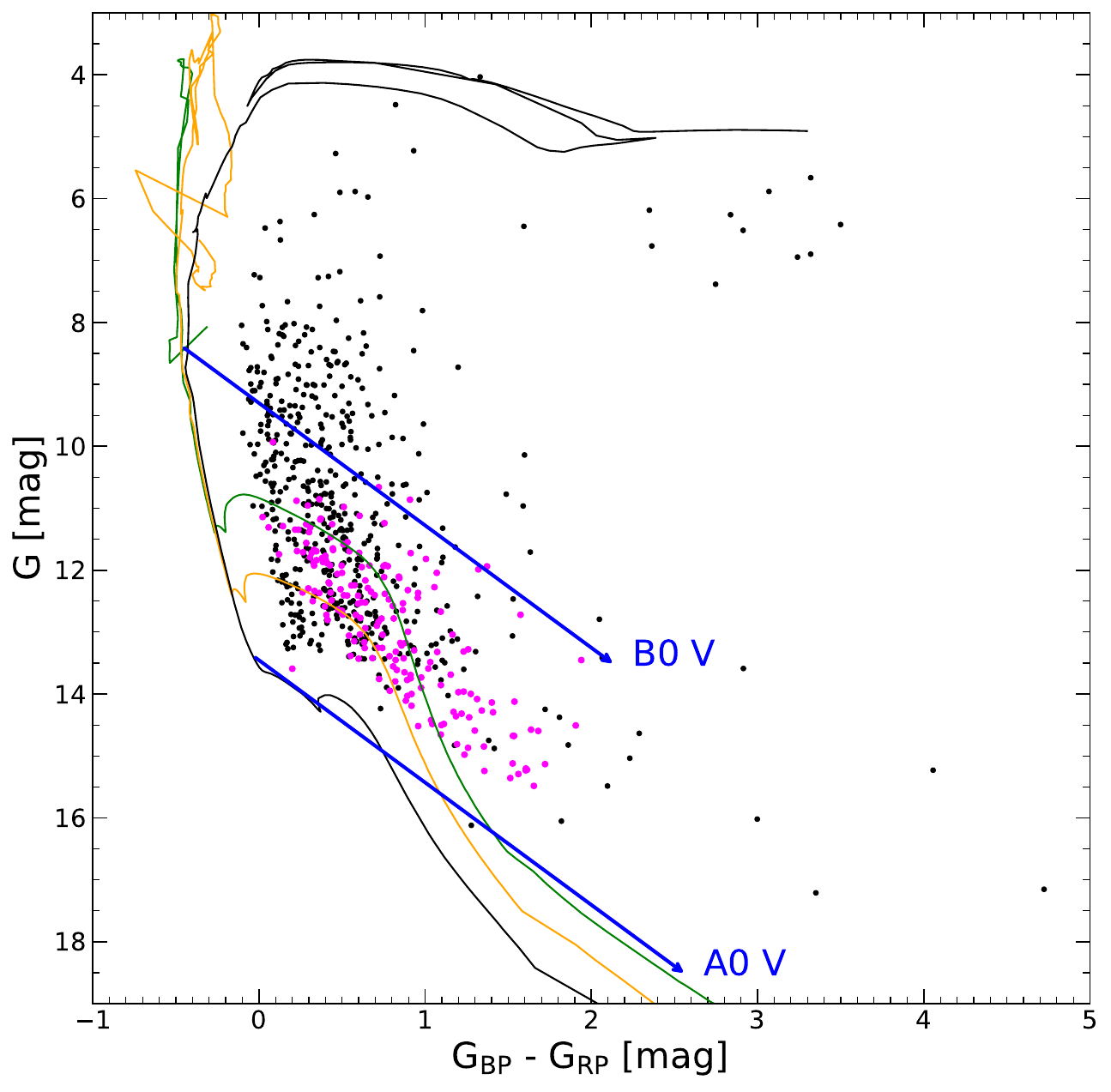}}
\caption{\textit{Gaia} CMD of the Car~OB1 high-mass stars and OB candidates. The Car~OB1 high-mass star sample is in black and new OB star candidates ($\texttt{teff\_esphs} \ge 17\,000$~K and identification with DBSCAN) are in magenta. PARSEC v3.7 isochrones \citep{2012MNRAS.427..127B} at ages 1~Myr (green), 3~Myr (red), and 10~Myr (black) are overplotted. The arrows show the reddening vectors of a 1~Myr old B0 and a 10~Myr old A0 star for $A_V = 6$~mag.}
\label{fig:CMDSpT}
\end{figure}

We started by selecting stars inside a box centered around the Carina Nebula with $25\degr$ side length side, \texttt{spectraltype\_esphs} O or B and \texttt{teff\_esphs} $ > 17\,000$~K in \textit{Gaia}~DR3. This leads to a sample containing 12\,915 stars. We then applied the clustering algorithm DBSCAN to the stars while using the same method as in Sect.~\ref{sec:FoundClusters} to determine the density parameter $\epsilon$.

Figure~\ref{fig:GaiaSpTDBSCAN} shows the stars that were identified by DBSCAN in an overdensity that has a similar distance as Car~OB1. DBSCAN recovered Car~OB1 as four overdensities with the Carina Nebula was as one large overdensity, the clusters NGC~3293 and IC~2581 as another one, while the clusters UBC~501 and NGC~3324 were identified as separate clusters. The stars of our Car~OB1 high-mass sample are plotted in black and show that the association found by DBSCAN extends farther north and north-west than our Car~OB1 high-mass sample. DBSCAN also identified Car~OB1 members in the region of the cluster UBC~501, which has only one cluster member in our Car~OB1 high-mass star sample.

In total 497 stars were identified by DBSCAN to be part of Car~OB1, out of which 472 (95\%) have a $2\sigma$ distance interval, which is consistent with the distance  of Car~OB1. 
Of these stars, 215 (46\%) lie inside one of the cluster circles defined in Sect.~\ref{sec:clusterprop} and can therefore be regarded as members in one of the clusters. The remaining 257 (54\%) stars can be interpreted as the distributed population in Car~OB1. This leads to a ratio of 46:54 of clustered versus distributed population. 

The 235 stars identified by DBSCAN as association members were already part of our Car~OB1 high-mass star sample and thus are spectroscopically confirmed high-mass stars. The other 237 stars have not been assigned to Car~OB1 before. We searched for spectral types for the 237 OB candidates and find spectral types for 38 stars (2 O-, 16 B-, 3 A-, and 17 OB-type), while 199 stars have no spectroscopically determined spectral type available in Simbad. 
Out of these 199 stars, 20 have a effective temperature in GES. 

We excluded ten stars as their GES effective temperature is below our cutoff temperature of 17\,000~K.
We classify stars with no available spectral type as O star candidates if $\texttt{teff\_esphs} > 31\,900$~K (effective temperature of a O9.5 V star according to \citep{2005A&A...436.1049M}), which leads to 5 new O-type and 184 new B-type candidates. 

Figure~\ref{fig:CMDSpT} shows a color-magnitude diagram of the Car~OB1 high-mass star sample and the DBSCAN identified \textit{Gaia} OB candidate population. The blue arrows show the reddening arrows of a B0 and an A0 star, we can see that all new \textit{Gaia} OB candidates in Car~OB1 lie above the A0 arrow and have photometry consistent with being an O- or B-type star.

\subsection{Revealing the distributed population of OB stars in Car OB1 with \textit{Gaia}}
\label{sec:OBstarsdistributed}

In the second approach, we considered all stars in the area with 
\texttt{spectraltype\_esphs} O or B and \texttt{teff\_esphs} $ \ge 17\,000$~K, and selected those stars that have parallaxes consistent with being members of Car~OB1.
This approach can also identify stars with proper motions deviating from the typical values of the association, as may be the case for ``run-away'' stars, for example, and it allows us to find more of the distributed Car~OB1 population.

\begin{figure}
    \resizebox{\hsize}{!}{\includegraphics{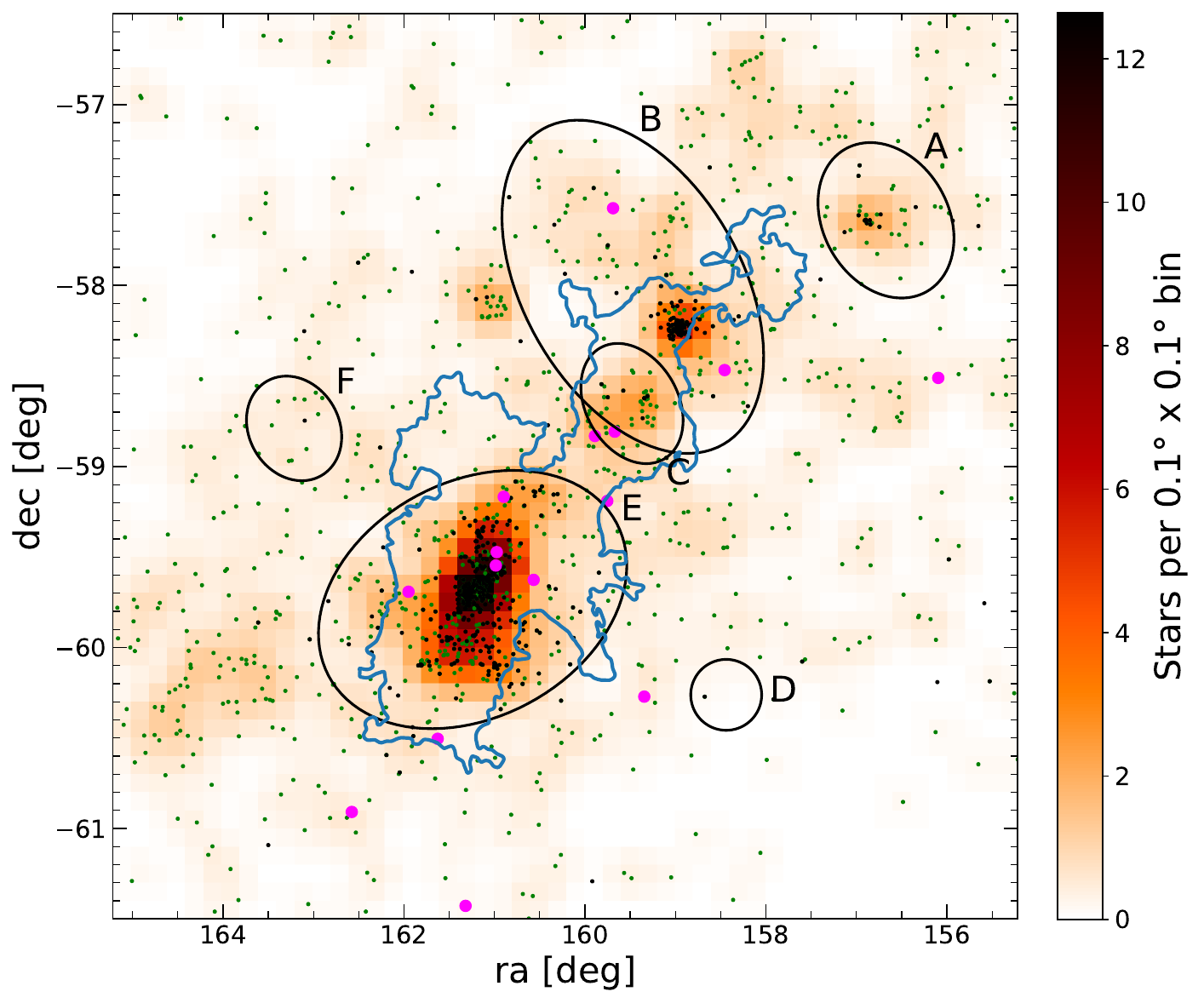}}
    \caption{Map of smoothed two-dimensional spatial histograms of stars with a compatible distance to Car~OB1 and \texttt{teff\_esphs} $ \ge 17\,000$~K. The Car~OB1 subgroups from \citet{1995AstL...21...10M} are shown as ellipses, and the CNC contour is in blue. Stars with compatible distance and \texttt{teff\_esphs} in green, stars from the Car~OB1 high-mass star sample in black, and new O-type candidates in magenta.}
    \label{fig:DensityStars}
\end{figure}

\begin{figure} 
 \resizebox{\hsize}{!}{\includegraphics{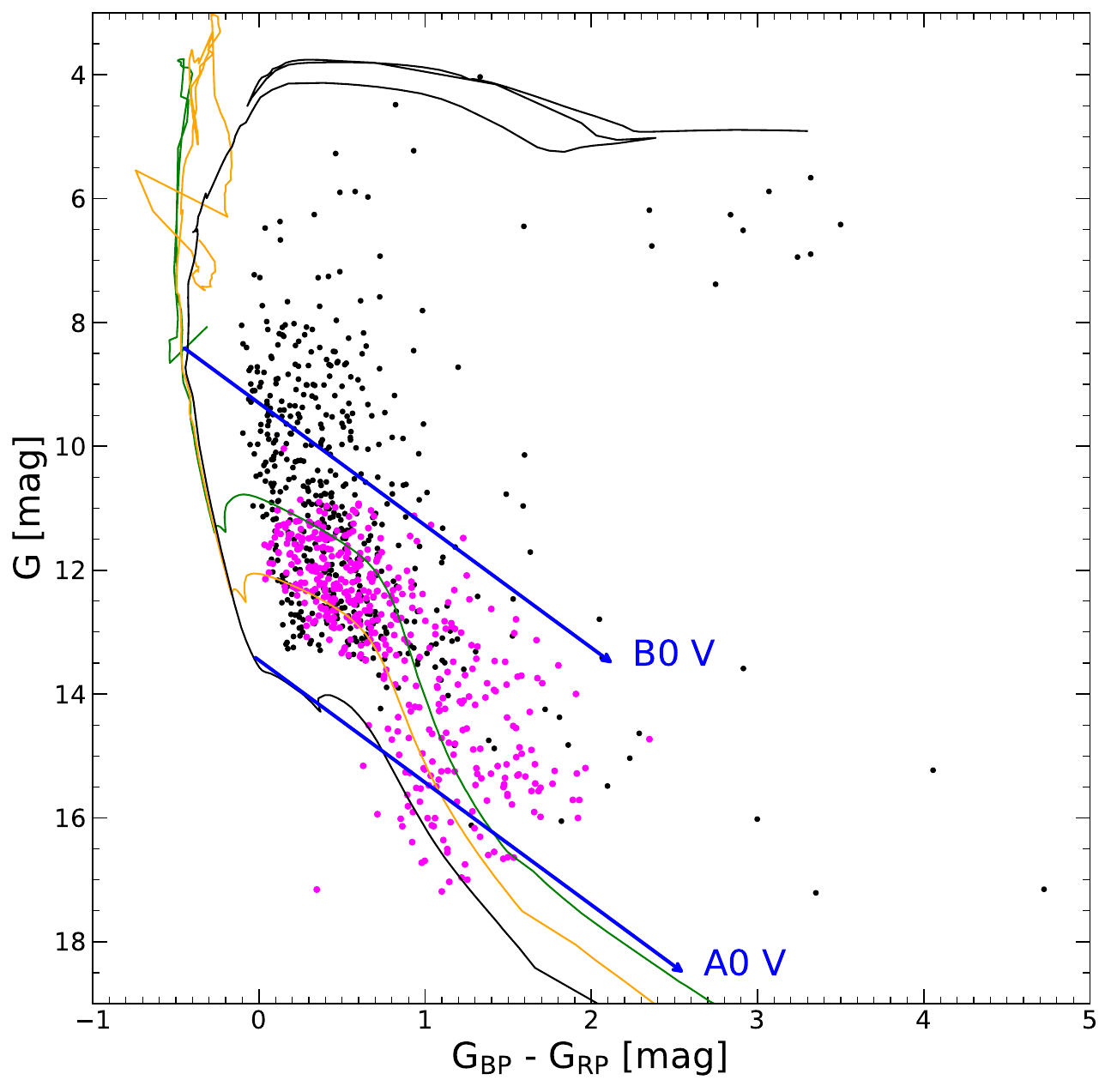}}
\caption{\textit{Gaia} CMD of the Car~OB1 high-mass stars and OB candidates. The Car~OB1 high-mass star sample is in black and new OB star candidates from Fig.~\ref{fig:DensityStars} are in magenta. PARSEC v3.7 isochrones \citep[see][]{2012MNRAS.427..127B} at ages 1~Myr (green), 3~Myr (red), and 10~Myr (black) are overplotted. The arrows show the reddening vectors of a 1~Myr old B0 and a 10~Myr old A0 star for $A_V = 6$~mag.}
\label{fig:CMDSpTDistributed}
\end{figure}

Figure~\ref{fig:DensityStars} shows a two dimensional spatial histograms of stars in \textit{Gaia} DR3 whose $2\sigma$ distance interval overlaps with the distance interval of Car~OB1 ([2.25, 2.45]~kpc) and which have \texttt{teff\_esphs} $> 17\,000$~K. The whole Car~OB1 can be seen as an overdensity with densities peaking in the central regions of Trumpler~14 and 16. 

A clear density decrease from the central Carina Nebula and the surrounding clusters toward the periphery of the field can be seen. The star density shows a strongly elongated shape, following the large-scale structure of the association (Fig.~\ref{fig:DensityStars}).
Figure~\ref{fig:DensityStars} reveals numerous new OB star candidates around the already known members of the association, and highlights the distributed population, that had remained largely unidentified so far.

The selection of stars in a box of $5\degr \times 5\degr$ centered around $(10^{\rm h}40^\mathrm{m}48^\mathrm{s},-59\degr)$ with a $2\sigma$ distance interval overlapping with the distance interval of Car~OB1 and \texttt{teff\_esphs}$> 17\,000$~K (as seen in Fig.~\ref{fig:DensityStars}) yields in total 1233 stars, with 306 being already a part of the Car~OB1 high-mass star sample, and 237 identified as Car~OB1 members in Sect.~\ref{sec:NewOBDBSCAN}. 
Out of the 690 stars new OB candidates, 233 have spectral types available in Simbad, including 2 O-, 186 B-, and 15 A-type stars and 30 stars with spectral type OB. Figure~\ref{fig:CMDSpTDistributed} shows a color-magnitude diagram for the 457 stars without available spectral type. Almost all stars (92\%) have optical photometry compatible with being an O- or B-type star. The 37 stars which are located below the reddening vector of an A0~V star are excluded from our list of possible OB candidates. The GES catalog contains effective temperature values for 7 stars in this sample which leads to the exclusion of 5 stars as their GES effective temperature is below 17\,000~K. This leads to 415 new OB star candidates in the region of Car~OB1 with 10 being O-type candidates and 405 B-type candidates and 218 spectroscopically identified stars in the Car~OB1 region.

If all of these 604 new OB candidates from Sects.~\ref{sec:NewOBDBSCAN} and \ref{sec:OBstarsdistributed} are actually high-mass stars in Car~OB1, then the ratio of clustered versus distributed population would be 26:74 (compared to 58:42 for the Car~OB1 high-mass star sample, and 46:54 for the DBSCAN-selected Car~OB1 population).
An $\sim 75\%$ fraction of distributed stars may appear quite high, but we note that similar fractions have been reported for other OB associations (e.g., for Sco~OB2, a ratio of 14.5:85.5 was reported by \citep{2019A&A...623A.112D}). All new OB star candidates are listed in Table~\ref{tab:CarOB1HighMass} and are labeled with 'TD' for stars from Sect.~\ref{sec:NewOBDBSCAN} and 'T' for stars from Sect.~\ref{sec:OBstarsdistributed} in the column 'Selection'.

\subsection{An estimate of the total stellar mass of Car~OB1}
\label{sec:IMF}

\begin{figure}
    \centering
\resizebox{\hsize}{!}{\includegraphics[width=8.0cm]{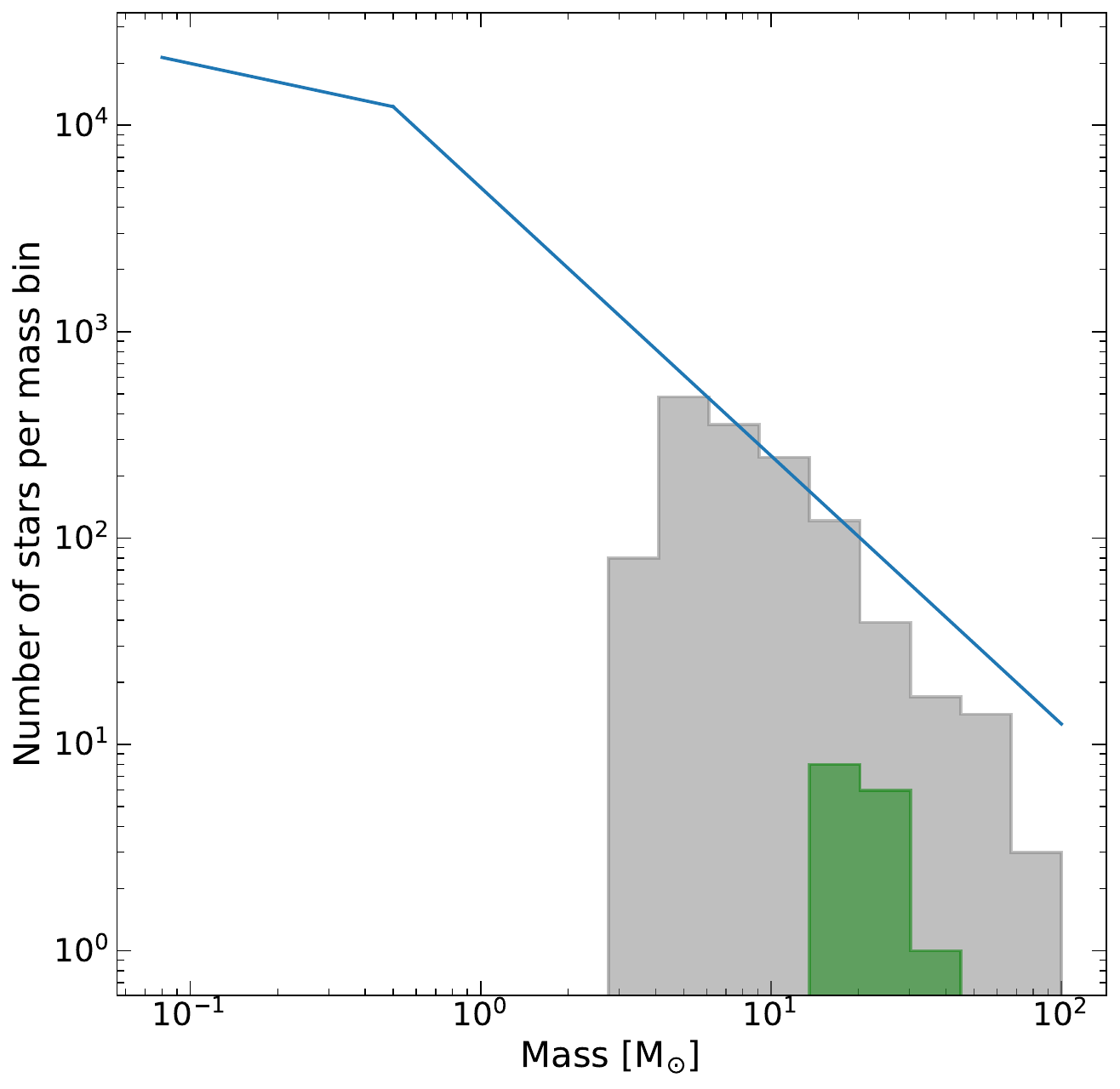}}
\caption{Histogram of the masses of the stars for our Car~OB1 high-mass star sample combined with the new identified OB star candidates from Sects.~\ref{sec:NewOBDBSCAN} and \ref{sec:OBstarsdistributed}. A fit of the \citet{2001MNRAS.322..231K} IMF is shown in blue. The contribution by our new O-type candidates is shown in green.}
\label{fig:IMF}
\end{figure}

In order to investigate the mass function and to obtain an estimate of the total number of stars in Car~OB1, we have combined our Car~OB1 high-mass star sample with our new OB star candidates in the Car~OB1 region from Sects.~\ref{sec:NewOBDBSCAN}~and~\ref{sec:OBstarsdistributed} which yields a total sample of 1374 stars which includes 92 O-type, 3 WR, 36 supergiants, and one luminous blue variable star ($\eta$ Car). 
We have converted spectral types to masses with the table from \citet{2013ApJS..208....9P}. Stars selected by \texttt{teff\_esphs} were taken into account by translating stellar temperatures to stellar masses with tables provided by \citet{2005A&A...436.1049M} and \citet{2013ApJS..208....9P}. We have used mass determinations provided in the literature \citep{2019A&A...625A..57H,2019A&A...621A..92S,2019A&A...621A..63G,2023MNRAS.519.5882S} for the 3 WR stars in our sample, HD~93129~A, and $\eta$ Car. We have not included supergiants in our IMF plot (Fig.~\ref{fig:IMF}) since the exact masses of supergiants are difficult to determine due to winds and high mass loss rates.

Figure~\ref{fig:IMF} shows a histogram of the star's masses and an approximate fit of a Kroupa IMF. The slope of the IMF agrees well with the histogram for masses between $\approx 4$ and $20.5\,M_{\odot}$ while it predicts more stars at higher masses than are present in our sample. This discrepancy is probably caused by the missing 36 supergiant stars due to their highly uncertain masses. 
Based on the observed mass function between $\approx 8$ and $20.5\,M_{\odot}$ with 367 stars in our combined Car~OB1 sample,
an extrapolation based on the Kroupa IMF leads to a total number of $\approx 79\,800$ stars with masses from $0.08\,M_{\odot}$ to $100\,M_{\odot}$, and a total stellar mass of $\approx 45\,800\,M_{\odot}$ in Car~OB1.
An extrapolation based on the 140 stars (including the 36 supergiants) with masses $> 18\,M_{\odot}$ leads to a total number of $\approx 69\,100$ stars with masses from $0.08\,M_{\odot}$--$100\,M_{\odot}$ and a total stellar mass of $\approx 39\,600\,M_{\odot}$ in Car~OB1.
It is important to note that both estimates are only lower limits, because the cloud complex contains numerous obscured stars. The sample of stars identified by optical spectroscopy, and also the Gaia-based sample,
are certainly incomplete to some degree.

Our estimate of a total population of $N \sim 8\times10^4$ stars in Car~OB1 is in good agreement with
the extrapolation based on the number of X-ray detected young stars by \cite{CCCP-Clusters}. It also confirms Car~OB1 as probably being the most massive of the well-studied OB associations in our galaxy. Our results suggest Car~OB1 being
substantially more massive than Cyg~OB2 \citep[with a total stellar mass of $\sim 16\,500\,M_{\odot}$;][]{2015MNRAS.449..741W}, which was considered to be the probably most massive OB association in our galaxy by \citet{2020NewAR..9001549W}, and also exceeding the total stellar mass of $\sim 36\,200\,M_{\odot}$ for the Per~OB1 association estimated by \citet{2020MNRAS.493.2339M}.
Car~OB1 is therefore comparable to the star-forming region 30 Doradus in the Large Magellanic Cloud, which is the most luminous star-forming region in our Local Group \citep{1984ApJ...287..116K}, with a total stellar mass of $110\,000\,M_{\odot}$ \citep{2013A&A...558A.134D}.

\section{Large-scale kinematics and expansion of the Car OB1 association}
\subsection{Expansion of Car~OB1}
\label{sec:ExpCarOB1}

\begin{figure}
    \centering
\resizebox{\hsize}{!}{\includegraphics[width=8.0cm]{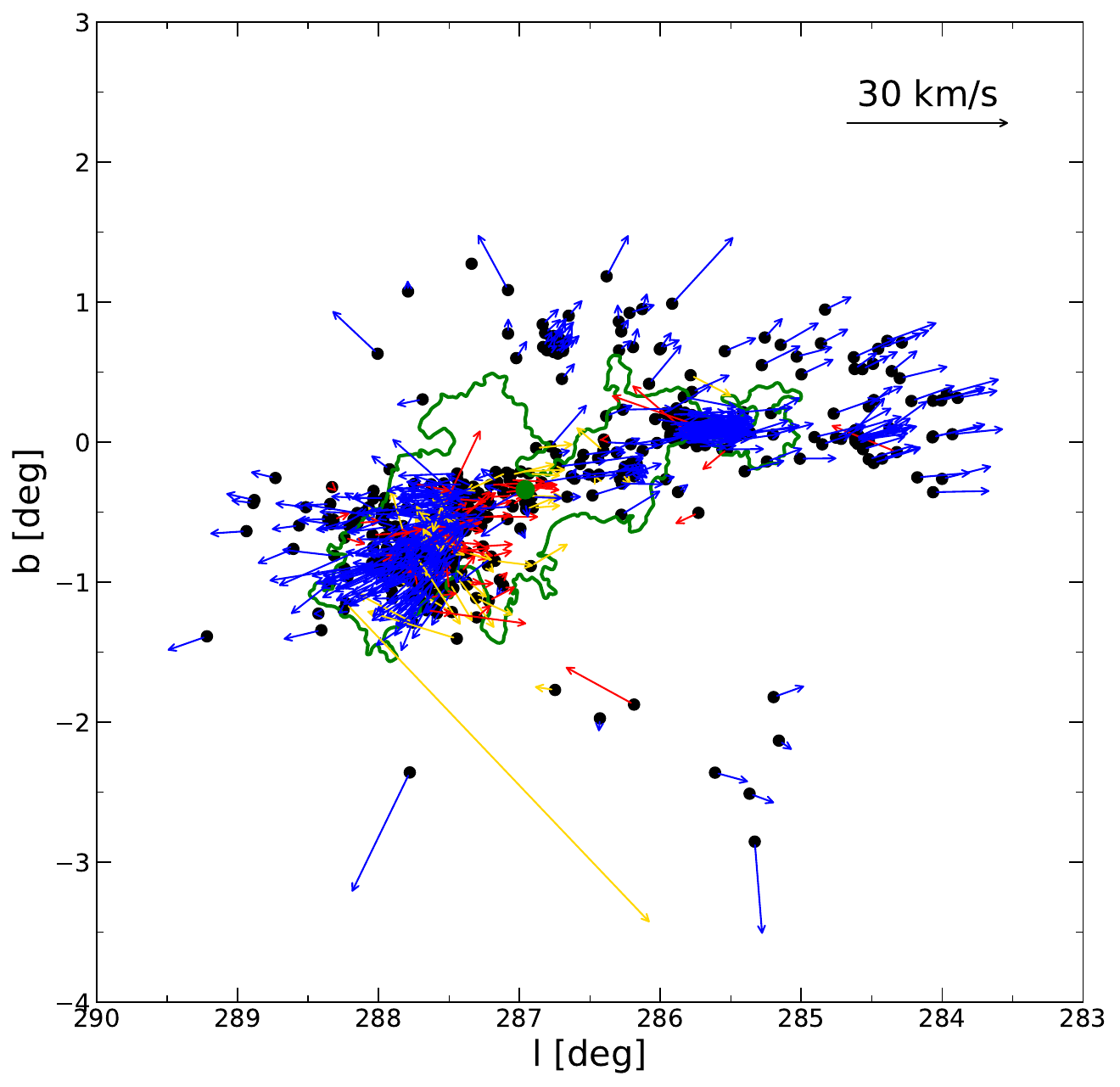}}
\caption{Relative proper motions of the stars in our combined Car~OB1 high-mass star sample. Motions pointing outward are shown with blue, inward motions with red, and rotating motions with yellow arrows. The green dot represents the association's center and the green contour the outline of the CNC.}
\label{fig:CarOB1mitPfeilen}
\end{figure} 

Previous studies of Car~OB1 \citep[see][]{2020MNRAS.493.2339M} found an expansion of the association. In order to confirm and quantify this expansion with our stellar samples, we analyze the kinematics of the stars in the Car~OB1 association. For this we combine the stars from our high-mass Car~OB1 sample with the kinematically coherent OB star candidates which were identified by DBSCAN in Sect.~\ref{sec:NewOBDBSCAN} which yields a total sample of 741 stars.
We plot the high-mass stars' proper motions relative to the center as arrows to visualize the kinematics of the association as shown in Fig.~\ref{fig:CarOB1mitPfeilen}. The stars are divided into three categories according to the angle between the vector of their relative proper motion and the vector of the star's position to the association's center. Stars whose proper motions are directed outward (angles between $120\degr$ and $240\degr$) are shown in blue, inward (angles between $300\degr$ and $60\degr$) in red, and moving tangentially around the cluster (angle between $240\degr$ and $300\degr$ or between $60\degr$ and $120\degr$) in yellow. Most arrows in Fig.~\ref{fig:CarOB1mitPfeilen} are in blue and pointing outward (572, 77\%) while there are fewer stars whose proper motions are pointing inward or rotating (101 and 63, respectively) which indicates that the whole Car~OB1 association is expanding. 
The relative proper motions show an expansion of the whole association and also confirm that members of the cluster UBC~501 have consistent proper motions with an expansion of the association which validates UBC~501's membership in the Car~OB1 association.

In order to determine the outward velocity of the association, we have calculated the projected outward velocity of each star with respect to the association's center and taken the mean which results in a velocity of $v_{\rm out} = 5.25\pm0.02~\rm km\,s^{-1}$ for Car~OB1. A large part of the stars with positive outward velocity (291 out of 612) move out at an angle between $12\pm 15 \degr$ with respect to the galactic plane which we assume to be the expansion angle of the whole association. Taking only stars which are contributing to the expansion, this results in a expansion velocity of $7.05\pm0.03~\rm km\,s^{-1}$ at an angle of $12\degr$ in galactic coordinates. This expansion follows along the elongation of the CNC cloud complex as shown in Fig.~\ref{fig:CarOB1mitPfeilen}.
Interestingly, this expansion angle of Car~OB1 at $42\degr$ (in ecliptic coordinates) is close to the angle of $38\degr$ at which NGC~3293 contracts.

\citet{2020MNRAS.493.2339M} investigated 28 OB associations using \textit{Gaia}~DR2 and found clear evidence of expansion (significance $> 3\sigma$) for 9 associations (32\%) including Car~OB1, Ori~OB1, and Sco~OB1. They determined a expansion velocity of $5.0\pm1.7~\rm km\,s^{-1}$ for Per~OB1 which is similar to the expansion velocity we estimated for Car~OB1.

\subsection{Traceback analysis of Car~OB1}
\label{sec:TracebackCarOB1}

In order to investigate the temporal evolution of the spatial extent of Car~OB1, we performed a traceback analysis for the stars in our combined Car~OB1 star sample.
We used the epicyclic equations from \citet{2006MNRAS.373..993F} to approximate the motion of the stars. They are given in galactocentric coordinates \textit{XYZ} with galactic space velocities \textit{UVW}: 

\begin{equation}
  \begin{aligned}
    X(t) = &\,X(0) -\frac{V(0)}{-2B}\,(1-\mathrm{cos}(\kappa\,t))+\frac{U(0)}{\kappa}\,\mathrm{sin}(\kappa\,t) \\
    Y(t) = &\,Y(0) +2\,A\,\left(X(0)-\frac{V(0)}{-2B}\right)\,t  \\
    &\,+\frac{\Omega_0}{-B\,\kappa}\,V(0)\,\mathrm{sin}(\kappa\,t)+\frac{2\Omega_0}{\kappa^2}\,U(0)\,(1-\mathrm{cos}(\kappa\,t)) \\ 
    Z(t) = &\,\frac{W(0)}{\nu}\, \mathrm{sin}(\nu\,t)+Z(0)\,\mathrm{cos}(\nu\,t).
  \end{aligned}
  \label{eq:XYZTraceback}
\end{equation}

Here \textit{A} and \textit{B} are the Oort constants, and $\Omega_0$ the angular velocity of the sun's circular orbit. $\kappa$ is the epicyclic frequency and is defined as $\kappa = \sqrt{-4\,\Omega_0\,B}$, while $\nu$ is the vertical oscillation frequency and is defined as $\nu = \sqrt{4\,\pi\,G\,\rho_0}$ with \textit{G} as the gravitational constant and $\rho_0$ as the local density. We have used values of $A = 15.1 \pm 0.1~\rm km\,s^{-1}\,kpc^{-1}$, $B = -13.4 \pm 0.1~\rm km\,s^{-1}\,kpc^{-1}$, and $\Omega_0 = 28.5\pm 0.1~\rm km\,s^{-1}\,kpc^{-1}$ \citep{2019ApJ...872..205L}, and $\rho_0 = 0.102~\rm M_{\odot}\,pc^{-3}$ \citep{2004MNRAS.352..440H}.
We used the median radial velocity of $-7.1\pm~5.4~\rm km \, s^{-1}$ and a distance of $2.35\pm0.1$~kpc for all stars for the conversion to galactocentric coordinates in order to minimize the effect of the greater uncertainty in measurements in the line-of-sight direction. 

\begin{figure}
    \centering
\begin{subfigure}[b]{8.cm}
   \includegraphics[width=7.5cm]{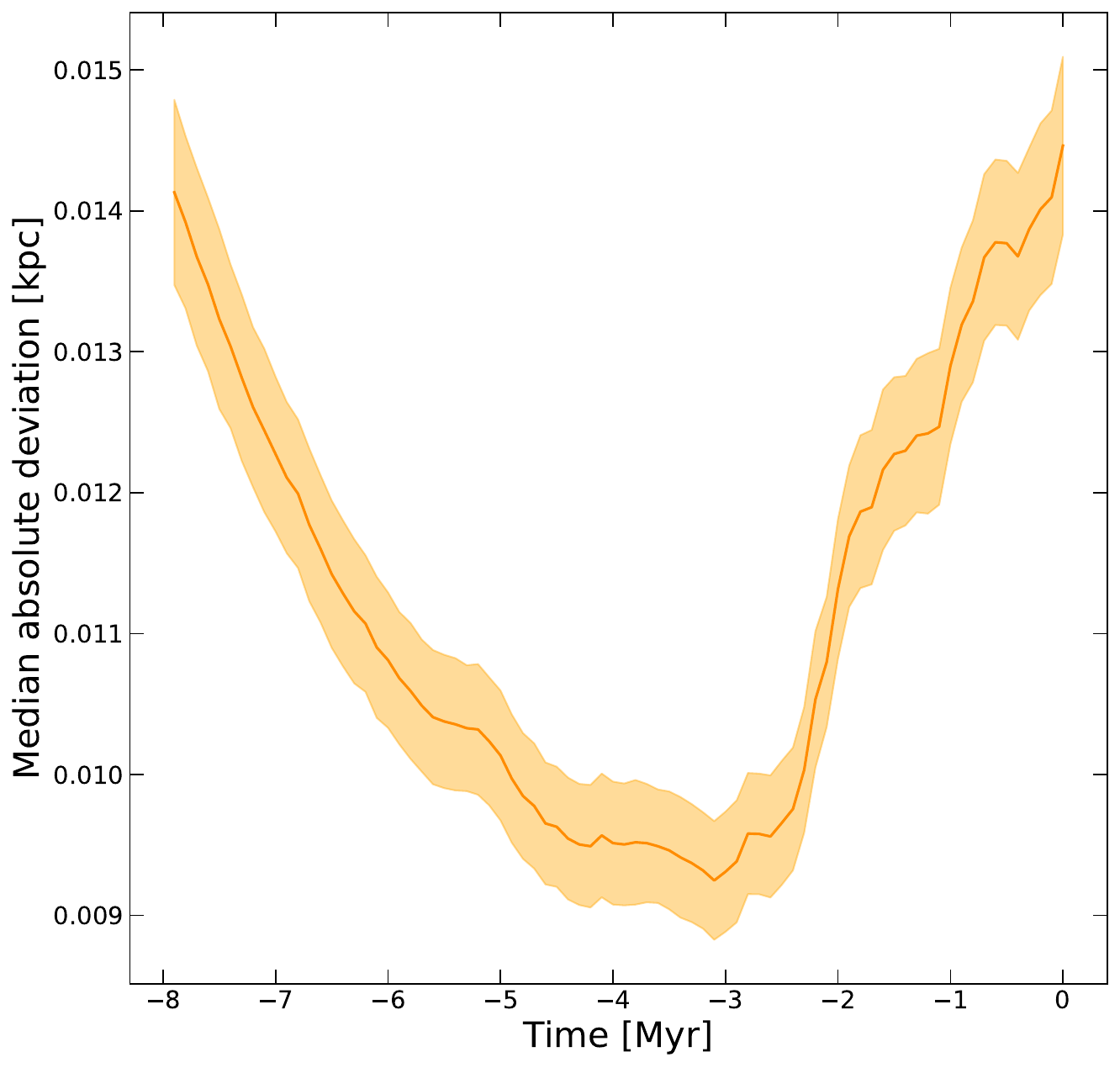}
    \end{subfigure}
\begin{subfigure}[b]{8.cm}
   \includegraphics[width=7.5cm]{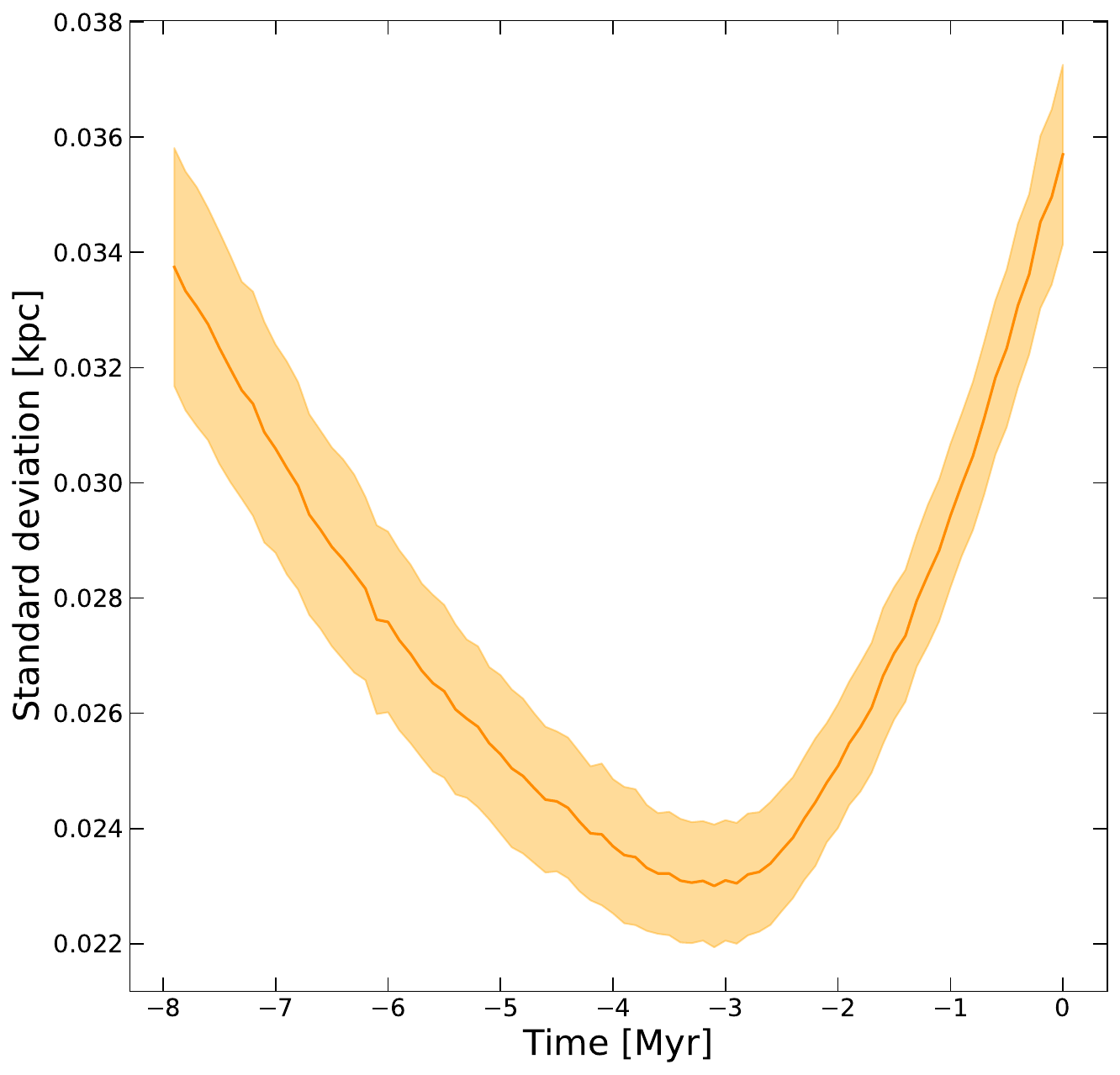}
    \end{subfigure}
        \begin{subfigure}[b]{8.cm}
   \includegraphics[width=7.5cm]{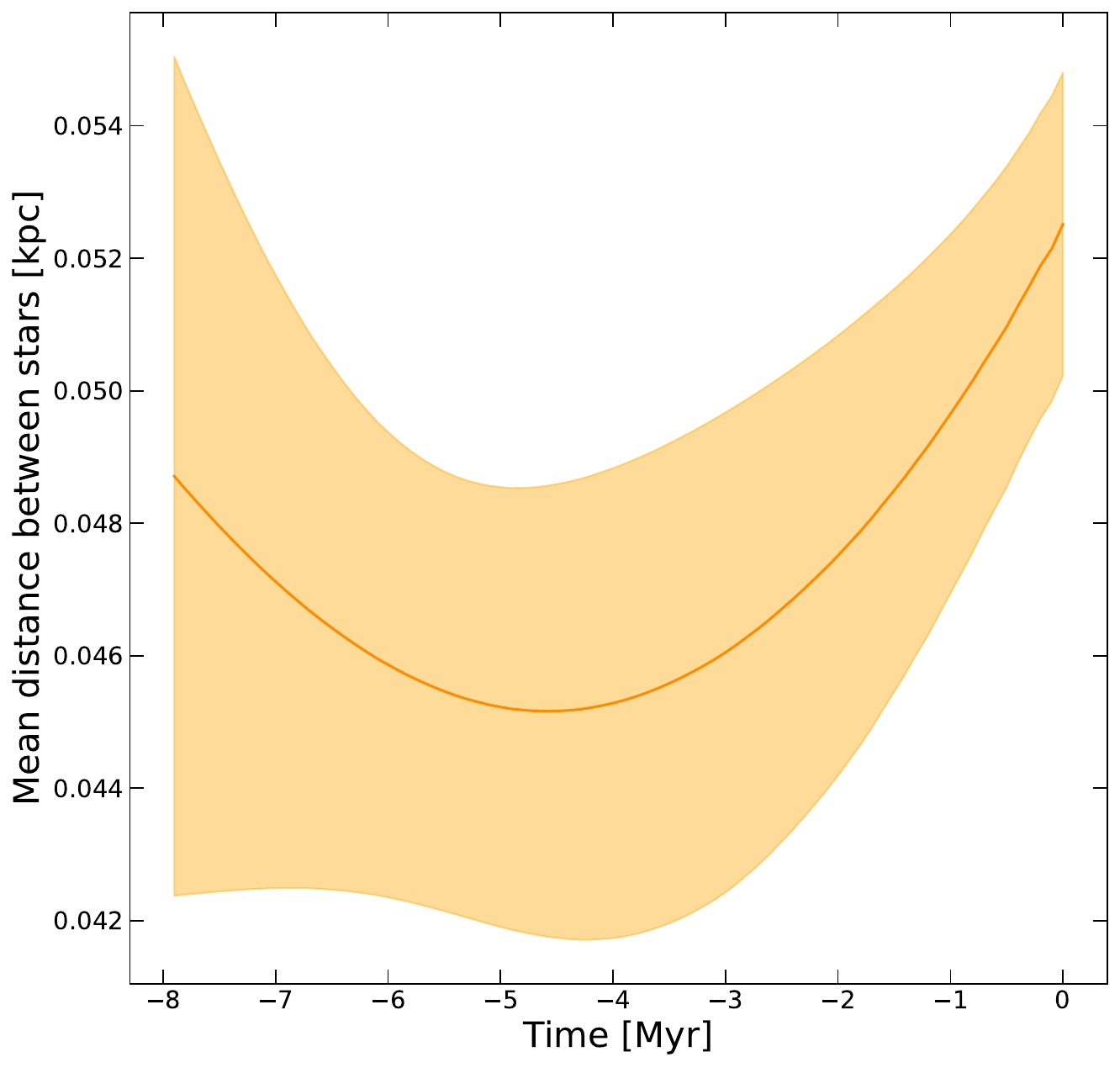}
    \end{subfigure}
    \caption{Three methods to trace the size of Car~OB1 over time. The methods are median absolute deviation of the stars' position to the association's center (upper plot), standard deviation of the association (middle), and mean distance between all stars (lower plot).}
    \label{fig:TracebackMAD}
\end{figure}

Following the method described in \citet{2023MNRAS.522.3124Q}, we explored at which time the association had its smallest extent. For this we traced back the stars' position using equations~\ref{eq:XYZTraceback} in steps of 0.1~Myr. At each time step we calculate and save the median absolute deviation to the associations center in \textit{XYZ} coordinates. The smaller the median absolute deviation, the compacter is the association. In order to derive uncertainties we add errors with random normally distributed weights to each variable and carry out the calculation 1000 times at each time step. We then save the median and standard deviation of the median absolute deviation at each time step and plot it over time as seen in the upper plot in Fig.~\ref{fig:TracebackMAD}.

We also quantified the spread of the association over time with two more measures of dispersion: the standard deviation of the association (middle plot in Fig.~\ref{fig:TracebackMAD}) and the mean distance between all stars in the sample (lower plot in Fig.~\ref{fig:TracebackMAD}).
All dispersion measures give similar results, with a minimum around 3--4~Myr. 

Although there is clear evidence for large-scale expansion of Car~OB1 during the last 3--4~Myr, we note that the absolute change of the spatial extension during this time period is relatively moderate: the current extent of the stellar populations is $\approx 220$~pc, and during the time of the minimum the extent was $\approx 190$~pc.
This result is in agreement with the conclusions from \citet{2020MNRAS.495..663W}, who found that the expansion seen in most OB associations cannot be interpreted as the expansion of one initial monolithic dense cluster. OB associations are thus not the product of an expanding cluster, but instead originate from a large-scale globally unbound and highly substructured initial configuration.
A more detailed analysis of the temporal evolution of the full Car~OB1 association and its individual components will be the topic of a separate study.

\section{Conclusions and summary}
\label{sec:Conclusions} 

The main results of our astrometric analysis of the high-mass stars over the full extent of the Car~OB1 association can be summarized as follows:
\begin{enumerate}

\item Applying a cluster-finding method to the \textit{Gaia}~DR3 data of the area, we have found 15 clusters and groups that have a spatial position, distance, and age compatible with Car~OB1. Four identified groups in Car~OB1 were not mentioned in the literature before. Seven of our recovered clusters, including the well-known clusters Trumpler 14, 15, and 16, NGC~3324, IC~2581, and Bochum~11, are split into subclusters. We also find that the cluster UBC~501 is part of Car~OB1. 
We find that subgroups A, B, C, and E of Car~OB1, which were identified by \citet{1995AstL...21...10M}, agree well with the our distribution of clusters, while we did not detect any clusters or groups with a compatible distance that are located in or near subgroups D and F.

\item We have assembled a new census and collected the largest sample of high-mass stars with a known spectral type in Car~OB1 so far, with 517 stars (including 88 O-type stars, three WR stars, and 36 supergiant stars) that have compatible distances. Our census of Car~OB1 provides a very substantial extension in the area covered and the star numbers compared to previously published censuses by \citet{Smith06} (70 O-type and WR stars) and by \citet{2023A&A...671A..20B} (80 O-type, WR, and supergiant stars), both of which were restricted to the central Carina Nebula.

\item We analyzed the kinematic properties of the clusters and found signs of expansion or contraction for 80\% of the clusters in Car~OB1 at a level of at least $1\sigma$ significance. The clusters Trumpler~14 and 16 show clear evidence of expansion at $5.2\sigma$ and $2.8\sigma$, respectively, while NGC~3293 contracts at a level of $2.3\sigma$. The derived kinematic ages for the clusters agree well with (previously estimated) ages in the literature and isochrone fittings to color-magnitude diagrams. 

\item We combined the full \textit{Gaia} astrometry with \textit{Gaia} spectral types (\texttt{spectraltype\_esphs}) and effective temperatures (\texttt{teff\_esphs}) in order to search for new OB star candidates and constrain the distributed OB population over the full spatial extent of the Car~OB1 association, including poorly studied regions outside of the prominent star clusters. This led to 604 new OB candidates, resulting in a total sample of 857 OB candidates and association members in Car~OB1. In a literature search of these candidates, we found a spectroscopic identification as an O- or B-type star for 253 (35 from Sect.~\ref{sec:NewOBDBSCAN} and 218 from Sect.~\ref{sec:OBstarsdistributed}) of the candidates, including four O-type stars. For the remaining 604 candidates, we classified 15 as O-type candidates and 589 as B-type candidates according to their effective temperatures (\texttt{teff\_esphs}) listed in the \textit{Gaia} data. 

The fraction of OB stars in a spatially non-clustered distributed population increases to 54\% (Sect.~\ref{sec:NewOBDBSCAN}) and 74\% (Sect.~\ref{sec:OBstarsdistributed}), respectively, when the new OB star candidates are taken into account.
Extrapolating the Kroupa IMF, based on the Car~OB1 high-mass star sample and OB candidates, down to $0.08\,M_{\odot}$, we  estimate that Car~OB1 contains a total population of at least $\sim 8\times10^4$ stars.

\item We also ascertained that the whole Car~OB1 association is expanding at a velocity of $v_{\rm out} = 5.25\pm0.02~\rm km\,s^{-1}$ by examining the kinematics of the Car~OB1 high-mass stars. Using the positions and proper motion of the high-mass Car~OB1 stars allowed us to kinematically trace back the association, and it shows that the spatial extent had a minimum about 3--4~Myr ago.

\end{enumerate}

In this work, we present a comprehensive OB star and cluster catalog of the full association and show that the association reaches beyond the Carina Nebula. To study the evolution of the association, it is therefore necessary to take the full extent of Car~OB1 into account.

\section*{Data availability}
Tables~\ref{tab:AllClusters} and \ref{tab:CarOB1HighMass} are only fully available in electronic form at the CDS via anonymous ftp to \url{cdsarc.u-strasbg.fr (130.79.128.5)} or via \url{http://cdsweb.u-strasbg.fr/cgi-bin/qcat?J/A+A/}.

\begin{acknowledgements}
This research was supported by the Excellence Cluster ORIGINS which is funded by the Deutsche Forschungsgemeinschaft (DFG, German Research Foundation) under Germany’s Excellence Strategy - EXC-2094 - 390783311.
This work has made use of data from the European Space Agency (ESA) mission
{\it Gaia} (\url{https://www.cosmos.esa.int/gaia}), processed by the {\it Gaia}
Data Processing and Analysis Consortium (DPAC,
\url{https://www.cosmos.esa.int/web/gaia/dpac/consortium}). Funding for the DPAC
has been provided by national institutions, in particular the institutions
participating in the {\it Gaia} Multilateral Agreement. 
This research has made use of the open-source Python packages \texttt{Astropy} \citep{2022ApJ...935..167A}, \texttt{NumPy} \citep{harris2020array}, \texttt{scipy} \citep{2020SciPy-NMeth}, \texttt{scikit-learn} \citep{scikit-learn}, \texttt{pandas} \citep{mckinney-proc-scipy-2010}, and \texttt{Matplotlib} \citep{2007CSE.....9...90H}.
\end{acknowledgements}

\bibliographystyle{aa}
\bibliography{ref}

\begin{thebibliography}{96}
\expandafter\ifx\csname natexlab\endcsname\relax\def\natexlab#1{#1}\fi

\bibitem[{{Adams}(2010)}]{2010ARA&A..48...47A}
{Adams}, F.~C. 2010, \araa, 48, 47

\bibitem[{{Alexander} {et~al.}(2016){Alexander}, {Hanes}, {Povich}, \&
  {McSwain}}]{2016AJ....152..190A}
{Alexander}, M.~J., {Hanes}, R.~J., {Povich}, M.~S., \& {McSwain}, M.~V. 2016,
  \aj, 152, 190

\bibitem[{{Arakawa} \& {Kokubo}(2023)}]{2023A&A...670A.105A}
{Arakawa}, S. \& {Kokubo}, E. 2023, \aap, 670, A105

\bibitem[{{Astropy Collaboration} {et~al.}(2022){Astropy Collaboration},
  {Price-Whelan}, {Lim}, {Earl}, {Starkman}, {Bradley}, {Shupe}, {Patil},
  {Corrales}, {Brasseur}, {N{\"o}the}, {Donath}, {Tollerud}, {Morris},
  {Ginsburg}, {Vaher}, {Weaver}, {Tocknell}, {Jamieson}, {van Kerkwijk},
  {Robitaille}, {Merry}, {Bachetti}, {G{\"u}nther}, {Aldcroft},
  {Alvarado-Montes}, {Archibald}, {B{\'o}di}, {Bapat}, {Barentsen},
  {Baz{\'a}n}, {Biswas}, {Boquien}, {Burke}, {Cara}, {Cara}, {Conroy},
  {Conseil}, {Craig}, {Cross}, {Cruz}, {D'Eugenio}, {Dencheva}, {Devillepoix},
  {Dietrich}, {Eigenbrot}, {Erben}, {Ferreira}, {Foreman-Mackey}, {Fox},
  {Freij}, {Garg}, {Geda}, {Glattly}, {Gondhalekar}, {Gordon}, {Grant},
  {Greenfield}, {Groener}, {Guest}, {Gurovich}, {Handberg}, {Hart},
  {Hatfield-Dodds}, {Homeier}, {Hosseinzadeh}, {Jenness}, {Jones}, {Joseph},
  {Kalmbach}, {Karamehmetoglu}, {Ka{\l}uszy{\'n}ski}, {Kelley}, {Kern},
  {Kerzendorf}, {Koch}, {Kulumani}, {Lee}, {Ly}, {Ma}, {MacBride}, {Maljaars},
  {Muna}, {Murphy}, {Norman}, {O'Steen}, {Oman}, {Pacifici}, {Pascual},
  {Pascual-Granado}, {Patil}, {Perren}, {Pickering}, {Rastogi}, {Roulston},
  {Ryan}, {Rykoff}, {Sabater}, {Sakurikar}, {Salgado}, {Sanghi}, {Saunders},
  {Savchenko}, {Schwardt}, {Seifert-Eckert}, {Shih}, {Jain}, {Shukla}, {Sick},
  {Simpson}, {Singanamalla}, {Singer}, {Singhal}, {Sinha}, {Sip{\H{o}}cz},
  {Spitler}, {Stansby}, {Streicher}, {{\v{S}}umak}, {Swinbank}, {Taranu},
  {Tewary}, {Tremblay}, {de Val-Borro}, {Van Kooten}, {Vasovi{\'c}}, {Verma},
  {de Miranda Cardoso}, {Williams}, {Wilson}, {Winkel}, {Wood-Vasey}, {Xue},
  {Yoachim}, {Zhang}, {Zonca}, \& {Astropy Project
  Contributors}}]{2022ApJ...935..167A}
{Astropy Collaboration}, {Price-Whelan}, A.~M., {Lim}, P.~L., {et~al.} 2022,
  \apj, 935, 167

\bibitem[{{Berlanas} {et~al.}(2023){Berlanas}, {Ma{\'\i}z Apell{\'a}niz},
  {Herrero}, {Mahy}, {Blomme}, {Negueruela}, {Dorda}, {Comer{\'o}n}, {Gosset},
  {Pantaleoni Gonz{\'a}lez}, {Molina Lera}, {Sota}, {Furst}, {Alfaro},
  {Bergemann}, {Carraro}, {Drew}, {Morbidelli}, \&
  {Vink}}]{2023A&A...671A..20B}
{Berlanas}, S.~R., {Ma{\'\i}z Apell{\'a}niz}, J., {Herrero}, A., {et~al.} 2023,
  \aap, 671, A20

\bibitem[{{Bica} {et~al.}(2019){Bica}, {Pavani}, {Bonatto}, \&
  {Lima}}]{2019AJ....157...12B}
{Bica}, E., {Pavani}, D.~B., {Bonatto}, C.~J., \& {Lima}, E.~F. 2019, \aj, 157,
  12

\bibitem[{{Bressan} {et~al.}(2012){Bressan}, {Marigo}, {Girardi}, {Salasnich},
  {Dal Cero}, {Rubele}, \& {Nanni}}]{2012MNRAS.427..127B}
{Bressan}, A., {Marigo}, P., {Girardi}, L., {et~al.} 2012, \mnras, 427, 127

\bibitem[{{Cantat-Gaudin} {et~al.}(2020){Cantat-Gaudin}, {Anders},
  {Castro-Ginard}, {Jordi}, {Romero-G{\'o}mez}, {Soubiran}, {Casamiquela},
  {Tarricq}, {Moitinho}, {Vallenari}, {Bragaglia}, {Krone-Martins}, \&
  {Kounkel}}]{2020A&A...640A...1C}
{Cantat-Gaudin}, T., {Anders}, F., {Castro-Ginard}, A., {et~al.} 2020, \aap,
  640, A1

\bibitem[{{Cantat-Gaudin} {et~al.}(2018){Cantat-Gaudin}, {Jordi}, {Vallenari},
  {Bragaglia}, {Balaguer-N{\'u}{\~n}ez}, {Soubiran}, {Bossini}, {Moitinho},
  {Castro-Ginard}, {Krone-Martins}, {Casamiquela}, {Sordo}, \&
  {Carrera}}]{Cantat-Gaudin.2018}
{Cantat-Gaudin}, T., {Jordi}, C., {Vallenari}, A., {et~al.} 2018, \aap, 618,
  A93

\bibitem[{{Carraro} {et~al.}(2004){Carraro}, {Romaniello}, {Ventura}, \&
  {Patat}}]{2004A&A...418..525C}
{Carraro}, G., {Romaniello}, M., {Ventura}, P., \& {Patat}, F. 2004, \aap, 418,
  525

\bibitem[{{Castro-Ginard} {et~al.}(2020){Castro-Ginard}, {Jordi}, {Luri},
  {{\'A}lvarez Cid-Fuentes}, {Casamiquela}, {Anders}, {Cantat-Gaudin},
  {Mongui{\'o}}, {Balaguer-N{\'u}{\~n}ez}, {Sol{\`a}}, \&
  {Badia}}]{2020A&A...635A..45C}
{Castro-Ginard}, A., {Jordi}, C., {Luri}, X., {et~al.} 2020, \aap, 635, A45

\bibitem[{{Castro-Ginard} {et~al.}(2022){Castro-Ginard}, {Jordi}, {Luri},
  {Cantat-Gaudin}, {Carrasco}, {Casamiquela}, {Anders},
  {Balaguer-N{\'u}{\~n}ez}, \& {Badia}}]{2022A&A...661A.118C}
{Castro-Ginard}, A., {Jordi}, C., {Luri}, X., {et~al.} 2022, \aap, 661, A118

\bibitem[{{Castro-Ginard} {et~al.}(2018){Castro-Ginard}, {Jordi}, {Luri},
  {Julbe}, {Morvan}, {Balaguer-N{\'u}{\~n}ez}, \&
  {Cantat-Gaudin}}]{2018A&A...618A..59C}
{Castro-Ginard}, A., {Jordi}, C., {Luri}, X., {et~al.} 2018, \aap, 618, A59

\bibitem[{{Crompvoets} {et~al.}(2024){Crompvoets}, {Di Francesco},
  {Teimoorinia}, \& {Preibisch}}]{2024AJ....168...63C}
{Crompvoets}, B.~L., {Di Francesco}, J., {Teimoorinia}, H., \& {Preibisch}, T.
  2024, \aj, 168, 63

\bibitem[{{Damiani} {et~al.}(2017){Damiani}, {Klutsch}, {Jeffries}, {Randich},
  {Prisinzano}, {Ma{\'\i}z Apell{\'a}niz}, {Micela}, {Kalari}, {Frasca},
  {Zwitter}, {Bonito}, {Gilmore}, {Flaccomio}, {Francois}, {Koposov},
  {Lanzafame}, {Sacco}, {Bayo}, {Carraro}, {Casey}, {Alfaro}, {Costado},
  {Donati}, {Franciosini}, {Hourihane}, {Jofr{\'e}}, {Lardo}, {Lewis},
  {Magrini}, {Monaco}, {Morbidelli}, {Worley}, {Vink}, \&
  {Zaggia}}]{2017A&A...603A..81D}
{Damiani}, F., {Klutsch}, A., {Jeffries}, R.~D., {et~al.} 2017, \aap, 603, A81

\bibitem[{{Damiani} {et~al.}(2019){Damiani}, {Prisinzano}, {Pillitteri},
  {Micela}, \& {Sciortino}}]{2019A&A...623A.112D}
{Damiani}, F., {Prisinzano}, L., {Pillitteri}, I., {Micela}, G., \&
  {Sciortino}, S. 2019, \aap, 623, A112

\bibitem[{{Della Croce} {et~al.}(2024){Della Croce}, {Dalessandro},
  {Livernois}, \& {Vesperini}}]{2024A&A...683A..10D}
{Della Croce}, A., {Dalessandro}, E., {Livernois}, A., \& {Vesperini}, E. 2024,
  \aap, 683, A10

\bibitem[{{Doran} {et~al.}(2013){Doran}, {Crowther}, {de Koter}, {Evans},
  {McEvoy}, {Walborn}, {Bastian}, {Bestenlehner}, {Gr{\"a}fener}, {Herrero},
  {K{\"o}hler}, {Ma{\'\i}z Apell{\'a}niz}, {Najarro}, {Puls}, {Sana},
  {Schneider}, {Taylor}, {van Loon}, \& {Vink}}]{2013A&A...558A.134D}
{Doran}, E.~I., {Crowther}, P.~A., {de Koter}, A., {et~al.} 2013, \aap, 558,
  A134

\bibitem[{Ester {et~al.}(1996)Ester, Kriegel, Sander, \&
  Xu}]{10.5555/3001460.3001507}
Ester, M., Kriegel, H.-P., Sander, J., \& Xu, X. 1996, in Proceedings of the
  Second International Conference on Knowledge Discovery and Data Mining,
  KDD'96 (AAAI Press), 226–231

\bibitem[{{Farias} {et~al.}(2024){Farias}, {Offner}, {Grudi{\'c}}, {Guszejnov},
  \& {Rosen}}]{2024MNRAS.527.6732F}
{Farias}, J.~P., {Offner}, S. S.~R., {Grudi{\'c}}, M.~Y., {Guszejnov}, D., \&
  {Rosen}, A.~L. 2024, \mnras, 527, 6732

\bibitem[{{Feigelson} {et~al.}(2011){Feigelson}, {Getman}, {Townsley}, {Broos},
  {Povich}, {Garmire}, {King}, {Montmerle}, {Preibisch}, {Smith}, {Stassun},
  {Wang}, {Wolk}, \& {Zinnecker}}]{CCCP-Clusters}
{Feigelson}, E.~D., {Getman}, K.~V., {Townsley}, L.~K., {et~al.} 2011, \apjs,
  194, 9

\bibitem[{{Fuchs} {et~al.}(2006){Fuchs}, {Breitschwerdt}, {de Avillez},
  {Dettbarn}, \& {Flynn}}]{2006MNRAS.373..993F}
{Fuchs}, B., {Breitschwerdt}, D., {de Avillez}, M.~A., {Dettbarn}, C., \&
  {Flynn}, C. 2006, \mnras, 373, 993

\bibitem[{{Gaia Collaboration} {et~al.}(2016){Gaia Collaboration}, {Prusti},
  {de Bruijne}, {Brown}, {Vallenari}, \& {Babusiaux}}]{Gaia}
{Gaia Collaboration}, {Prusti}, T., {de Bruijne}, J.~H.~J., {et~al.} 2016,
  \aap, 595, A1

\bibitem[{{Gaia Collaboration} {et~al.}(2023){Gaia Collaboration}, {Vallenari},
  {Brown}, {Prusti}, {de Bruijne}, {Arenou}, {Babusiaux}, {Biermann},
  {Creevey}, {Ducourant}, {Evans}, {Eyer}, {Guerra}, {Hutton}, {Jordi},
  {Klioner}, {Lammers}, {Lindegren}, {Luri}, {Mignard}, {Panem}, {Pourbaix},
  {Randich}, {Sartoretti}, {Soubiran}, {Tanga}, {Walton}, {Bailer-Jones},
  {Bastian}, {Drimmel}, {Jansen}, {Katz}, {Lattanzi}, {van Leeuwen}, {Bakker},
  {Cacciari}, {Casta{\~n}eda}, {De Angeli}, {Fabricius}, {Fouesneau},
  {Fr{\'e}mat}, {Galluccio}, {Guerrier}, {Heiter}, {Masana}, {Messineo},
  {Mowlavi}, {Nicolas}, {Nienartowicz}, {Pailler}, {Panuzzo}, {Riclet}, {Roux},
  {Seabroke}, {Sordo}, {Th{\'e}venin}, {Gracia-Abril}, {Portell}, {Teyssier},
  {Altmann}, {Andrae}, {Audard}, {Bellas-Velidis}, {Benson}, {Berthier},
  {Blomme}, {Burgess}, {Busonero}, {Busso}, {C{\'a}novas}, {Carry}, {Cellino},
  {Cheek}, {Clementini}, {Damerdji}, {Davidson}, {de Teodoro}, {Nu{\~n}ez
  Campos}, {Delchambre}, {Dell'Oro}, {Esquej}, {Fern{\'a}ndez-Hern{\'a}ndez},
  {Fraile}, {Garabato}, {Garc{\'\i}a-Lario}, {Gosset}, {Haigron}, {Halbwachs},
  {Hambly}, {Harrison}, {Hern{\'a}ndez}, {Hestroffer}, {Hodgkin}, {Holl},
  {Jan{\ss}en}, {Jevardat de Fombelle}, {Jordan}, {Krone-Martins}, {Lanzafame},
  {L{\"o}ffler}, {Marchal}, {Marrese}, {Moitinho}, {Muinonen}, {Osborne},
  {Pancino}, {Pauwels}, {Recio-Blanco}, {Reyl{\'e}}, {Riello}, {Rimoldini},
  {Roegiers}, {Rybizki}, {Sarro}, {Siopis}, {Smith}, {Sozzetti}, {Utrilla},
  {van Leeuwen}, {Abbas}, {{\'A}brah{\'a}m}, {Abreu Aramburu}, {Aerts},
  {Aguado}, {Ajaj}, {Aldea-Montero}, {Altavilla}, {{\'A}lvarez}, {Alves},
  {Anders}, {Anderson}, {Anglada Varela}, {Antoja}, {Baines}, {Baker},
  {Balaguer-N{\'u}{\~n}ez}, {Balbinot}, {Balog}, {Barache}, {Barbato},
  {Barros}, {Barstow}, {Bartolom{\'e}}, {Bassilana}, {Bauchet}, {Becciani},
  {Bellazzini}, {Berihuete}, {Bernet}, {Bertone}, {Bianchi}, {Binnenfeld},
  {Blanco-Cuaresma}, {Blazere}, {Boch}, {Bombrun}, {Bossini}, {Bouquillon},
  {Bragaglia}, {Bramante}, {Breedt}, {Bressan}, {Brouillet}, {Brugaletta},
  {Bucciarelli}, {Burlacu}, {Butkevich}, {Buzzi}, {Caffau}, {Cancelliere},
  {Cantat-Gaudin}, {Carballo}, {Carlucci}, {Carnerero}, {Carrasco},
  {Casamiquela}, {Castellani}, {Castro-Ginard}, {Chaoul}, {Charlot}, {Chemin},
  {Chiaramida}, {Chiavassa}, {Chornay}, {Comoretto}, {Contursi}, {Cooper},
  {Cornez}, {Cowell}, {Crifo}, {Cropper}, {Crosta}, {Crowley}, {Dafonte},
  {Dapergolas}, {David}, {David}, {de Laverny}, {De Luise}, {De March}, {De
  Ridder}, {de Souza}, {de Torres}, {del Peloso}, {del Pozo}, {Delbo},
  {Delgado}, {Delisle}, {Demouchy}, {Dharmawardena}, {Di Matteo}, {Diakite},
  {Diener}, {Distefano}, {Dolding}, {Edvardsson}, {Enke}, {Fabre}, {Fabrizio},
  {Faigler}, {Fedorets}, {Fernique}, {Fienga}, {Figueras}, {Fournier},
  {Fouron}, {Fragkoudi}, {Gai}, {Garcia-Gutierrez}, {Garcia-Reinaldos},
  {Garc{\'\i}a-Torres}, {Garofalo}, {Gavel}, {Gavras}, {Gerlach}, {Geyer},
  {Giacobbe}, {Gilmore}, {Girona}, {Giuffrida}, {Gomel}, {Gomez},
  {Gonz{\'a}lez-N{\'u}{\~n}ez}, {Gonz{\'a}lez-Santamar{\'\i}a},
  {Gonz{\'a}lez-Vidal}, {Granvik}, {Guillout}, {Guiraud},
  {Guti{\'e}rrez-S{\'a}nchez}, {Guy}, {Hatzidimitriou}, {Hauser}, {Haywood},
  {Helmer}, {Helmi}, {Sarmiento}, {Hidalgo}, {Hilger}, {H{\l}adczuk}, {Hobbs},
  {Holland}, {Huckle}, {Jardine}, {Jasniewicz}, {Jean-Antoine Piccolo},
  {Jim{\'e}nez-Arranz}, {Jorissen}, {Juaristi Campillo}, {Julbe}, {Karbevska},
  {Kervella}, {Khanna}, {Kontizas}, {Kordopatis}, {Korn}, {K{\'o}sp{\'a}l},
  {Kostrzewa-Rutkowska}, {Kruszy{\'n}ska}, {Kun}, {Laizeau}, {Lambert},
  {Lanza}, {Lasne}, {Le Campion}, {Lebreton}, {Lebzelter}, {Leccia}, {Leclerc},
  {Lecoeur-Taibi}, {Liao}, {Licata}, {Lindstr{\o}m}, {Lister}, {Livanou},
  {Lobel}, {Lorca}, {Loup}, {Madrero Pardo}, {Magdaleno Romeo}, {Managau},
  {Mann}, {Manteiga}, {Marchant}, {Marconi}, {Marcos}, {Marcos Santos},
  {Mar{\'\i}n Pina}, {Marinoni}, {Marocco}, {Marshall}, {Martin Polo},
  {Mart{\'\i}n-Fleitas}, {Marton}, {Mary}, {Masip}, {Massari},
  {Mastrobuono-Battisti}, {Mazeh}, {McMillan}, {Messina}, {Michalik}, {Millar},
  {Mints}, {Molina}, {Molinaro}, {Moln{\'a}r}, {Monari}, {Mongui{\'o}},
  {Montegriffo}, {Montero}, {Mor}, {Mora}, {Morbidelli}, {Morel}, {Morris},
  {Muraveva}, {Murphy}, {Musella}, {Nagy}, {Noval}, {Oca{\~n}a}, {Ogden},
  {Ordenovic}, {Osinde}, {Pagani}, {Pagano}, {Palaversa}, {Palicio},
  {Pallas-Quintela}, {Panahi}, {Payne-Wardenaar}, {Pe{\~n}alosa Esteller},
  {Penttil{\"a}}, {Pichon}, {Piersimoni}, {Pineau}, {Plachy}, {Plum}, {Poggio},
  {Pr{\v{s}}a}, {Pulone}, {Racero}, {Ragaini}, {Rainer}, {Raiteri}, {Rambaux},
  {Ramos}, {Ramos-Lerate}, {Re Fiorentin}, {Regibo}, {Richards}, {Rios Diaz},
  {Ripepi}, {Riva}, {Rix}, {Rixon}, {Robichon}, {Robin}, {Robin}, {Roelens},
  {Rogues}, {Rohrbasser}, {Romero-G{\'o}mez}, {Rowell}, {Royer}, {Ruz Mieres},
  {Rybicki}, {Sadowski}, {S{\'a}ez N{\'u}{\~n}ez}, {Sagrist{\`a} Sell{\'e}s},
  {Sahlmann}, {Salguero}, {Samaras}, {Sanchez Gimenez}, {Sanna},
  {Santove{\~n}a}, {Sarasso}, {Schultheis}, {Sciacca}, {Segol}, {Segovia},
  {S{\'e}gransan}, {Semeux}, {Shahaf}, {Siddiqui}, {Siebert}, {Siltala},
  {Silvelo}, {Slezak}, {Slezak}, {Smart}, {Snaith}, {Solano}, {Solitro},
  {Souami}, {Souchay}, {Spagna}, {Spina}, {Spoto}, {Steele},
  {Steidelm{\"u}ller}, {Stephenson}, {S{\"u}veges}, {Surdej}, {Szabados},
  {Szegedi-Elek}, {Taris}, {Taylor}, {Teixeira}, {Tolomei}, {Tonello}, {Torra},
  {Torra}, {Torralba Elipe}, {Trabucchi}, {Tsounis}, {Turon}, {Ulla}, {Unger},
  {Vaillant}, {van Dillen}, {van Reeven}, {Vanel}, {Vecchiato}, {Viala},
  {Vicente}, {Voutsinas}, {Weiler}, {Wevers}, {Wyrzykowski}, {Yoldas}, {Yvard},
  {Zhao}, {Zorec}, {Zucker}, \& {Zwitter}}]{Gaia-DR3}
{Gaia Collaboration}, {Vallenari}, A., {Brown}, A.~G.~A., {et~al.} 2023, \aap,
  674, A1

\bibitem[{{Gilmore} {et~al.}(2012){Gilmore}, {Randich}, {Asplund}, {Binney},
  {Bonifacio}, {Drew}, {Feltzing}, {Ferguson}, {Jeffries}, {Micela},
  {Negueruela}, {Prusti}, {Rix}, {Vallenari}, {Alfaro}, {Allende-Prieto},
  {Babusiaux}, {Bensby}, {Blomme}, {Bragaglia}, {Flaccomio}, {Fran{\c{c}}ois},
  {Irwin}, {Koposov}, {Korn}, {Lanzafame}, {Pancino}, {Paunzen},
  {Recio-Blanco}, {Sacco}, {Smiljanic}, {Van Eck}, {Walton}, {Aden}, {Aerts},
  {Affer}, {Alcala}, {Altavilla}, {Alves}, {Antoja}, {Arenou}, {Argiroffi},
  {Asensio Ramos}, {Bailer-Jones}, {Balaguer-Nunez}, {Bayo}, {Barbuy},
  {Barisevicius}, {Barrado y Navascues}, {Battistini}, {Bellas Velidis},
  {Bellazzini}, {Belokurov}, {Bergemann}, {Bertelli}, {Biazzo}, {Bienayme},
  {Bland-Hawthorn}, {Boeche}, {Bonito}, {Boudreault}, {Bouvier}, {Brandao},
  {Brown}, {de Bruijne}, {Burleigh}, {Caballero}, {Caffau}, {Calura},
  {Capuzzo-Dolcetta}, {Caramazza}, {Carraro}, {Casagrande}, {Casewell},
  {Chapman}, {Chiappini}, {Chorniy}, {Christlieb}, {Cignoni}, {Cocozza},
  {Colless}, {Collet}, {Collins}, {Correnti}, {Covino}, {Crnojevic}, {Cropper},
  {Cunha}, {Damiani}, {David}, {Delgado}, {Duffau}, {Edvardsson}, {Eldridge},
  {Enke}, {Eriksson}, {Evans}, {Eyer}, {Famaey}, {Fellhauer}, {Ferreras},
  {Figueras}, {Fiorentino}, {Flynn}, {Folha}, {Franciosini}, {Frasca},
  {Freeman}, {Fremat}, {Friel}, {Gaensicke}, {Gameiro}, {Garzon}, {Geier},
  {Geisler}, {Gerhard}, {Gibson}, {Gomboc}, {Gomez}, {Gonzalez-Fernandez},
  {Gonzalez Hernandez}, {Gosset}, {Grebel}, {Greimel}, {Groenewegen},
  {Grundahl}, {Guarcello}, {Gustafsson}, {Hadrava}, {Hatzidimitriou}, {Hambly},
  {Hammersley}, {Hansen}, {Haywood}, {Heber}, {Heiter}, {Held}, {Helmi},
  {Hensler}, {Herrero}, {Hill}, {Hodgkin}, {Huelamo}, {Huxor}, {Ibata},
  {Jackson}, {de Jong}, {Jonker}, {Jordan}, {Jordi}, {Jorissen}, {Katz},
  {Kawata}, {Keller}, {Kharchenko}, {Klement}, {Klutsch}, {Knude}, {Koch},
  {Kochukhov}, {Kontizas}, {Koubsky}, {Lallement}, {de Laverny}, {van Leeuwen},
  {Lemasle}, {Lewis}, {Lind}, {Lindstrom}, {Lobel}, {Lopez Santiago}, {Lucas},
  {Ludwig}, {Lueftinger}, {Magrini}, {Maiz Apellaniz}, {Maldonado}, {Marconi},
  {Marino}, {Martayan}, {Martinez-Valpuesta}, {Matijevic}, {McMahon},
  {Messina}, {Meyer}, {Miglio}, {Mikolaitis}, {Minchev}, {Minniti}, {Moitinho},
  {Momany}, {Monaco}, {Montalto}, {Monteiro}, {Monier}, {Montes}, {Mora},
  {Moraux}, {Morel}, {Mowlavi}, {Mucciarelli}, {Munari}, {Napiwotzki},
  {Nardetto}, {Naylor}, {Naze}, {Nelemans}, {Okamoto}, {Ortolani}, {Pace},
  {Palla}, {Palous}, {Parker}, {Penarrubia}, {Pillitteri}, {Piotto}, {Posbic},
  {Prisinzano}, {Puzeras}, {Quirrenbach}, {Ragaini}, {Read}, {Read}, {Reyle},
  {De Ridder}, {Robichon}, {Robin}, {Roeser}, {Romano}, {Royer}, {Ruchti},
  {Ruzicka}, {Ryan}, {Ryde}, {Santos}, {Sanz Forcada}, {Sarro Baro},
  {Sbordone}, {Schilbach}, {Schmeja}, {Schnurr}, {Schoenrich}, {Scholz},
  {Seabroke}, {Sharma}, {De Silva}, {Smith}, {Solano}, {Sordo}, {Soubiran},
  {Sousa}, {Spagna}, {Steffen}, {Steinmetz}, {Stelzer}, {Stempels},
  {Tabernero}, {Tautvaisiene}, {Thevenin}, {Torra}, {Tosi}, {Tolstoy}, {Turon},
  {Walker}, {Wambsganss}, {Worley}, {Venn}, {Vink}, {Wyse}, {Zaggia},
  {Zeilinger}, {Zoccali}, {Zorec}, {Zucker}, {Zwitter}, \& {Gaia-ESO Survey
  Team}}]{2012Msngr.147...25G}
{Gilmore}, G., {Randich}, S., {Asplund}, M., {et~al.} 2012, The Messenger, 147,
  25

\bibitem[{{G{\"o}ppl} \& {Preibisch}(2022)}]{2022A&A...660A..11G}
{G{\"o}ppl}, C. \& {Preibisch}, T. 2022, \aap, 660, A11

\bibitem[{{Graci{\'a}-Carpio} {et~al.}(2017){Graci{\'a}-Carpio}, {Wetzstein},
  {Roussel}, \& {PACS Instrument Control Centre Team}}]{2017ASPC..512..379G}
{Graci{\'a}-Carpio}, J., {Wetzstein}, M., {Roussel}, H., \& {PACS Instrument
  Control Centre Team}. 2017, in Astronomical Society of the Pacific Conference
  Series, Vol. 512, Astronomical Data Analysis Software and Systems XXV, ed.
  N.~P.~F. {Lorente}, K.~{Shortridge}, \& R.~{Wayth}, 379

\bibitem[{{Gruner} {et~al.}(2019){Gruner}, {Hainich}, {Sander}, {Shenar},
  {Todt}, {Oskinova}, {Ramachandran}, {Ayres}, \&
  {Hamann}}]{2019A&A...621A..63G}
{Gruner}, D., {Hainich}, R., {Sander}, A.~A.~C., {et~al.} 2019, \aap, 621, A63

\bibitem[{{Hamann} {et~al.}(2019){Hamann}, {Gr{\"a}fener}, {Liermann},
  {Hainich}, {Sander}, {Shenar}, {Ramachandran}, {Todt}, \&
  {Oskinova}}]{2019A&A...625A..57H}
{Hamann}, W.~R., {Gr{\"a}fener}, G., {Liermann}, A., {et~al.} 2019, \aap, 625,
  A57

\bibitem[{{Hao} {et~al.}(2022){Hao}, {Xu}, {Wu}, {Lin}, {Liu}, \&
  {Li}}]{2022A&A...660A...4H}
{Hao}, C.~J., {Xu}, Y., {Wu}, Z.~Y., {et~al.} 2022, \aap, 660, A4

\bibitem[{Harris {et~al.}(2020)Harris, Millman, van~der Walt, Gommers,
  Virtanen, Cournapeau, Wieser, Taylor, Berg, Smith, Kern, Picus, Hoyer, van
  Kerkwijk, Brett, Haldane, del R{\'{i}}o, Wiebe, Peterson,
  G{\'{e}}rard-Marchant, Sheppard, Reddy, Weckesser, Abbasi, Gohlke, \&
  Oliphant}]{harris2020array}
Harris, C.~R., Millman, K.~J., van~der Walt, S.~J., {et~al.} 2020, Nature, 585,
  357

\bibitem[{{Holmberg} \& {Flynn}(2004)}]{2004MNRAS.352..440H}
{Holmberg}, J. \& {Flynn}, C. 2004, \mnras, 352, 440

\bibitem[{{Hourihane} {et~al.}(2023){Hourihane}, {Francois}, {Worley},
  {Magrini}, {Gonneau}, {Casey}, {Gilmore}, {Randich}, {Sacco}, {Recio-Blanco},
  {Korn}, {Allende Prieto}, {Smiljanic}, {Blomme}, {Bragaglia}, {Walton}, {van
  Eck}, {Bensby}, {Lanzafame}, {Frasca}, {Franciosini}, {Damiani}, {Lind},
  {Bergemann}, {Bonifacio}, {Hill}, {Lobel}, {Montes}, {Feuillet},
  {Tautvaisiene}, {Guiglion}, {Tabernero}, {Gonzalez Hernandez}, {Gebran}, {van
  der Swaelmen}, {Mikolaitis}, {Daflon}, {Merle}, {Morel}, {Lewis}, {Gonzalez
  Solares}, {Murphy}, {Jeffries}, {Jackson}, {Feltzing}, {Prusti}, {Carraro},
  {Biazzo}, {Prisinzano}, {Jofre}, {Zaggia}, {Drazdauskas}, {Stonkute},
  {Marfil}, {Jimenez-Esteban}, {Mahy}, {Gutierrez Albarran}, {Berlanas},
  {Santos}, {Morbidelli}, {Spina}, \& {Minkeviciute}}]{2023yCat..36760129H}
{Hourihane}, A., {Francois}, P., {Worley}, C.~C., {et~al.} 2023, VizieR Online
  Data Catalog, J/A+A/676/A129

\bibitem[{{Humphreys}(1978)}]{1978ApJS...38..309H}
{Humphreys}, R.~M. 1978, \apjs, 38, 309

\bibitem[{{Hunt} \& {Reffert}(2023)}]{2023A&A...673A.114H}
{Hunt}, E.~L. \& {Reffert}, S. 2023, \aap, 673, A114

\bibitem[{{Hunter}(2007)}]{2007CSE.....9...90H}
{Hunter}, J.~D. 2007, Computing in Science and Engineering, 9, 90

\bibitem[{{Hur} {et~al.}(2023){Hur}, {Lim}, \& {Chun}}]{2023JKAS...56...97H}
{Hur}, H., {Lim}, B., \& {Chun}, M.-Y. 2023, Journal of Korean Astronomical
  Society, 56, 97

\bibitem[{{Hur} {et~al.}(2015){Hur}, {Park}, {Sung}, {Bessell}, {Lim}, {Chun},
  \& {Sohn}}]{2015MNRAS.446.3797H}
{Hur}, H., {Park}, B.-G., {Sung}, H., {et~al.} 2015, \mnras, 446, 3797

\bibitem[{{Itrich} {et~al.}(2024){Itrich}, {Testi}, {Beccari}, {Manara},
  {Reiter}, {Preibisch}, {McLeod}, {Rosotti}, {Klessen}, {Molinari}, \&
  {Hennebelle}}]{2024A&A...685A.100I}
{Itrich}, D., {Testi}, L., {Beccari}, G., {et~al.} 2024, \aap, 685, A100

\bibitem[{{Kennicutt}(1984)}]{1984ApJ...287..116K}
{Kennicutt}, R.~C., J. 1984, \apj, 287, 116

\bibitem[{{Kharchenko} {et~al.}(2016){Kharchenko}, {Piskunov}, {Schilbach},
  {R{\"o}ser}, \& {Scholz}}]{2016A&A...585A.101K}
{Kharchenko}, N.~V., {Piskunov}, A.~E., {Schilbach}, E., {R{\"o}ser}, S., \&
  {Scholz}, R.~D. 2016, \aap, 585, A101

\bibitem[{{Kroupa}(2001)}]{2001MNRAS.322..231K}
{Kroupa}, P. 2001, \mnras, 322, 231

\bibitem[{{Kuhn} {et~al.}(2014){Kuhn}, {Feigelson}, {Getman}, {Baddeley},
  {Broos}, {Sills}, {Bate}, {Povich}, {Luhman}, {Busk}, {Naylor}, \&
  {King}}]{2014ApJ...787..107K}
{Kuhn}, M.~A., {Feigelson}, E.~D., {Getman}, K.~V., {et~al.} 2014, \apj, 787,
  107

\bibitem[{{Kuhn} {et~al.}(2019){Kuhn}, {Hillenbrand}, {Sills}, {Feigelson}, \&
  {Getman}}]{2019ApJ...870...32K}
{Kuhn}, M.~A., {Hillenbrand}, L.~A., {Sills}, A., {Feigelson}, E.~D., \&
  {Getman}, K.~V. 2019, \apj, 870, 32

\bibitem[{{Li} {et~al.}(2019){Li}, {Zhao}, \& {Yang}}]{2019ApJ...872..205L}
{Li}, C., {Zhao}, G., \& {Yang}, C. 2019, \apj, 872, 205

\bibitem[{{Lim} {et~al.}(2019){Lim}, {Naz{\'e}}, {Gosset}, \&
  {Rauw}}]{2019MNRAS.490..440L}
{Lim}, B., {Naz{\'e}}, Y., {Gosset}, E., \& {Rauw}, G. 2019, \mnras, 490, 440

\bibitem[{{Lindegren} {et~al.}(2021){Lindegren}, {Bastian}, {Biermann},
  {Bombrun}, {de Torres}, {Gerlach}, {Geyer}, {Hern{\'a}ndez}, {Hilger},
  {Hobbs}, {Klioner}, {Lammers}, {McMillan}, {Ramos-Lerate},
  {Steidelm{\"u}ller}, {Stephenson}, \& {van Leeuwen}}]{Lindegren.2021}
{Lindegren}, L., {Bastian}, U., {Biermann}, M., {et~al.} 2021, \aap, 649, A4

\bibitem[{{Lindegren} {et~al.}(2018){Lindegren}, {Hern{\'a}ndez}, {Bombrun},
  {Klioner}, {Bastian}, {Ramos-Lerate}, {de Torres}, {Steidelm{\"u}ller},
  {Stephenson}, {Hobbs}, {Lammers}, {Biermann}, {Geyer}, {Hilger}, {Michalik},
  {Stampa}, {McMillan}, {Casta{\~n}eda}, {Clotet}, {Comoretto}, {Davidson},
  {Fabricius}, {Gracia}, {Hambly}, {Hutton}, {Mora}, {Portell}, {van Leeuwen},
  {Abbas}, {Abreu}, {Altmann}, {Andrei}, {Anglada}, {Balaguer-N{\'u}{\~n}ez},
  {Barache}, {Becciani}, {Bertone}, {Bianchi}, {Bouquillon}, {Bourda},
  {Br{\"u}semeister}, {Bucciarelli}, {Busonero}, {Buzzi}, {Cancelliere},
  {Carlucci}, {Charlot}, {Cheek}, {Crosta}, {Crowley}, {de Bruijne}, {de
  Felice}, {Drimmel}, {Esquej}, {Fienga}, {Fraile}, {Gai}, {Garralda},
  {Gonz{\'a}lez-Vidal}, {Guerra}, {Hauser}, {Hofmann}, {Holl}, {Jordan},
  {Lattanzi}, {Lenhardt}, {Liao}, {Licata}, {Lister}, {L{\"o}ffler},
  {Marchant}, {Martin-Fleitas}, {Messineo}, {Mignard}, {Morbidelli}, {Poggio},
  {Riva}, {Rowell}, {Salguero}, {Sarasso}, {Sciacca}, {Siddiqui}, {Smart},
  {Spagna}, {Steele}, {Taris}, {Torra}, {van Elteren}, {van Reeven}, \&
  {Vecchiato}}]{Lindegren.2018}
{Lindegren}, L., {Hern{\'a}ndez}, J., {Bombrun}, A., {et~al.} 2018, \aap, 616,
  A2

\bibitem[{{Liu} \& {Pang}(2019)}]{2019ApJS..245...32L}
{Liu}, L. \& {Pang}, X. 2019, \apjs, 245, 32

\bibitem[{{Martins} {et~al.}(2005){Martins}, {Schaerer}, \&
  {Hillier}}]{2005A&A...436.1049M}
{Martins}, F., {Schaerer}, D., \& {Hillier}, D.~J. 2005, \aap, 436, 1049

\bibitem[{{Melnik} \& {Dambis}(2020)}]{2020MNRAS.493.2339M}
{Melnik}, A.~M. \& {Dambis}, A.~K. 2020, \mnras, 493, 2339

\bibitem[{{Mel'Nik} \& {Efremov}(1995)}]{1995AstL...21...10M}
{Mel'Nik}, A.~M. \& {Efremov}, Y.~N. 1995, Astronomy Letters, 21, 10

\bibitem[{{Morel} {et~al.}(2022){Morel}, {Blaz{\`e}re}, {Semaan}, {Gosset},
  {Zorec}, {Fr{\'e}mat}, {Blomme}, {Daflon}, {Lobel}, {Nieva}, {Przybilla},
  {Gebran}, {Herrero}, {Mahy}, {Santos}, {Tautvai{\v{s}}ien{\.{e}}}, {Gilmore},
  {Randich}, {Alfaro}, {Bergemann}, {Carraro}, {Damiani}, {Franciosini},
  {Morbidelli}, {Pancino}, {Worley}, \& {Zaggia}}]{2022A&A...665A.108M}
{Morel}, T., {Blaz{\`e}re}, A., {Semaan}, T., {et~al.} 2022, \aap, 665, A108

\bibitem[{{Ohlendorf} {et~al.}(2013){Ohlendorf}, {Preibisch}, {Gaczkowski},
  {Ratzka}, {Ngoumou}, {Roccatagliata}, \& {Grellmann}}]{Ohlendorf13}
{Ohlendorf}, H., {Preibisch}, T., {Gaczkowski}, B., {et~al.} 2013, \aap, 552,
  A14

\bibitem[{{Olivares} {et~al.}(2020){Olivares}, {Sarro}, {Bouy}, {Miret-Roig},
  {Casamiquela}, {Galli}, {Berihuete}, \& {Tarricq}}]{2020A&A...644A...7O}
{Olivares}, J., {Sarro}, L.~M., {Bouy}, H., {et~al.} 2020, \aap, 644, A7

\bibitem[{{Oliveira} {et~al.}(2018){Oliveira}, {Bica}, \&
  {Bonatto}}]{2018MNRAS.476..842O}
{Oliveira}, R.~A.~P., {Bica}, E., \& {Bonatto}, C. 2018, \mnras, 476, 842

\bibitem[{{Patat} \& {Carraro}(2001)}]{2001MNRAS.325.1591P}
{Patat}, F. \& {Carraro}, G. 2001, \mnras, 325, 1591

\bibitem[{{Pecaut} \& {Mamajek}(2013)}]{2013ApJS..208....9P}
{Pecaut}, M.~J. \& {Mamajek}, E.~E. 2013, \apjs, 208, 9

\bibitem[{Pedregosa {et~al.}(2011)Pedregosa, Varoquaux, Gramfort, Michel,
  Thirion, Grisel, Blondel, Prettenhofer, Weiss, Dubourg, Vanderplas, Passos,
  Cournapeau, Brucher, Perrot, \& Duchesnay}]{scikit-learn}
Pedregosa, F., Varoquaux, G., Gramfort, A., {et~al.} 2011, Journal of Machine
  Learning Research, 12, 2825

\bibitem[{{Povich} {et~al.}(2019){Povich}, {Maldonado}, {Haze Nu{\~n}ez}, \&
  {Robitaille}}]{Povich19}
{Povich}, M.~S., {Maldonado}, J.~T., {Haze Nu{\~n}ez}, E., \& {Robitaille},
  T.~P. 2019, \apj, 881, 37

\bibitem[{{Preibisch} {et~al.}(2017){Preibisch}, {Flaischlen}, {Gaczkowski},
  {Townsley}, \& {Broos}}]{2017A&A...605A..85P}
{Preibisch}, T., {Flaischlen}, S., {Gaczkowski}, B., {Townsley}, L., \&
  {Broos}, P. 2017, \aap, 605, A85

\bibitem[{{Preibisch} {et~al.}(2021){Preibisch}, {Flaischlen}, {G{\"o}ppl},
  {Ercolano}, \& {Roccatagliata}}]{Tr16-SE-KMOS}
{Preibisch}, T., {Flaischlen}, S., {G{\"o}ppl}, C., {Ercolano}, B., \&
  {Roccatagliata}, V. 2021, \aap, 648, A34

\bibitem[{{Preibisch} {et~al.}(2011{\natexlab{a}}){Preibisch}, {Hodgkin},
  {Irwin}, {Lewis}, {King}, {McCaughrean}, {Zinnecker}, {Townsley}, \&
  {Broos}}]{CCCP-HAWKI}
{Preibisch}, T., {Hodgkin}, S., {Irwin}, M., {et~al.} 2011{\natexlab{a}},
  \apjs, 194, 10

\bibitem[{{Preibisch} {et~al.}(2014{\natexlab{a}}){Preibisch}, {Mehlhorn},
  {Townsley}, {Broos}, \& {Ratzka}}]{Preibisch14}
{Preibisch}, T., {Mehlhorn}, M., {Townsley}, L., {Broos}, P., \& {Ratzka}, T.
  2014{\natexlab{a}}, \aap, 564, A120

\bibitem[{{Preibisch} {et~al.}(2011{\natexlab{b}}){Preibisch}, {Ratzka},
  {Kuderna}, {Ohlendorf}, {King}, {Hodgkin}, {Irwin}, {Lewis}, {McCaughrean},
  \& {Zinnecker}}]{HAWKI-survey}
{Preibisch}, T., {Ratzka}, T., {Kuderna}, B., {et~al.} 2011{\natexlab{b}},
  \aap, 530, A34

\bibitem[{{Preibisch} {et~al.}(2012){Preibisch}, {Roccatagliata}, {Gaczkowski},
  \& {Ratzka}}]{Preibisch12}
{Preibisch}, T., {Roccatagliata}, V., {Gaczkowski}, B., \& {Ratzka}, T. 2012,
  \aap, 541, A132

\bibitem[{{Preibisch} {et~al.}(2014{\natexlab{b}}){Preibisch}, {Zeidler},
  {Ratzka}, {Roccatagliata}, \& {Petr-Gotzens}}]{VISTA1}
{Preibisch}, T., {Zeidler}, P., {Ratzka}, T., {Roccatagliata}, V., \&
  {Petr-Gotzens}, M.~G. 2014{\natexlab{b}}, \aap, 572, A116

\bibitem[{{Quintana} {et~al.}(2023){Quintana}, {Wright}, \&
  {Jeffries}}]{2023MNRAS.522.3124Q}
{Quintana}, A.~L., {Wright}, N.~J., \& {Jeffries}, R.~D. 2023, \mnras, 522,
  3124

\bibitem[{{Ram{\'\i}rez-Tannus} {et~al.}(2023){Ram{\'\i}rez-Tannus}, {Bik},
  {Cuijpers}, {Waters}, {G{\"o}ppl}, {Henning}, {Kamp}, {Preibisch}, {Getman},
  {Chaparro}, {Cuartas-Restrepo}, {de Koter}, {Feigelson}, {Grant}, {Haworth},
  {Hern{\'a}ndez}, {Kuhn}, {Perotti}, {Povich}, {Reiter}, {Roccatagliata},
  {Sabbi}, {Tabone}, {Winter}, {McLeod}, {van Boekel}, \& {van
  Terwisga}}]{2023ApJ...958L..30R}
{Ram{\'\i}rez-Tannus}, M.~C., {Bik}, A., {Cuijpers}, L., {et~al.} 2023, \apjl,
  958, L30

\bibitem[{{Rebolledo} {et~al.}(2016){Rebolledo}, {Burton}, {Green}, {Braiding},
  {Molinari}, {Wong}, {Blackwell}, {Elia}, \& {Schisano}}]{2016MNRAS.456.2406R}
{Rebolledo}, D., {Burton}, M., {Green}, A., {et~al.} 2016, \mnras, 456, 2406

\bibitem[{{Sanchawala} {et~al.}(2007){Sanchawala}, {Chen}, {Lee}, {Chu},
  {Nakajima}, {Tamura}, {Baba}, \& {Sato}}]{2007ApJ...656..462S}
{Sanchawala}, K., {Chen}, W.-P., {Lee}, H.-T., {et~al.} 2007, \apj, 656, 462

\bibitem[{{Sander} {et~al.}(2019){Sander}, {Hamann}, {Todt}, {Hainich},
  {Shenar}, {Ramachandran}, \& {Oskinova}}]{2019A&A...621A..92S}
{Sander}, A.~A.~C., {Hamann}, W.~R., {Todt}, H., {et~al.} 2019, \aap, 621, A92

\bibitem[{{Sander} {et~al.}(1998){Sander}, {Ester}, {Kriegel}, \&
  {Xu}}]{1998DMKD....2..169S}
{Sander}, J., {Ester}, M., {Kriegel}, H.-P., \& {Xu}, X. 1998, Data Mining and
  Knowledge Discovery, 2, 169

\bibitem[{{Smith}(2006)}]{Smith06}
{Smith}, N. 2006, \mnras, 367, 763

\bibitem[{{Smith} \& {Brooks}(2007)}]{SB07}
{Smith}, N. \& {Brooks}, K.~J. 2007, \mnras, 379, 1279

\bibitem[{{Smith} \& {Brooks}(2008)}]{SB08}
{Smith}, N. \& {Brooks}, K.~J. 2008, ASP Monograph Publications, Vol.~5, {The
  Carina Nebula: A Laboratory for Feedback and Triggered Star Formation}, ed.
  B.~{Reipurth} (Astronomical Society of the Pacific), 138

\bibitem[{{Smith} {et~al.}(2000){Smith}, {Egan}, {Carey}, {Price}, {Morse}, \&
  {Price}}]{2000ApJ...532L.145S}
{Smith}, N., {Egan}, M.~P., {Carey}, S., {et~al.} 2000, \apjl, 532, L145

\bibitem[{{Smith} {et~al.}(2010){Smith}, {Povich}, {Whitney}, {Churchwell},
  {Babler}, {Meade}, {Bally}, {Gehrz}, {Robitaille}, \& {Stassun}}]{Smith10b}
{Smith}, N., {Povich}, M.~S., {Whitney}, B.~A., {et~al.} 2010, \mnras, 406, 952

\bibitem[{{Smith} {et~al.}(2005){Smith}, {Stassun}, \&
  {Bally}}]{2005AJ....129..888S}
{Smith}, N., {Stassun}, K.~G., \& {Bally}, J. 2005, \aj, 129, 888

\bibitem[{{Strawn} {et~al.}(2023){Strawn}, {Richardson}, {Moffat}, {Ibrahim},
  {Lane}, {Pickett}, {Chen{\'e}}, {Corcoran}, {Damineli}, {Gull}, {Hillier},
  {Morris}, {Pablo}, {Thomas}, {Stevens}, {Teodoro}, \&
  {Weigelt}}]{2023MNRAS.519.5882S}
{Strawn}, E., {Richardson}, N.~D., {Moffat}, A. F.~J., {et~al.} 2023, \mnras,
  519, 5882

\bibitem[{{Tarricq} {et~al.}(2021){Tarricq}, {Soubiran}, {Casamiquela},
  {Cantat-Gaudin}, {Chemin}, {Anders}, {Antoja}, {Romero-G{\'o}mez},
  {Figueras}, {Jordi}, {Bragaglia}, {Balaguer-N{\'u}{\~n}ez}, {Carrera},
  {Castro-Ginard}, {Moitinho}, {Ramos}, \& {Bossini}}]{2021A&A...647A..19T}
{Tarricq}, Y., {Soubiran}, C., {Casamiquela}, L., {et~al.} 2021, \aap, 647, A19

\bibitem[{{Townsley} {et~al.}(2011){Townsley}, {Broos}, {Corcoran},
  {Feigelson}, {Gagn{\'e}}, {Montmerle}, {Oey}, {Smith}, {Garmire}, {Getman},
  {Povich}, {Remage Evans}, {Naz{\'e}}, {Parkin}, {Preibisch}, {Wang}, {Wolk},
  {Chu}, {Cohen}, {Gruendl}, {Hamaguchi}, {King}, {Mac Low}, {McCaughrean},
  {Moffat}, {Oskinova}, {Pittard}, {Stassun}, {ud-Doula}, {Walborn}, {Waldron},
  {Churchwell}, {Nichols}, {Owocki}, \& {Schulz}}]{CCCP-intro}
{Townsley}, L.~K., {Broos}, P.~S., {Corcoran}, M.~F., {et~al.} 2011, \apjs,
  194, 1

\bibitem[{{Ulla} {et~al.}(2022){Ulla}, {Creevey}, {{\'A}lvarez}, {Andrae},
  {Bailer-Jones}, {Bellas-Velidis}, {Brugaletta}, {Carballo}, {Dafonte},
  {Delchambre}, {Dharmawardena}, {Drimmel}, {Fouesneau}, {Fr{\'e}mat},
  {Garabato}, {Hatzidimitriou}, {Heiter}, {Kordopatis}, {Korn}, {Lanzafame},
  {Lobel}, {Manteiga}, {Marshall}, {Pailler}, {Pallas-Quintela},
  {Recio-Blanco}, {Rybizki}, {Sarro Baro}, {Schultheis}, {Sordo}, {Soubiran},
  {Th{\'e}venin}, \& {Vallenari}}]{2022gdr3.reptE..11U}
{Ulla}, A., {Creevey}, O.~L., {{\'A}lvarez}, M.~A., {et~al.} 2022, {Gaia DR3
  documentation Chapter 11: Astrophysical parameters}, Gaia DR3 documentation,
  European Space Agency; Gaia Data Processing and Analysis Consortium.

\bibitem[{{van Leeuwen}(2009)}]{2009A&A...497..209V}
{van Leeuwen}, F. 2009, \aap, 497, 209

\bibitem[{Virtanen {et~al.}(2020)Virtanen, Gommers, Oliphant, Haberland, Reddy,
  Cournapeau, Burovski, Peterson, Weckesser, Bright, {van der Walt}, Brett,
  Wilson, Millman, Mayorov, Nelson, Jones, Kern, Larson, Carey, Polat, Feng,
  Moore, {VanderPlas}, Laxalde, Perktold, Cimrman, Henriksen, Quintero, Harris,
  Archibald, Ribeiro, Pedregosa, {van Mulbregt}, \& {SciPy 1.0
  Contributors}}]{2020SciPy-NMeth}
Virtanen, P., Gommers, R., Oliphant, T.~E., {et~al.} 2020, Nature Methods, 17,
  261

\bibitem[{{Wang} {et~al.}(2011){Wang}, {Feigelson}, {Townsley}, {Broos},
  {Getman}, {Wolk}, {Preibisch}, {Stassun}, {Moffat}, {Garmire}, {King},
  {McCaughrean}, \& {Zinnecker}}]{2011ApJS..194...11W}
{Wang}, J., {Feigelson}, E.~D., {Townsley}, L.~K., {et~al.} 2011, \apjs, 194,
  11

\bibitem[{{Ward} {et~al.}(2020){Ward}, {Kruijssen}, \&
  {Rix}}]{2020MNRAS.495..663W}
{Ward}, J.~L., {Kruijssen}, J.~M.~D., \& {Rix}, H.-W. 2020, \mnras, 495, 663

\bibitem[{{W}es {M}c{K}inney(2010)}]{mckinney-proc-scipy-2010}
{W}es {M}c{K}inney. 2010, in {P}roceedings of the 9th {P}ython in {S}cience
  {C}onference, ed. {S}t\'efan van~der {W}alt \& {J}arrod {M}illman, 56 -- 61

\bibitem[{{Winter} \& {Haworth}(2022)}]{2022EPJP..137.1132W}
{Winter}, A.~J. \& {Haworth}, T.~J. 2022, European Physical Journal Plus, 137,
  1132

\bibitem[{{Wolk} {et~al.}(2011){Wolk}, {Broos}, {Getman}, {Feigelson},
  {Preibisch}, {Townsley}, {Wang}, {Stassun}, {King}, {McCaughrean}, {Moffat},
  \& {Zinnecker}}]{2011ApJS..194...12W}
{Wolk}, S.~J., {Broos}, P.~S., {Getman}, K.~V., {et~al.} 2011, \apjs, 194, 12

\bibitem[{{Wright}(2020)}]{2020NewAR..9001549W}
{Wright}, N.~J. 2020, \nar, 90, 101549

\bibitem[{{Wright} {et~al.}(2015){Wright}, {Drew}, \&
  {Mohr-Smith}}]{2015MNRAS.449..741W}
{Wright}, N.~J., {Drew}, J.~E., \& {Mohr-Smith}, M. 2015, \mnras, 449, 741

\bibitem[{{Wright} {et~al.}(2024){Wright}, {Jeffries}, {Jackson}, {Sacco},
  {Arnold}, {Franciosini}, {Gilmore}, {Gonneau}, {Morbidelli}, {Prisinzano},
  {Randich}, \& {Worley}}]{2024MNRAS.533..705W}
{Wright}, N.~J., {Jeffries}, R.~D., {Jackson}, R.~J., {et~al.} 2024, \mnras,
  533, 705

\bibitem[{{Wright} {et~al.}(2023){Wright}, {Kounkel}, {Zari}, {Goodwin}, \&
  {Jeffries}}]{2023ASPC..534..129W}
{Wright}, N.~J., {Kounkel}, M., {Zari}, E., {Goodwin}, S., \& {Jeffries}, R.~D.
  2023, in Astronomical Society of the Pacific Conference Series, Vol. 534,
  Protostars and Planets VII, ed. S.~{Inutsuka}, Y.~{Aikawa}, T.~{Muto},
  K.~{Tomida}, \& M.~{Tamura}, 129

\bibitem[{{Zeidler} {et~al.}(2016){Zeidler}, {Preibisch}, {Ratzka},
  {Roccatagliata}, \& {Petr-Gotzens}}]{VISTA2}
{Zeidler}, P., {Preibisch}, T., {Ratzka}, T., {Roccatagliata}, V., \&
  {Petr-Gotzens}, M.~G. 2016, \aap, 585, A49

\bibitem[{{Zeidler} {et~al.}(2018){Zeidler}, {Sabbi}, {Nota}, {Pasquali},
  {Grebel}, {McLeod}, {Kamann}, {Tosi}, {Cignoni}, \&
  {Ramsay}}]{2018AJ....156..211Z}
{Zeidler}, P., {Sabbi}, E., {Nota}, A., {et~al.} 2018, \aj, 156, 211

\end{thebibliography}

\begin{appendix} 
\section{Kinematical analysis without correction factor}

Table~\ref{tab:compare} shows a comparison of the kinematical analysis from Sect.~\ref{sec:Expansion} with and without the correction factor. It can be seen that the correction factor has only a small influence on the end result.

\begingroup
\renewcommand{\arraystretch}{1.5}
\begin{table}[h]
     \caption{Comparison of kinematical results for clusters with and without correction factor.}
    \label{tab:compare}
     \resizebox{\hsize}{!}{\begin{tabular}{lcccccc}
    \hline \hline
     Cluster& \multicolumn{3}{c}{With correction} & \multicolumn{3}{c}{Without correction} \\
     & Angle [$\degr$]& \multicolumn{1}{c}{Expansion} &   Contraction &  Angle [$\degr$]& \multicolumn{1}{c}{Expansion} &   Contraction\\
     & &   Significance  & Significance &&  Significance & Significance\\
        \hline
NGC~3293 &38 & & $2.3\,\sigma$  & 38 & & $2.3\,\sigma$\\ 
Trumpler~14& 90 & $5.2\,\sigma$ & &90 &  $5.2\,\sigma$\\
Trumpler~16 & 168 & $2.8\,\sigma$& &  176 &  $2.8\,\sigma$\\
Trumpler 15 &  148 &  $0.9\,\sigma$& &149 &  $0.9\,\sigma$\\ 
Collinder~228 & 54 &  $0.8\,\sigma$&& 58 &  $0.8\,\sigma$\\ 
\end{tabular}}
\end{table}
\endgroup

\section{Supplemental plots}

\begin{figure*}
\centering
\begin{subfigure}[b]{4cm}
   \includegraphics[width=4.cm]{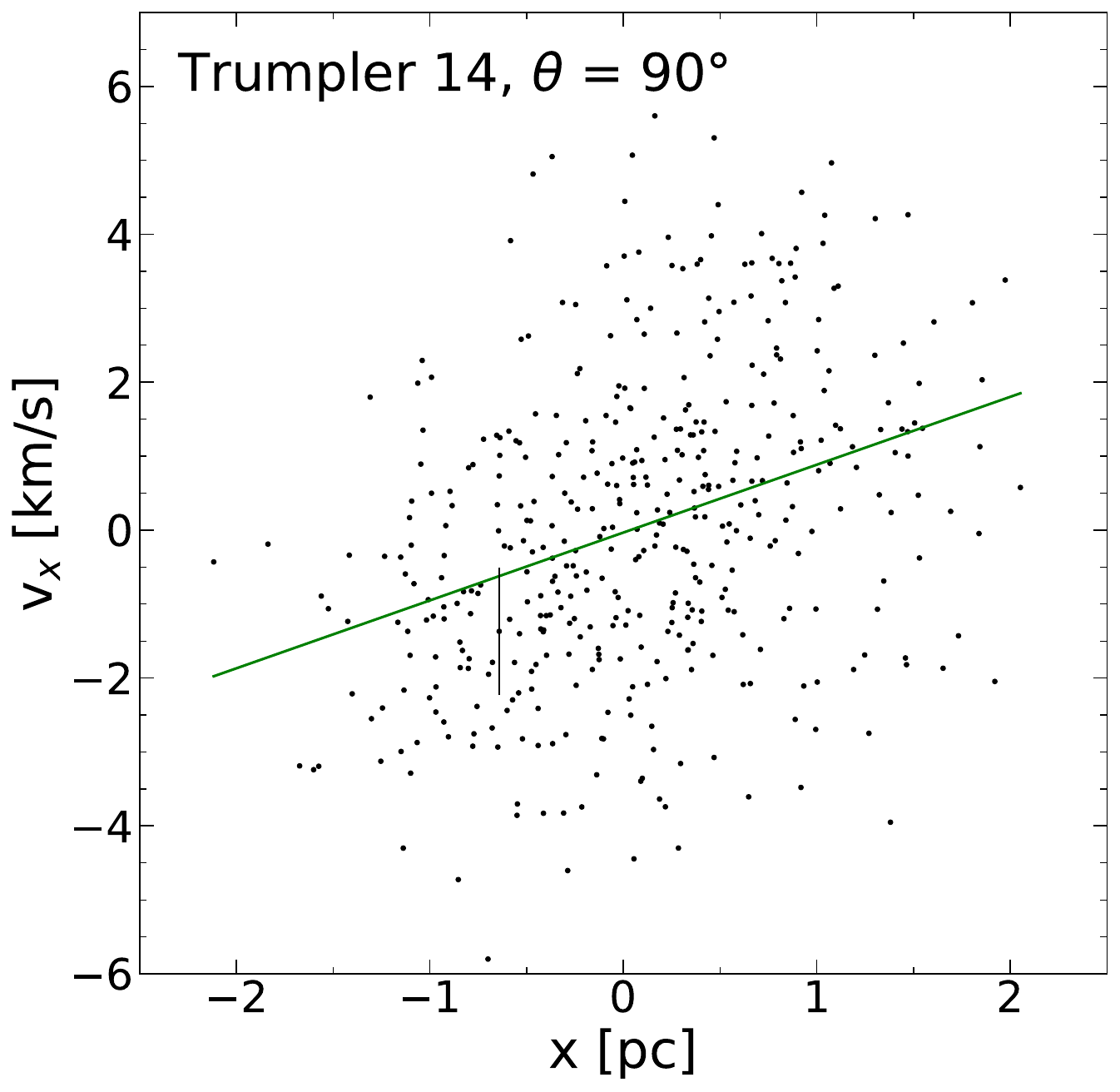}
\end{subfigure}
\begin{subfigure}[b]{4cm}
   \includegraphics[width=4cm]{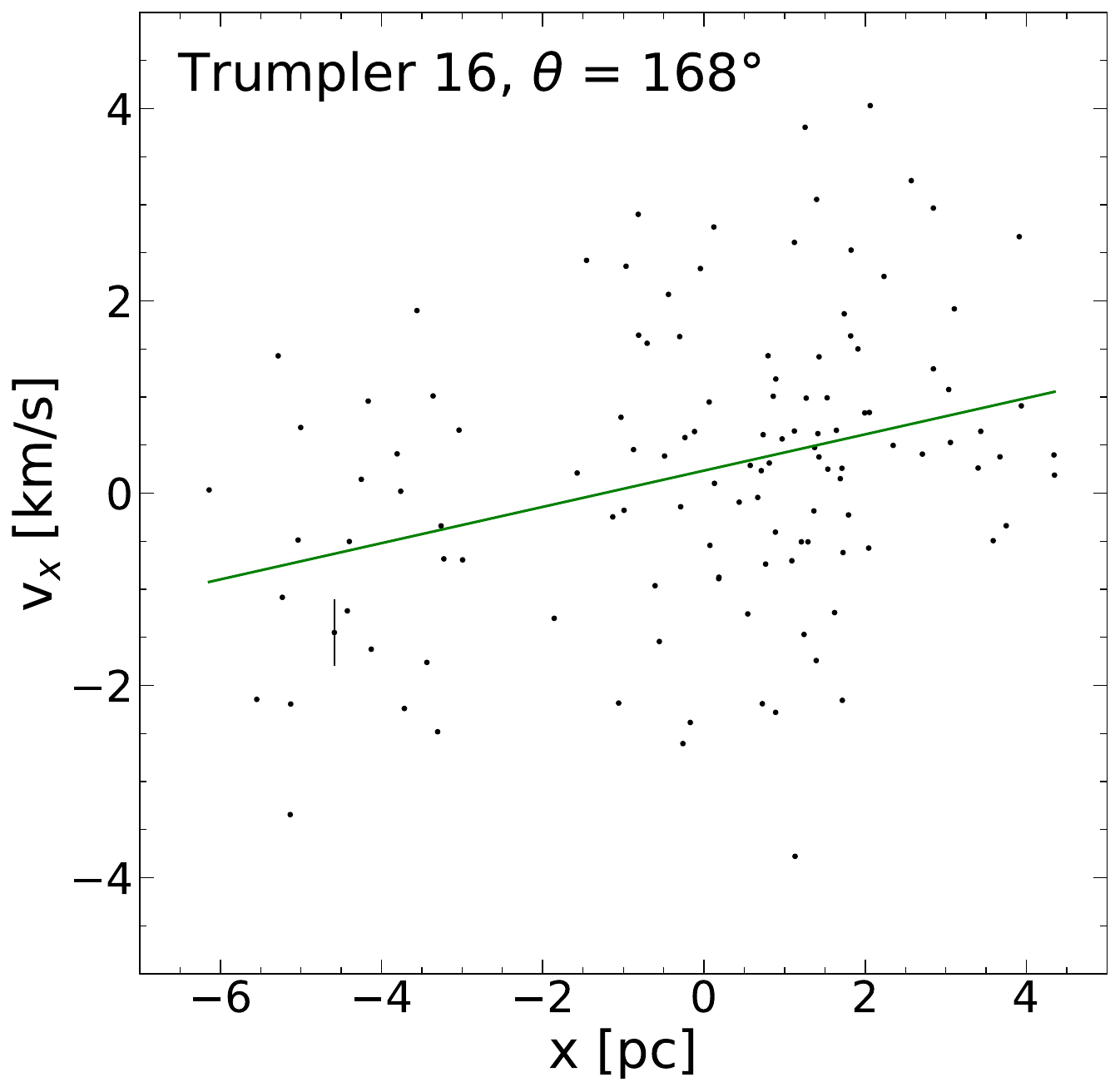}
\end{subfigure}
\begin{subfigure}[b]{4cm}
   \includegraphics[width=4cm]{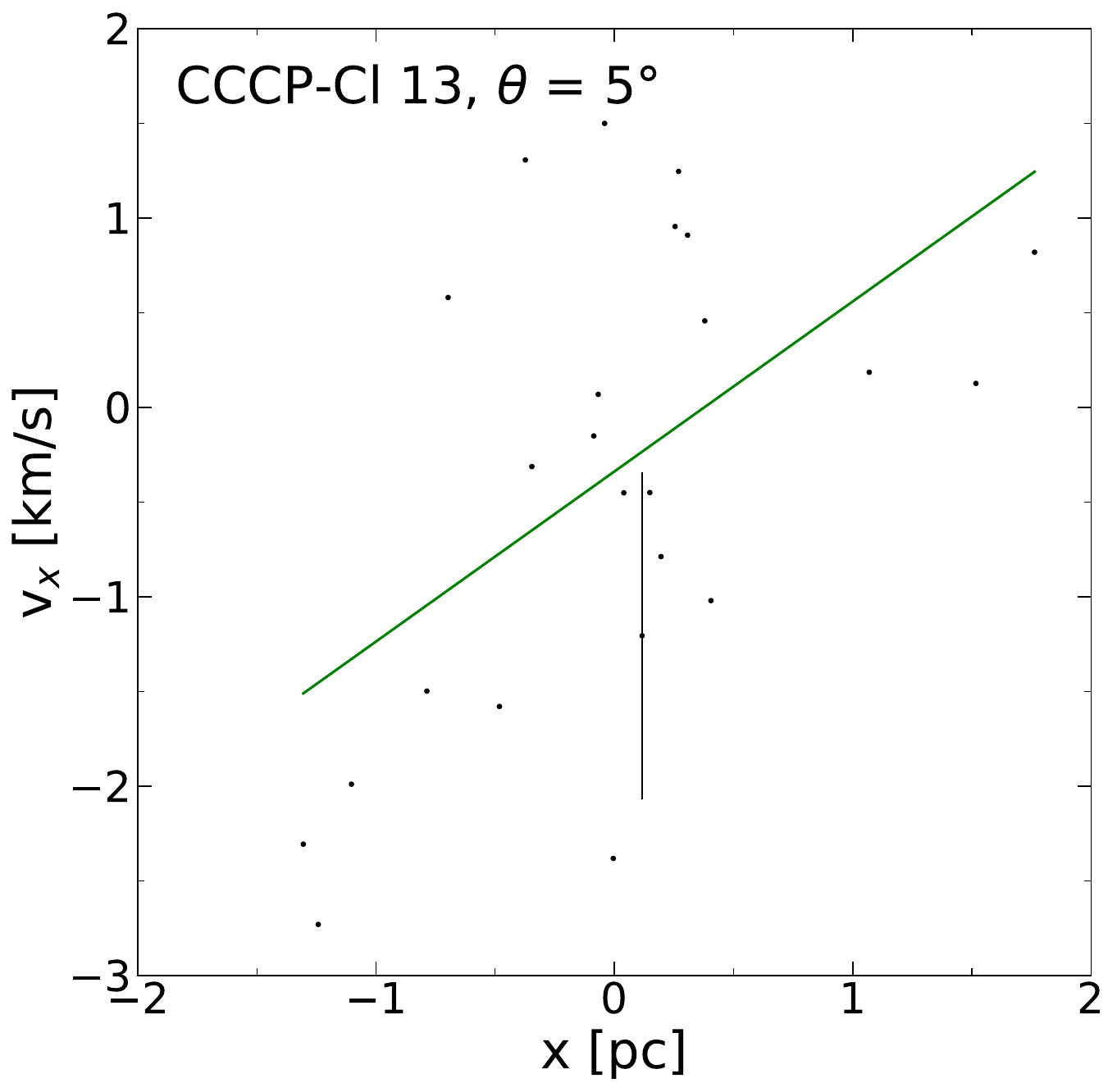}
\end{subfigure}
\begin{subfigure}[b]{4cm}
   \includegraphics[width=4cm]{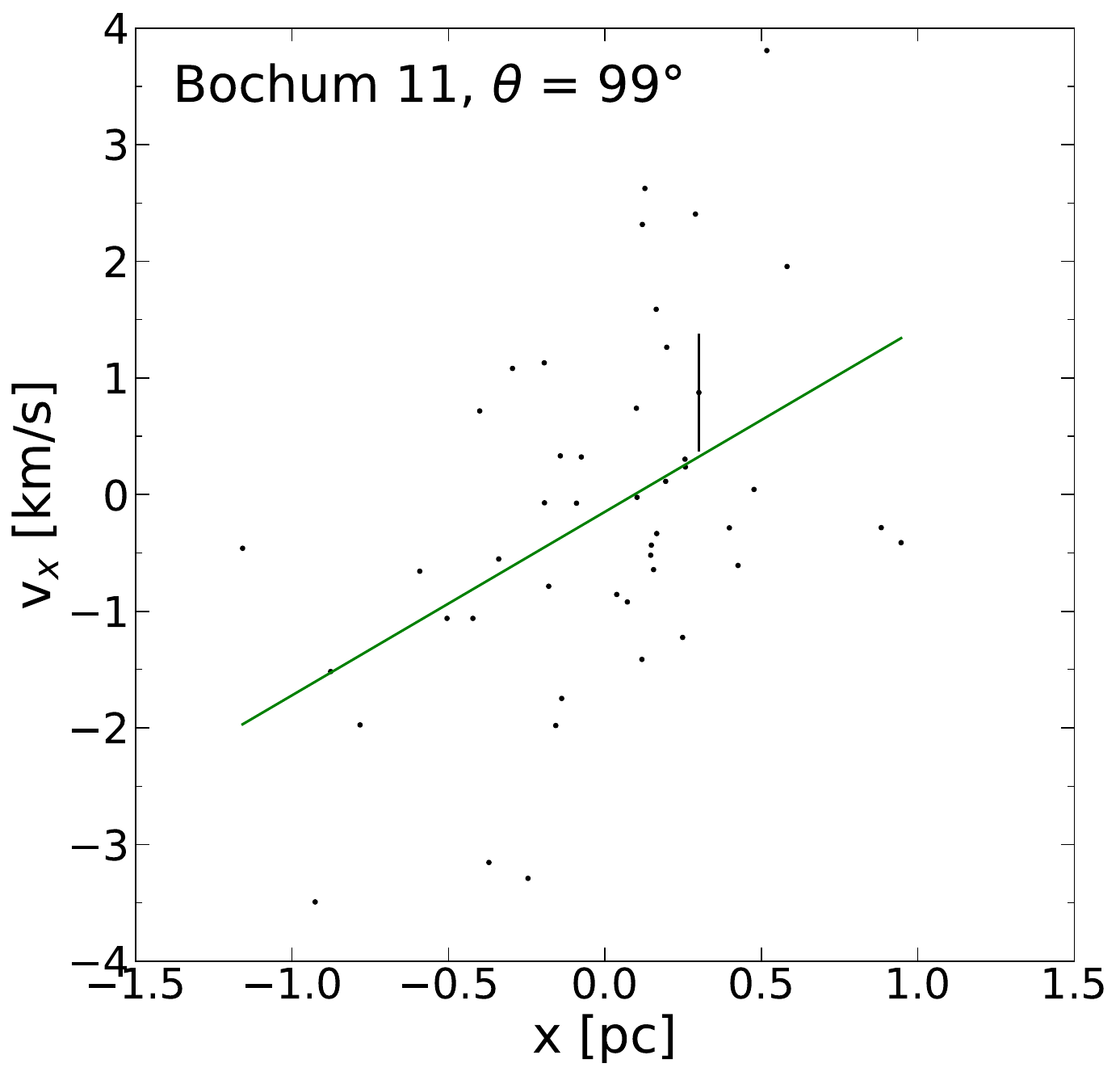}
\end{subfigure}
\begin{subfigure}[b]{4cm}
   \includegraphics[width=4cm]{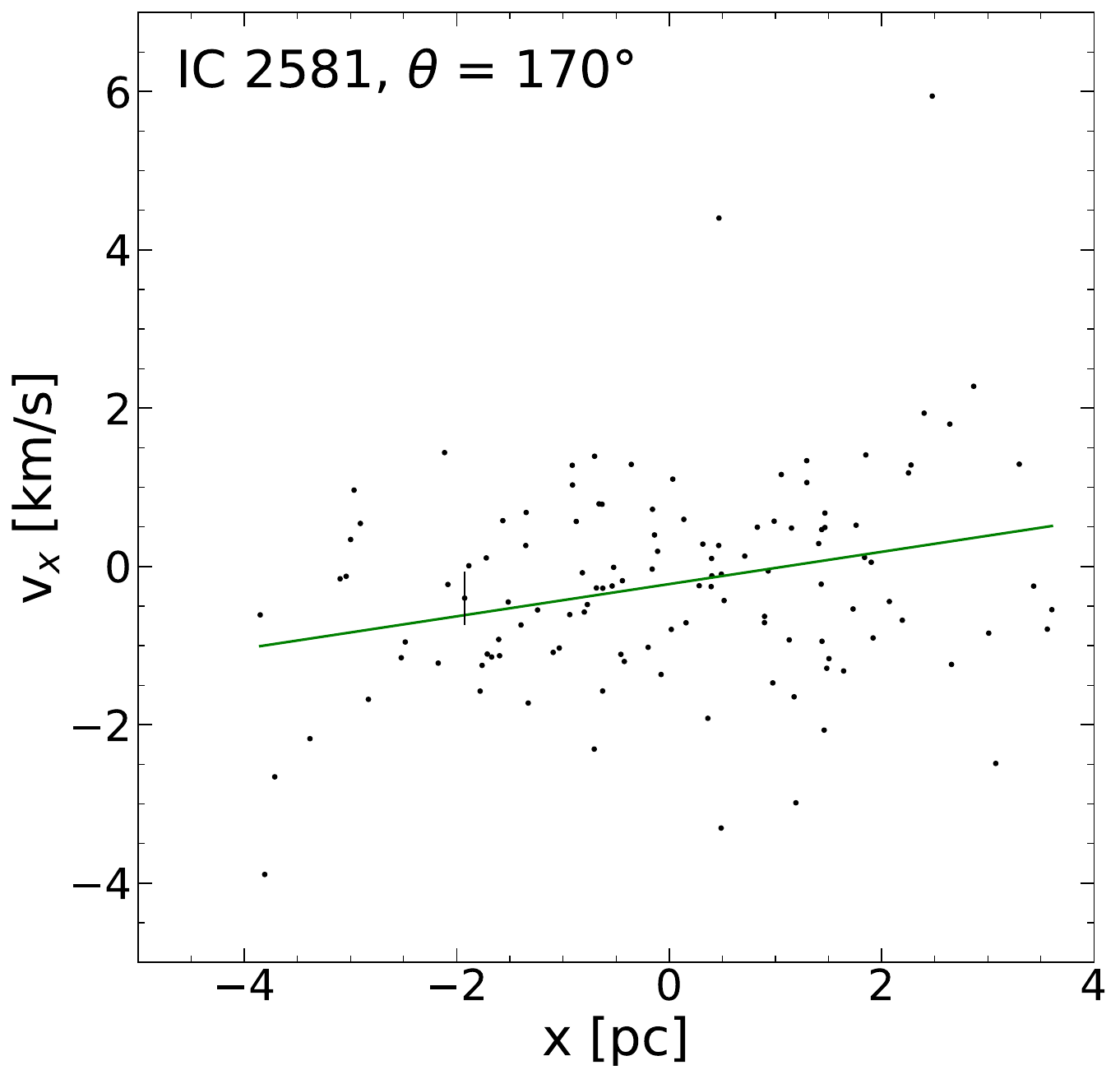}
\end{subfigure}
\begin{subfigure}[b]{4cm}
   \includegraphics[width=4cm]{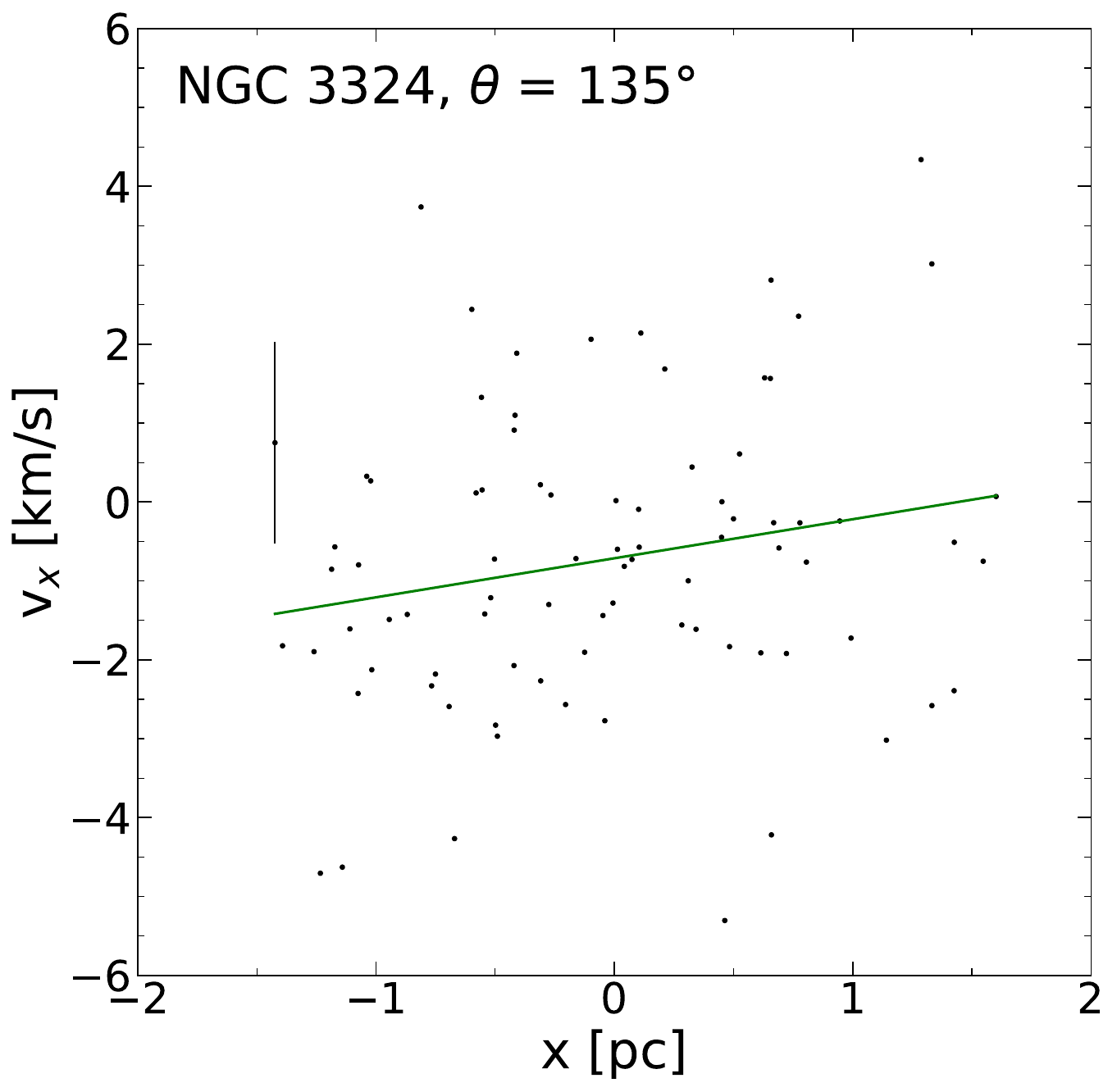}
\end{subfigure}
\begin{subfigure}[b]{4cm}
   \includegraphics[width=4cm]{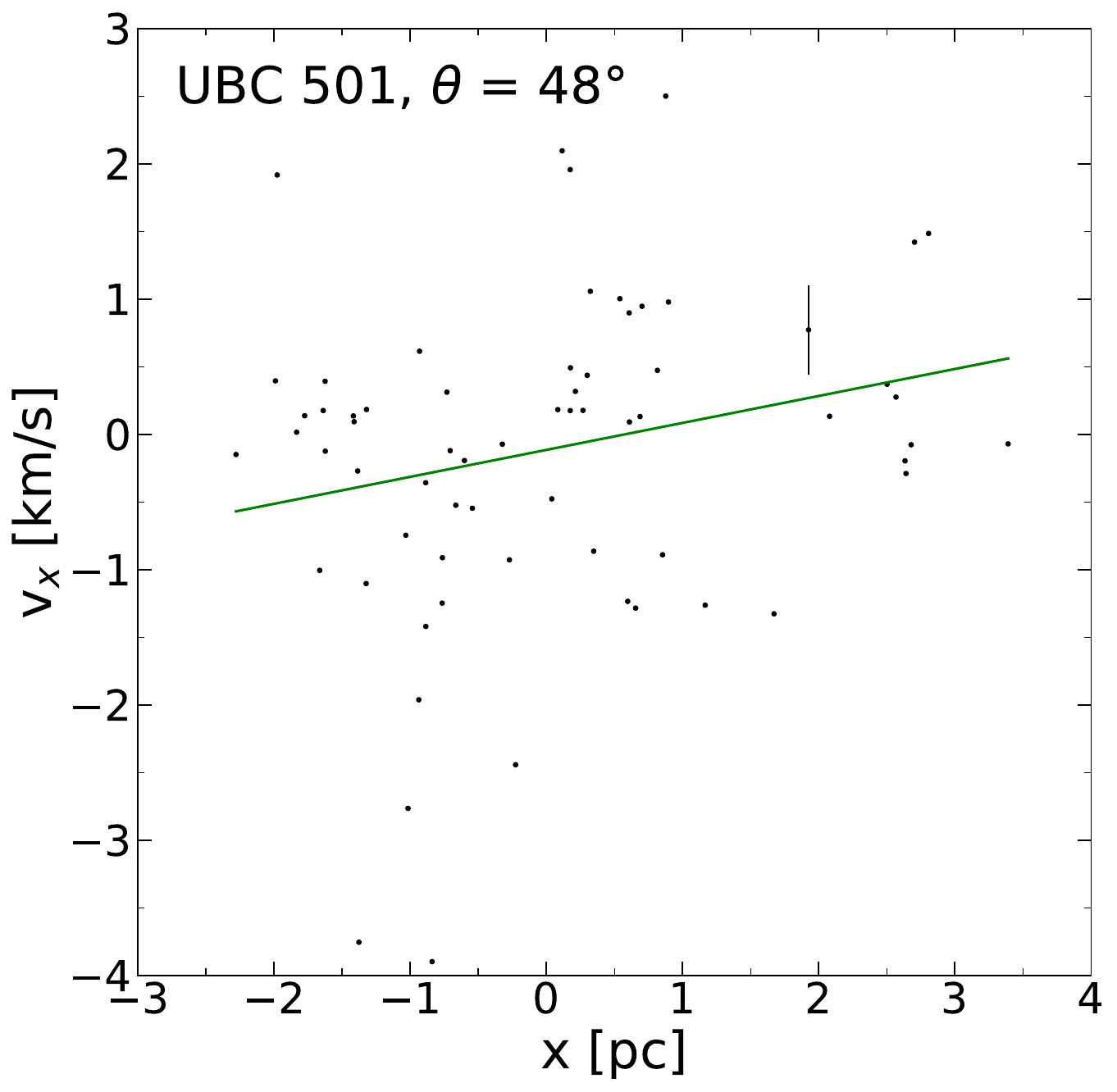}
\end{subfigure}
\caption{Cluster members' relative position versus relative velocity with its slope at the angle of highest significance. The vertical black line shows the cluster's median uncertainty in velocity. Panel (a): Trumpler~14, (b): Trumpler~16, (c): CCCP-Cl~13, (d): Bochum~11, (e): IC~2581, (f): NGC~3324, (g): UBC~501}
\label{fig:expansionplots2}
\end{figure*}

\begin{figure*} 
\centering
\begin{subfigure}[b]{4cm}
   \includegraphics[width=4cm]{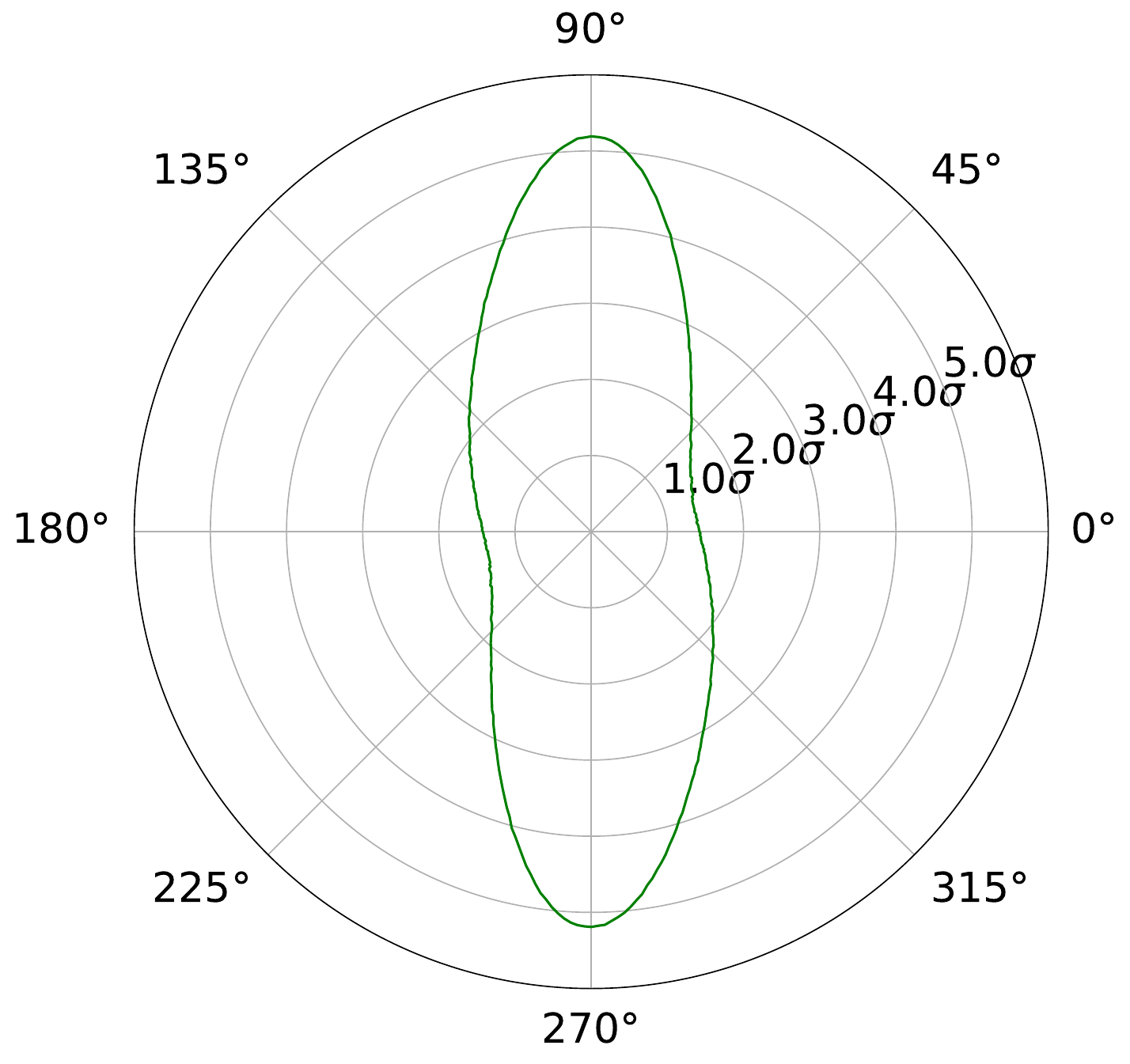}
\end{subfigure}
\begin{subfigure}[b]{4cm}
   \includegraphics[width=4cm]{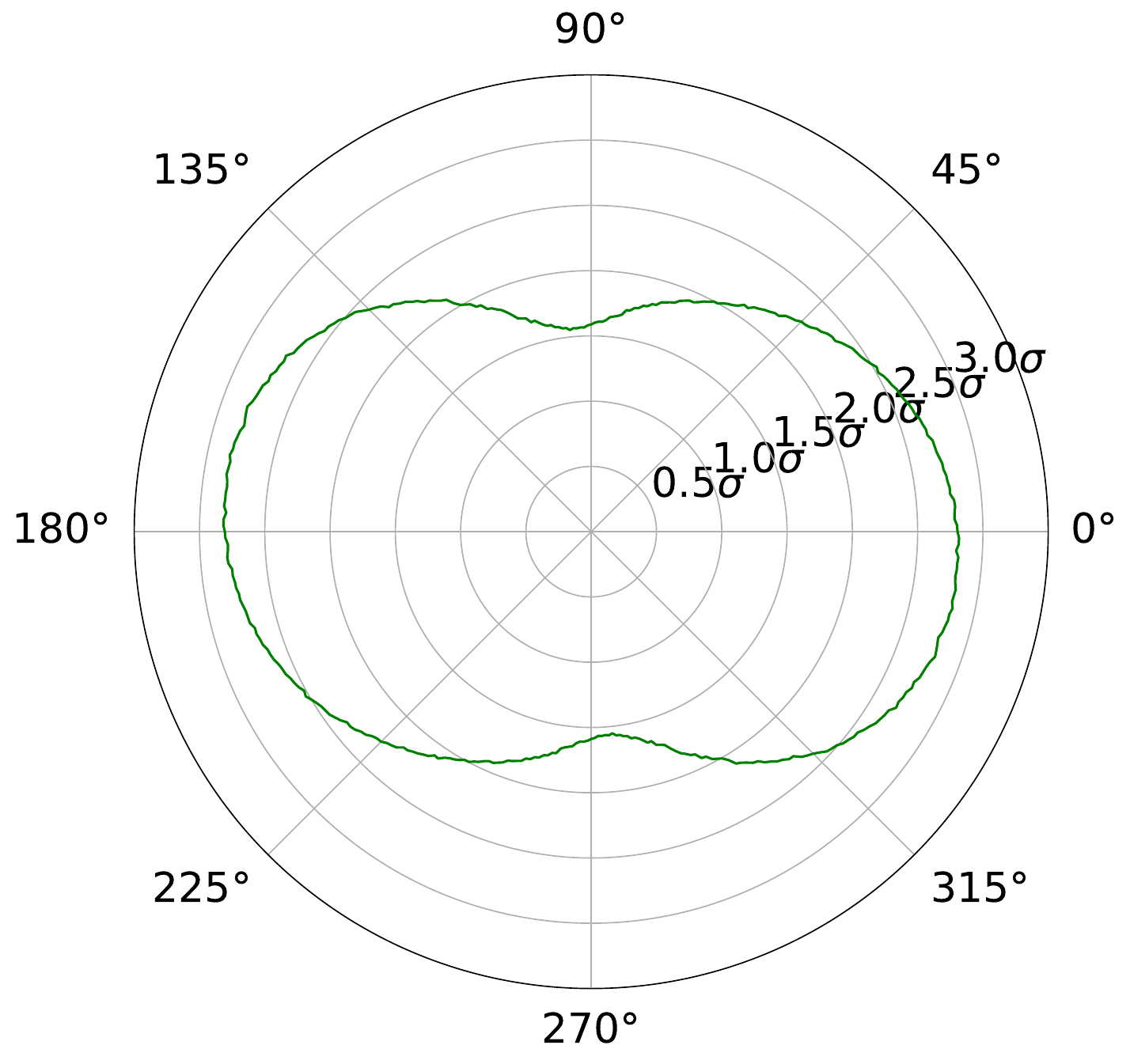}
\end{subfigure}
\begin{subfigure}[b]{4cm}
   \includegraphics[width=4cm]{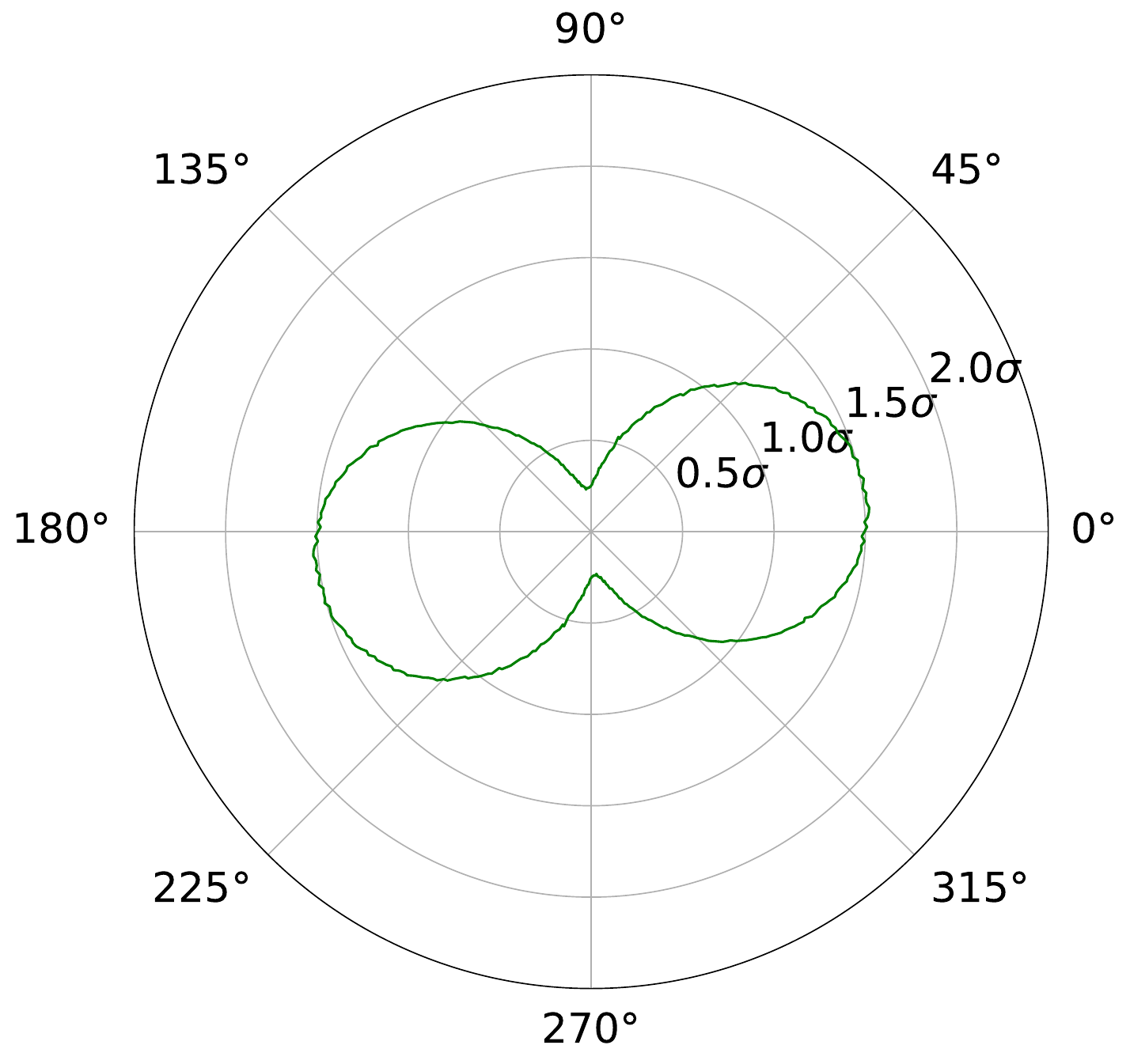}
\end{subfigure}
\begin{subfigure}[b]{4cm}
   \includegraphics[width=4cm]{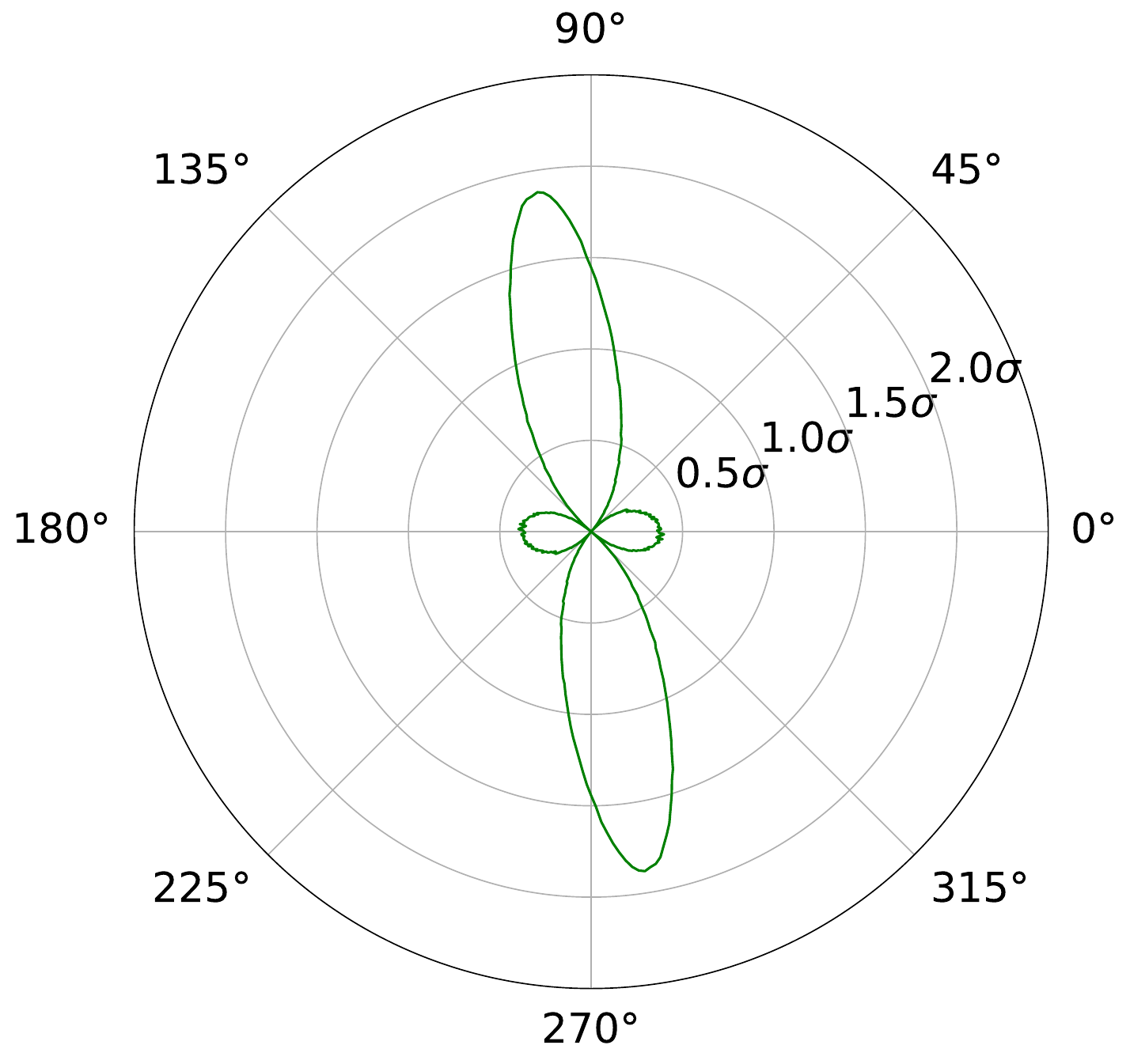}
\end{subfigure}
\begin{subfigure}[b]{4cm}
   \includegraphics[width=4cm]{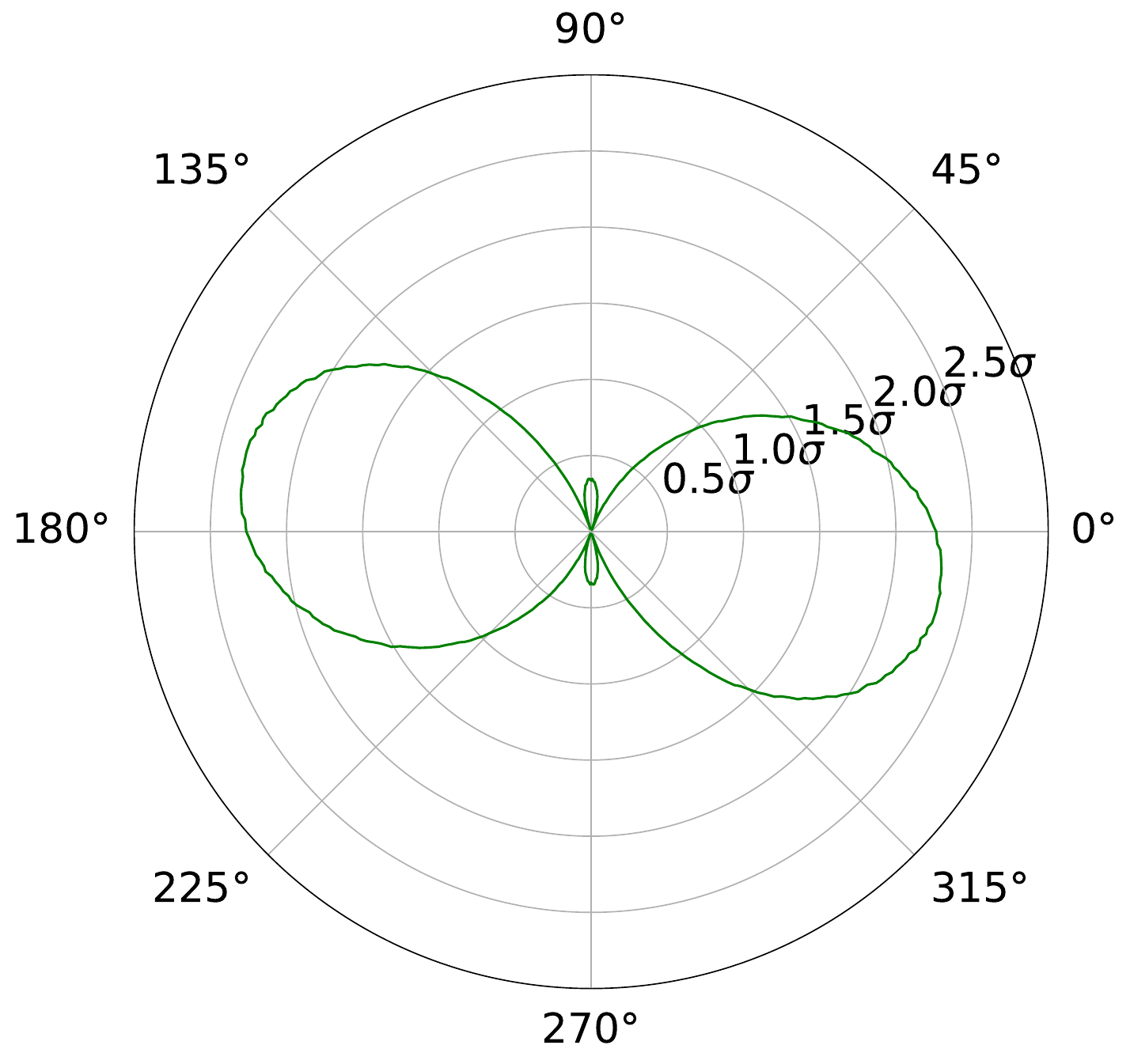}
\end{subfigure}
\begin{subfigure}[b]{4cm}
   \includegraphics[width=4cm]{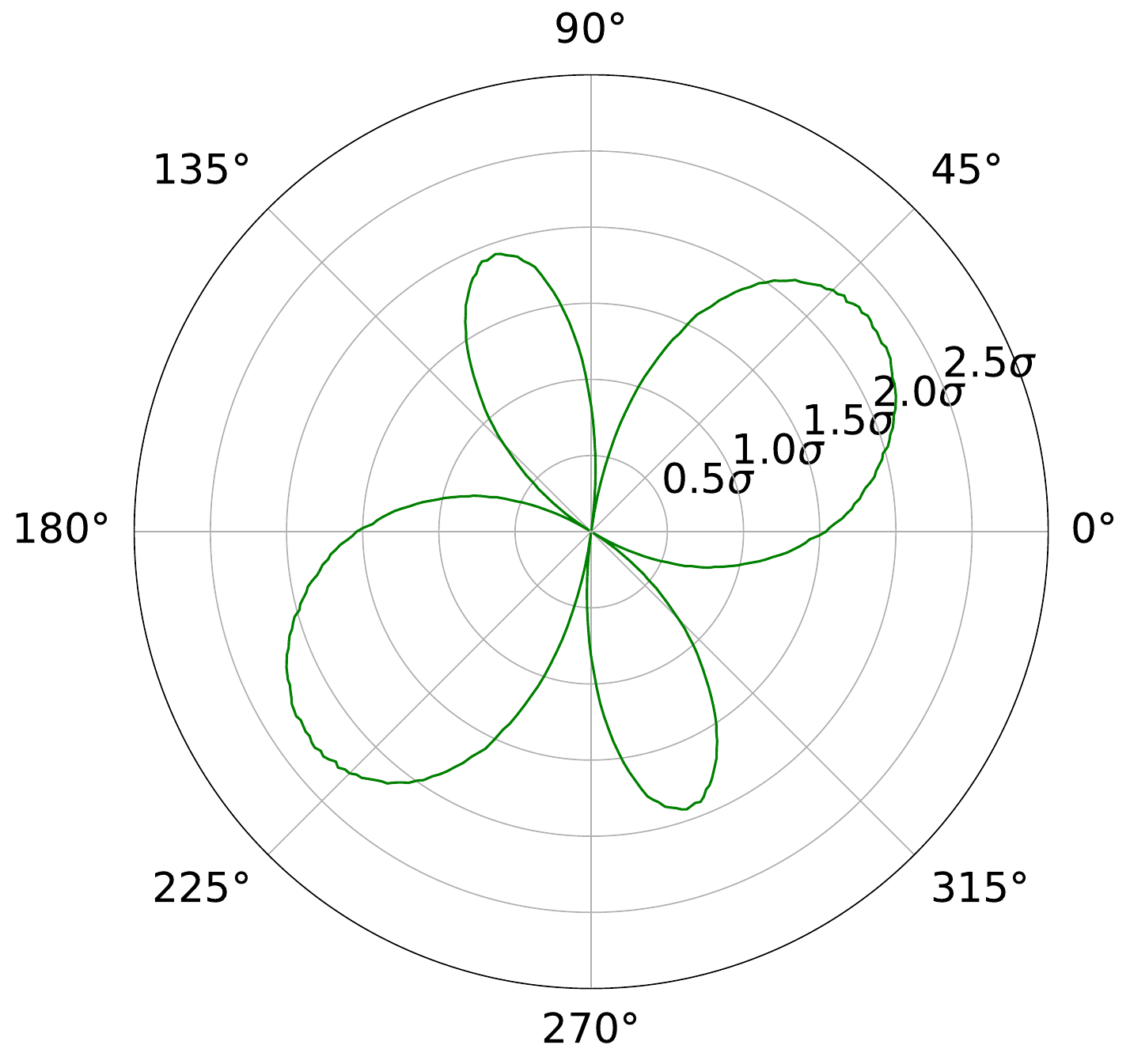}
\end{subfigure}
\begin{subfigure}[b]{4cm}
   \includegraphics[width=4cm]{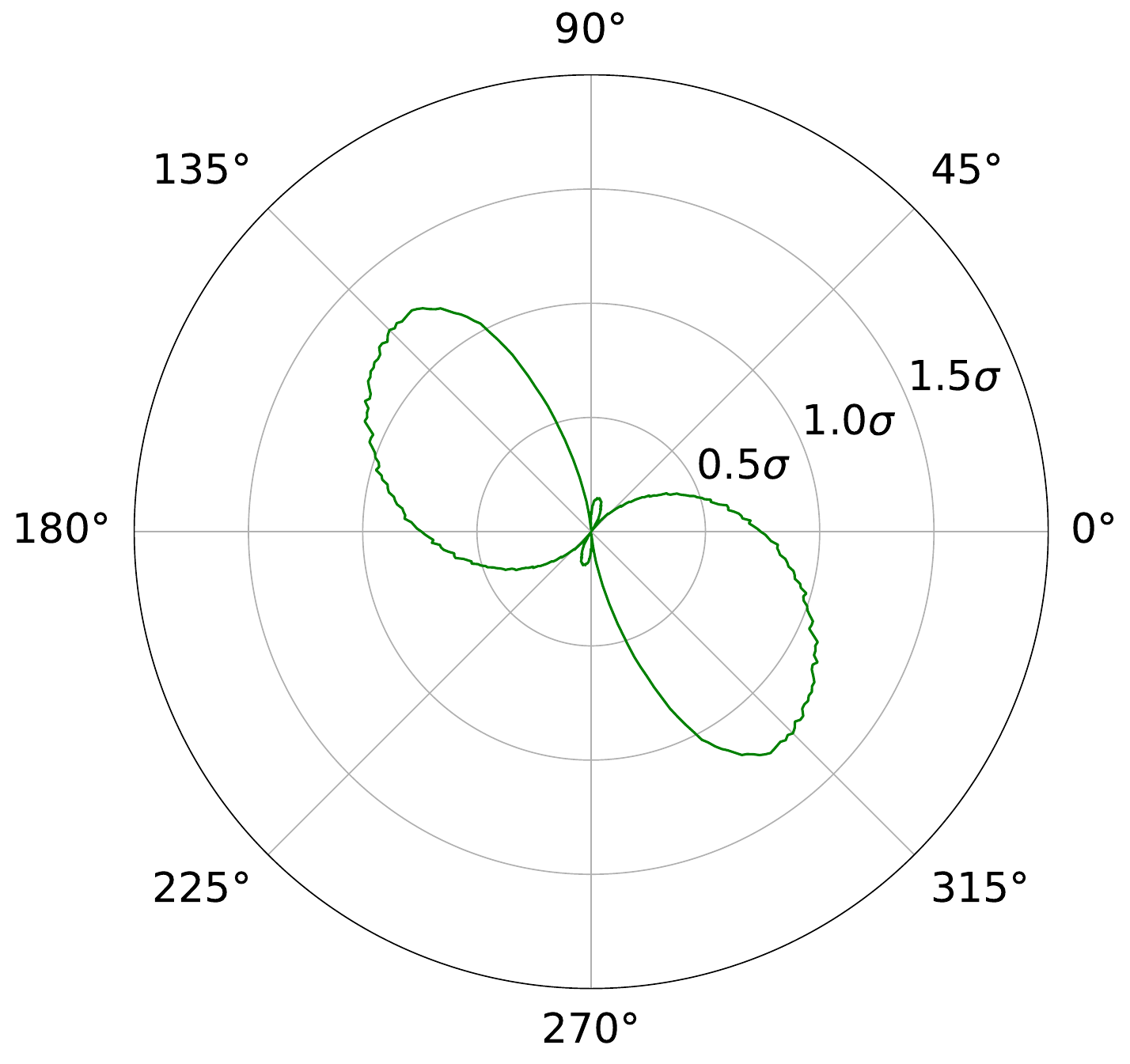}
\end{subfigure}
\begin{subfigure}[b]{4cm}
   \includegraphics[width=4cm]{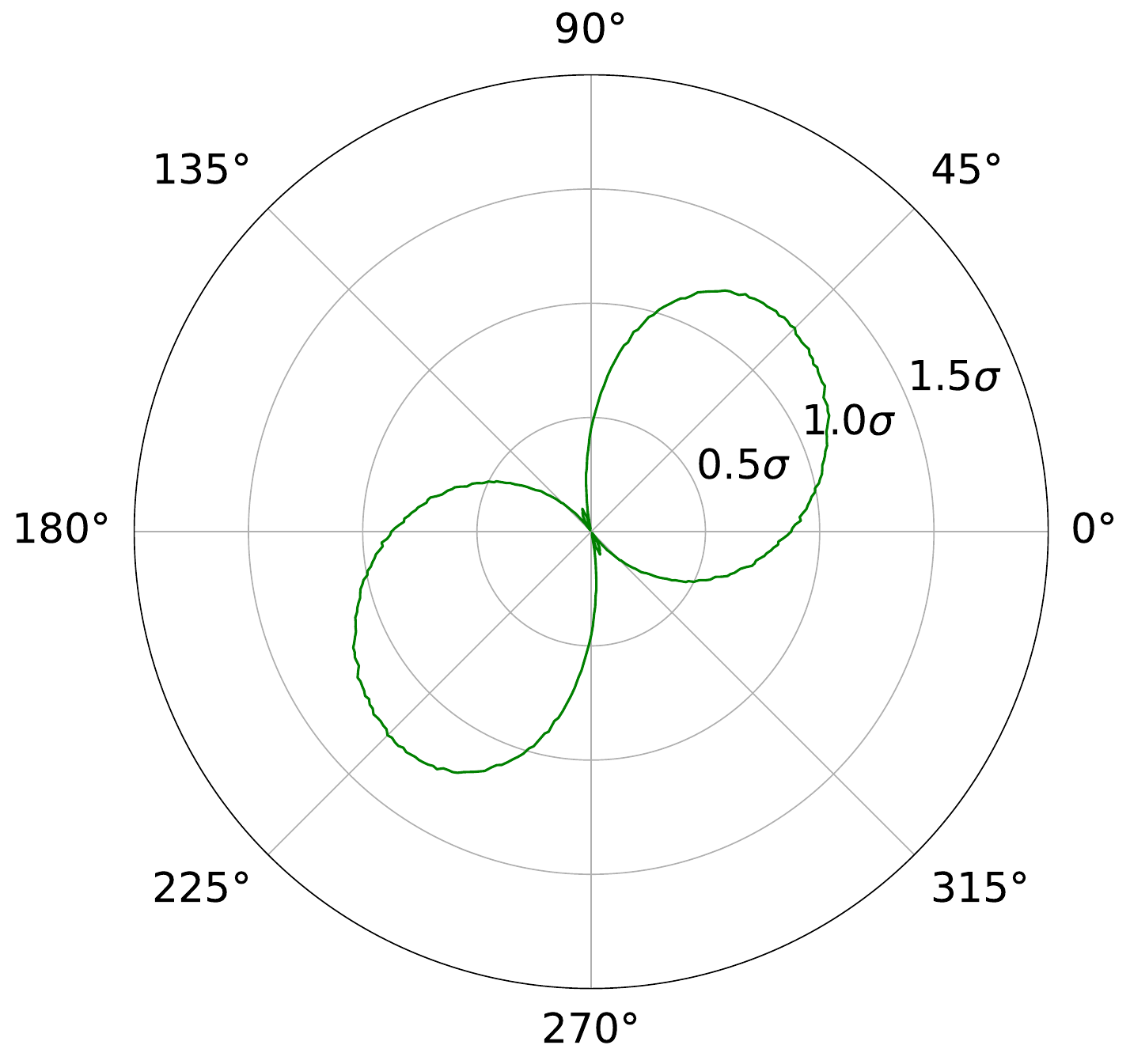}
\end{subfigure}
\caption{Significance of the correlation between position and velocity for angles between $0\degr$ and $359\degr$. Panel (a): Trumpler~14, (b): Trumpler~16, (c): CCCP-Cl~13, (d): Bochum~11, (e): IC~2581, (f): NGC~3293, (g): NGC~3324, (h): UBC~501.}
\label{fig:polarplots2}
\end{figure*}

\FloatBarrier

\section{Previews of the supplemental tables}

\begin{table*}[b]
\caption{Clusters found with DBSCAN in the region of the CNC. The clusters are listed with their member size, weighted position, weighted proper motion, weighted mean parallax with its uncertainty, mean \textit{Kalkyotl} distance with its central 68.3\% quantile, their counterpart in literature with reference, if the cluster is part of the Car~OB1, and if the cluster is substructure, if available. This table is available in its entirety in electronic form at the CDS.}
\resizebox{\hsize}{!}{\begin{tabular}{llccccccclcc}
\hline\hline
Cluster & NMem & RA & Dec & $\mu_\alpha^*$ [mas$\,\rm yr^{-1}$] & $\mu_\delta$ [mas$\,\rm yr^{-1}$]& $\varpi \pm \sigma_{\varpi}$ [mas] & $D_{Kalkayotl}$ [kpc] & Counterparts &Reference Counterpart&Part of Car~OB1? & Clusters associated\\
\hline
1& 11 &10:17:34.52& $-56$:52:40.9 &$ -5.800$&$ 3.267$ &$0.279\pm0.032$&3.706 [3.275,4.135] && & N\\
2& 15 &10:17:52.57& $-57$:15:53.0 &$ -5.435$&$ 3.600$ &$0.238\pm0.010$&4.253 [3.971,4.539] &&& N &3,8,10\\
3& 17 &10:17:53.14& $-57$:20:20.6 &$ -5.595$&$ 3.020$ &$0.252\pm0.011$&4.022 [3.773,4.271] &&& N &2,8,10\\
4& 10 &10:18:03.98& $-57$:25:52.3 &$ -5.556$&$ 3.431$ &$0.209\pm0.015$&4.884 [4.455,5.312] & &&  N\\
5& 10 &10:18:13.79& $-61$:40:20.0 &$ -3.740$&$ 2.443$ &$0.058\pm0.031$& & &&  N\\
6& 11 &10:18:14.06& $-57$:21:02.8 &$ -6.901$&$ 3.842$ &$0.350\pm0.014$&2.874 [2.739,3.009] && &N\\
7& 10 &10:18:18.93& $-59$:46:56.7 &$ -4.838$&$ 2.866$ &$0.115\pm0.033$& & & & N&2,3,10\\
8& 10 &10:18:20.41& $-57$:16:01.8 &$ -5.551$&$ 3.287$ &$0.273\pm0.020$&3.730 [3.423,4.031] && & N\\
9& 36 &10:18:21.48& $-59$:49:18.1 &$ -5.029$&$ 2.757$ &$0.252\pm0.007$&3.994 [3.778,4.211] & CWNU 2300&2023A\&A...673A.114H&  N\\
10& 28 &10:18:25.80& $-57$:19:03.5 &$ -5.529$&$ 3.178$&$0.225\pm0.014$ &4.530 [4.166,4.895] & HSC 2325 &2023A\&A...673A.114H&N&2,3,8\\
\hline
\end{tabular}}
\label{tab:AllClusters}
\end{table*}

\begin{table*}
\caption{Stars in our Car~OB1 high-mass star and OB star candidates sample. This table contains the stars' astrometric parameters, \texttt{teff\_esphs}, spectral type, to which population (clustered or distributed) a star belongs, and how the star was selected. The table is available in its entirety in electronic form at the CDS.}
\resizebox{\hsize}{!}{\begin{tabular}{llccccllcl} 
\hline\hline
Gaia DR3 Designation & Name  & RA & Dec & $\varpi \pm \sigma_{\varpi}$ [mas] & \texttt{teff\_esphs} [K]& Spectraltype & SpT Reference&Population & Selection\\\hline

5350357519345171200&HD 93162&10:44:10.37&$-59$:43:11.1&$0.456\pm0.020$&  &O2.5 If*/WN6&2014ApJS..211...10S&C&L\\ 
5254268071479968512&HD 93131&10:43:52.24&$-60$:07:04.0&$0.380\pm0.023$&  &WN6ha-w&2006A\&A...457.1015H&C&L\\ 
5350370026290390912&HD 92740&10:41:17.50&$-59$:40:36.8&$0.402\pm0.022$&  &WN7h + O9III-V&2006A\&A...457.1015H&D&L\\ 
5350357313186767104&HD 93205&10:44:33.72&$-59$:44:15.4&$0.443\pm0.025$&42623.5&O3.5 V((f)) + O8 V&2014ApJS..211...10S&C&L\\ 
5350383460949215232&HD 93250&10:44:45.01&$-59$:33:54.6&$0.424\pm0.020$&  &O4 III((fc))&2014ApJS..211...10S&C&L\\ 
5350358683250920704&HDE 303308&10:45:05.90&$-59$:40:05.9&$0.455\pm0.021$&47500.0&O4.5 V((fc))&2014ApJS..211...10S&C&L\\ 
5350357205782177664&HD 93204&10:44:32.32&$-59$:44:31.0&$0.436\pm0.023$&40080.8&O5.5 V((f))&2014ApJS..211...10S&C&L\\ 
5350357519345176192&ALS 15210&10:44:13.18&$-59$:43:10.2&$0.416\pm0.014$&  &O3.5 If*Nwk&2014ApJS..211...10S&C&L\\ 
5350356419833915904&CPD $-$59 2600&10:44:41.78&$-59$:46:56.4&$0.403\pm0.023$&  &O6 V((f))&2014ApJS..211...10S&C&L\\ 
5350358069101053184&CPD $-$559 2603&10:44:47.29&$-59$:43:53.2&$0.381\pm0.021$&31747.8&O7.5 V(n)z + B0 V(n)&2014ApJS..211...10S&C&L\\ 
\hline
\end{tabular}}
\label{tab:CarOB1HighMass}
\end{table*}

\end{appendix}
\end{document}